\documentclass[journal]{IEEEtran}
\usepackage{graphicx}
\usepackage{amsmath}
\usepackage{amssymb}
\usepackage{afterpage}
\usepackage{verbatim}
\usepackage{psfrag}
\usepackage{color}
\usepackage{cite}
\usepackage{setspace}
\usepackage{epsfig}
\usepackage{epstopdf}
\usepackage{footmisc}
\usepackage{fmtcount}
\usepackage{stackrel}
\usepackage{float}
\usepackage{breqn}
\usepackage{enumerate}
\usepackage{graphicx}
\usepackage{subcaption}

\begin{document}

\title{Underwater Optical Wireless Communications, Networking, and Localization: A Survey}

\author{Nasir Saeed,~\IEEEmembership{Member,~IEEE}, Abdulkadir Celik,~\IEEEmembership{Member,~IEEE},  Tareq Y. Al-Naffouri,~\IEEEmembership{Member,~IEEE}, Mohamed-Slim Alouini,~\IEEEmembership{Fellow,~IEEE}
\thanks{The authors are with the Department of Electrical Engineering, Computer Electrical and Mathematical Sciences \& Engineering (CEMSE) Division, King Abdullah University of Science and Technology (KAUST), Thuwal, Makkah Province, Kingdom of Saudi Arabia, 23955-6900.}
}
\maketitle{}
\begin{abstract}
Underwater wireless communications can be carried out through acoustic, radio frequency (RF), and optical waves. Compared to its bandwidth limited  acoustic and RF counterparts, underwater optical wireless communications (UOWCs) can support higher data rates at low latency levels. However, severe aquatic channel conditions (e.g., absorption, scattering, turbulence, etc.) pose great challenges for UOWCs and significantly reduce the attainable communication ranges, which necessitates efficient networking and localization solutions. Therefore, we provide a comprehensive survey on the challenges, advances, and prospects of underwater optical wireless networks (UOWNs) from a layer by layer perspective which includes: 1) Potential network architectures; 2) Physical layer issues including propagation characteristics, channel modeling, and modulation techniques 3) Data link layer problems covering link configurations, link budgets, performance metrics, and multiple access schemes; 4) Network layer topics containing relaying techniques and potential routing algorithms; 5) Transport layer subjects such as connectivity, reliability, flow and congestion control; 6) Application layer goals and state-of-the-art UOWN applications, and 7) Localization and its impacts on UOWN layers. Finally, we outline the open research challenges and point out the future directions for underwater optical wireless communications, networking, and localization research.
\end{abstract}


\maketitle
\section{Introduction}
\label{sec:intro}

According to a recent survey by the United States national oceanic and atmospheric administration, about 97 percent of the Earth's water covers the surface of the earth in the form of oceans \cite{noaa2017}. The early study of oceans (oceanography), extends back to tens of thousands of years, which includes acquiring the knowledge of ocean tides, currents, and waves. However, it was not until late 18th century that the British government announced an expedition to conduct appropriate oceans scientific investigation. The results of this expedition were published in 1882  as ``\textit{Report Of The Scientific Results of the Exploring Voyage of H.M.S. Challenger during the years 1873-76} \cite{‎Charles1873}." After this expedition number of books have been published on modern oceanography which includes `` \textit{Geography of the oceans}  \cite{Laloe2016}", ``\textit{Handbuch der Ozeanographie} \cite{Otto2014}", ``\textit{The depths of the oceans}\cite{Johan2016}", ``\textit{The Oceans} \cite{Prager2001}", ``\textit{The Sea} \cite{Hill2005}", and ``\textit{Encyclopedia of Oceanography} \cite{Rhodes1966}". More recently, there has been a growing interest to explore the underwater environment for numerous applications such as climate change, the study of oceanic animals, monitoring of oil rigs, surveillance, and unmanned operations. All of these applications require a medium to communicate in the underwater environment and to the outside world. In recent past years, the study of underwater wireless media has attracted much attention for underwater communications.

Today, underwater wireless communications (UWCs) are implemented using communication systems based on acoustic waves, radio frequency (RF) waves, and optical waves. Underwater acoustic wireless communications (UAWCs) have been one of the most used UWC technology as it provides communication over very long distances. In 1995, an UAWC system was proposed in  \cite{Stojanovic1995} with the data rate of 40 kbps. In 1996, an 8 kbps  UAWC system was developed for a depth of 20 m and horizontal distance of 13 km \cite{Zielinski1995}. In 2005, a more high-speed UAWC system was proposed in \cite{Hiroshi2005} which records a data rate of 125 kbps using 32 quadrature amplitude modulation technique (QAM) with symbol error rate of $10^{-4}$. Furthermore, a 60 kbps UAWC system was demonstrated in \cite{Song2013} using 32 QAM which can support communication over depth of 100 m and horizontal distance of 3 km. However, acoustic waves still have many drawbacks including scattering, high delay due to the low propagation speeds, high attenuation, low bandwidth, and bad impacts on the underwater mammals and fishes.

To alleviate the insufficient data rate of UAWC systems, research has been carried out in the past to use low frequency RF waves, e.g., the authors in \cite{Moore1967} proposed microwaves based wireless communication system over the surface of the ocean water which can transmit data over tens of kilometers.  An underwater microwaves based wireless communication system was employed in \cite{Shamma2004}, which can communicate over a horizontal distance of 85 m. A similar approach was followed in \cite{Shaw2006} with the data rate of 500 kbps over a horizontal distance of 90 m. The authors in  \cite{Uribe2009} have improved the capacity of underwater microwaves based wireless communication system further to 10 Mbps over the distance of 100 m. However, RF waves including microwaves suffer from serious attenuation in water, e.g., the attenuation in the ocean is about 169 dB/m for 2.4 GHz band while the attenuation in freshwater is much higher, i.e., 189 dB/m \cite{Zeng2017}. Moreover, RF based UWC requires huge antennas and is limited to the shallow areas of the sea. On the other hand, operating at ultra-low frequencies yields reduced attenuation levels, at the expense of high hardware costs and low data rates.  

\begin{figure}
\begin{center}  
\includegraphics[width=1\columnwidth]{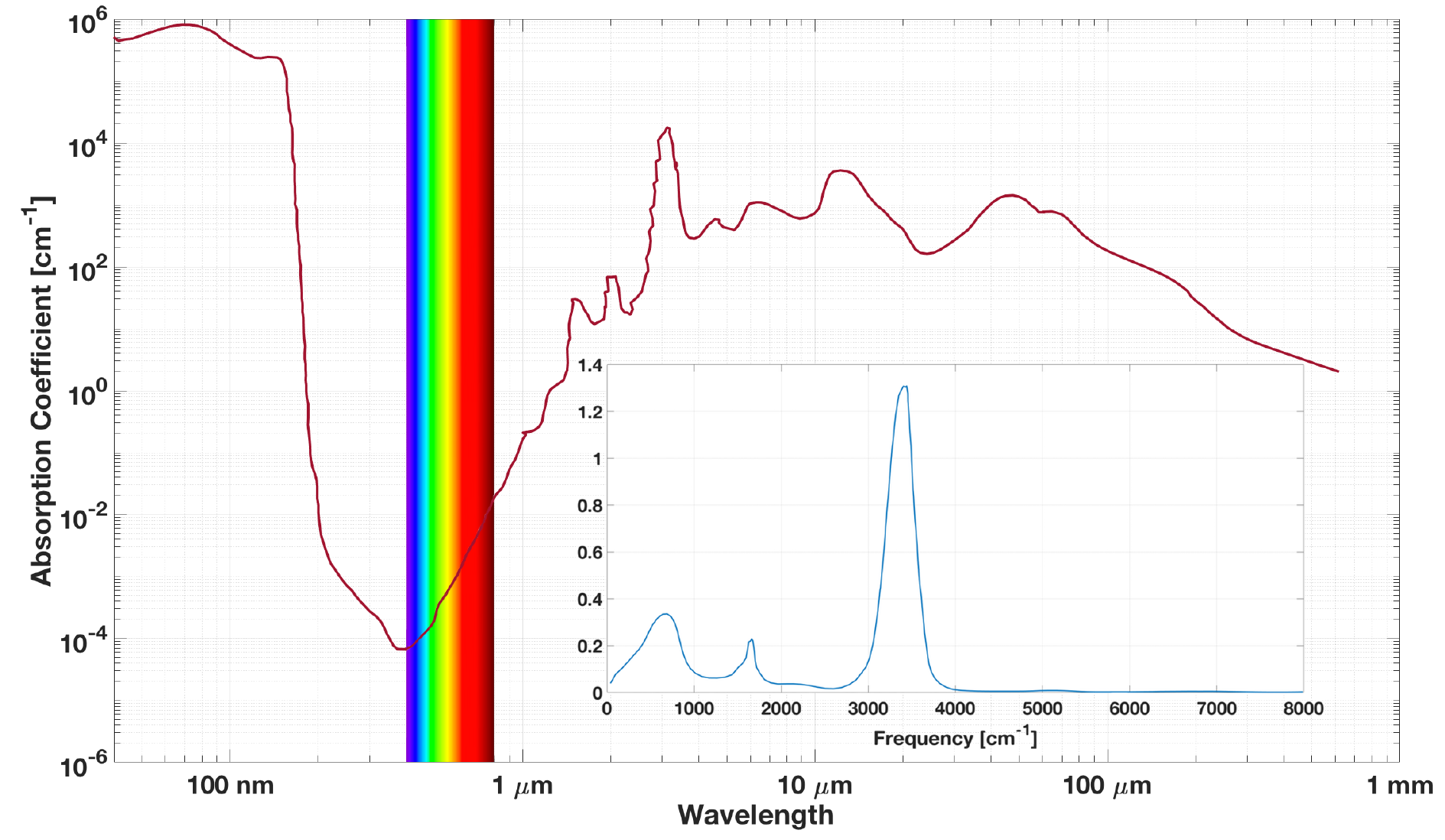}  
\caption{Attenuation of optical waves in aquatic medium. \label{absorption}}  
\end{center}  
\end{figure}

\begin{figure*}[htb!]
\begin{center}  
\includegraphics[width=2\columnwidth]{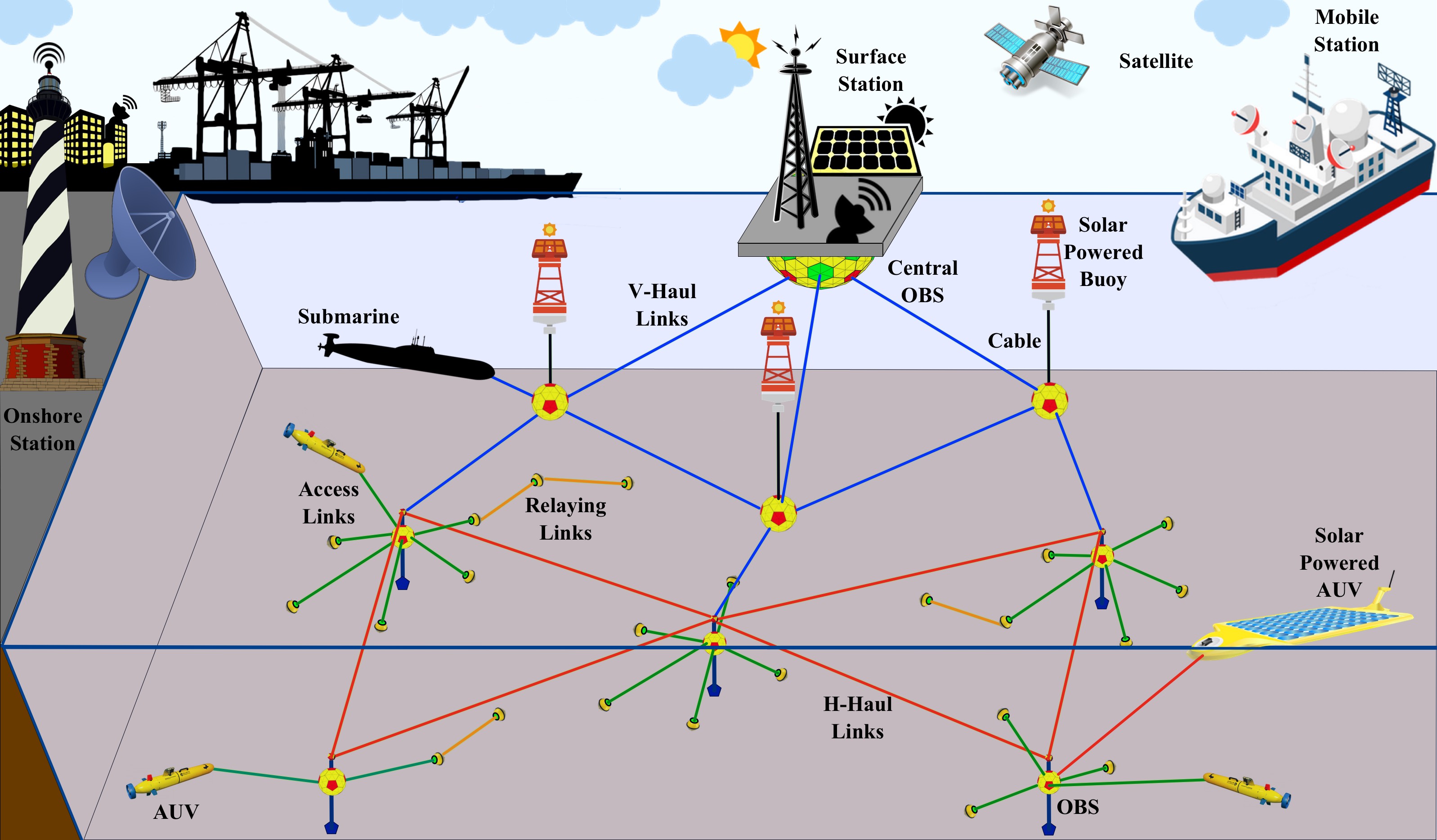}  
\caption{Illustration of a generic underwater optical wireless network (UOWN) architecture.\label{fig:uwsn}}  
\end{center}  
\end{figure*}

Due to the limitations of low bandwidth and low data rate of underwater acoustic and RF waves, an alternative approach is to use optical waves which can provide high-speed underwater optical wireless communication (UOWC) at low latencies in return for a limited communication range. Underwater propagation of optical waves also exhibits distinctive characteristics in different wavelengths as shown in Fig.~\ref{absorption}. In 1963, the authors in \cite{Duntley:63} found that attenuation within the range of 450-550 nm wavelengths (blue and green lights) is much smaller compared to the other wavelengths. In 1966, Gilbert et al. \cite{Gilbert1966} experimentally confirmed this behavior of optical waves, which provided the foundation of UOWC systems. The research on UOWC is mainly focused on increasing the transmission range and data rate of UOWC systems. The trend to improve the data rate of UOWC systems by using light emitting diodes (LEDs) has been followed in \cite{Hanson2008, Shlomi2010, Doniec2013, Jaffe2015, XU2016, Wang2018, Cossu2018}.

All of these LED-based UOWC systems have insufficient bandwidth, thus have low achievable data rate and low transmission distance. Therefore, laser-based UOWC systems were proposed in \cite{Atef2012, Cochenour2013, Tsonev2015} which provide large bandwidth and high-speed data rate. A green laser with 532 nm wavelength was employed in \cite{Hanson2008} to provide a UOWC link which covers a distance of 2 m with the data rate of 1 Gbps. In 2015, the authors in \cite{Nakamura2015} used a blue laser with 405 nm wavelength to provide a UOWC link with 1.45 Gbps and transmission distance of 4.8 m. To further improve the transmission distance and data rate, the authors in \cite{Oubei2015} and \cite{Najda2016} employed a UOWC link with 2.3 Gbps and 2.488 Gbps over a transmission distance of 7 m and 1 m, respectively. Subsequently, the authors in \cite{Oubei22015} demonstrated a UOWC system with the data rate of 4.8 Gbps using 16 quadrature amplitude modulation-orthogonal frequency division multiplexing (QAM-OFDM). A UOWC system with the data rate of 4.88 Gbps was proposed in \cite{Xu22016} by using 32 QAM-OFDM to get the transmission distance of 6 m. Recently, a 7.2 Gbps UOWC system has been proposed in \cite{Tsai2017} for 450 nm blue laser with the transmission distance of 6 m. Table \ref{Tableuwcsystems} summarizes the comparison between three different kinds of underwater wireless communication systems.

Optical waves have the advantage of higher data rate, low latency, and power efficiency at the expense of  limited communication ranges. Therefore, networking  solutions are crucial for mitigating range related deficiencies in order to employ optical waves for underwater wireless communications applications. Furthermore, accurate and precise localization schemes are also essential for developing effective networking protocols. We should also note that some application types heavily depend upon the sensing location since the obtained measurements are meaningful only if it refers to an accurate location. Localization in terrestrial wireless networks has been studied widely and detailed surveys are presented on this topic \cite{Santosh2006, MAO2007, Kulaib2011, Nabil2013, Kuriakose2014, Khan2017}. However, global positioning system (GPS) and RF-based localization schemes cannot work in the underwater environment as a result of hostile aquatic channel conditions. Thus, many researchers developed localization schemes for the underwater environment based on acoustic waves. Localization of underwater acoustic networks have also been studied widely in the past and a number of surveys are written on this subject \cite{Chandrasekhar2006, kantarci2011survey, TAN20111663, Tuna2017, Luo2018}.  Since the underwater optical wireless channel poses new challenges, the existing localization techniques used for terrestrial wireless networks and underwater acoustic networks are not directly applicable to underwater optical wireless networks (UOWNs). Therefore, novel time of arrival (ToA) and received signal strength (RSS) based distributed localization schemes are developed in \cite{Akhoundi2017underwater} for UOWNs.  Recently, RSS based centralized localization schemes for UOWNs are proposed in \cite{Nasir2018limited, Saeed2017, Nasir2018twc}.


\subsection{Related Surveys on UOWNs}
\label{sec:related}
With the increasing demands for UOWN applications, quite a few brief survey articles have been published to discuss  physical layer aspects of UOWNs.  A recent survey on UOWC systems has been proposed in \cite{Zeng2017}, where the authors have discussed different modulation schemes, channel models, link management, and coding techniques along with the possible practical implementations of UOWC systems. The link performance of UOWC systems was evaluated in \cite{Shlomi2010} and the authors have introduced different challenges associated with link developments of UOWC systems. In  \cite{Kaushal2016underwater}, the authors have reviewed UOWC systems in terms of modulation schemes, channel models, and coding schemes. The channel models of UOWC systems have also been surveyed in \cite{Johnson2013} and \cite{Johnson2014}, where the authors have considered vector radiative transfer theory, variable water composition, and inherent properties of light. The inherent features of underwater wireless communications including UOWC have been briefly surveyed in \cite{Camila2016}. The recent advances in system analysis and channel modeling of UOWC systems have been summarized in \cite{Gkoura2017}. In \cite{Maria2017} the future vision of UOWC systems and some of its challenges were presented. 


Unlike the surveys above which mainly tackle the physical layer aspects of UOWNs, this paper provides a comprehensive survey on the challenges, advances, and prospects of UOWNs from a layer by layer perspective which includes: 1) Potential network architectures; 2) Physical layer issues including propagation characteristics, channel modeling, and modulation techniques; 3) Data link layer problems covering link configurations, link budgets, performance metrics, and multiple access schemes; 4) Network layer topics containing relaying techniques and potential routing algorithms; 5) Transport layer subjects such as connectivity, reliability, flow and congestion control; 6) Application layer goals and state-of-the-art UOWNs applications, and 7) Localization and its impacts on UOWNs layers. Furthermore, we outline the open research challenges and point out the future directions in UOWNs research.
\subsection{Survey Organization}
\label{sec:org}
The rest of this survey is organized as follows:  In Section \ref{sec:arch}, possible architectures for UOWNs are presented. Section \ref{sec:PHY}, addresses the physical layer aspects of UOWNs such as  underwater propagation characteristics of optical waves, channel modeling, and UOWC modulation techniques. The data link layer issues such as fundamental tradeoff between transmission angle and range, link configurations, error and data rate performance, and multiple access schemes are covered in Section \ref{sec:Data_Link}.  Section \ref{sec:Network} discusses network layer problems including relaying techniques and routing protocols. Section \ref{sec:transport} covers the transport layers topics including connectivity, reliability, flow control, and congestion control for UOWNs. Application layer goals and a number of UOWN applications are presented in Section \ref{sec:applications}. Different localization techniques for UOWNs are presented in Section \ref{sec:loc}. Section \ref{sec:challenges} outlines the open research challenges and point out the future directions in UOWNs research. Finally, Section \ref{sec:conc} concludes the survey with a few remarks.

\begin{figure*}[htb!]
\begin{center}  
\includegraphics[width=2\columnwidth]{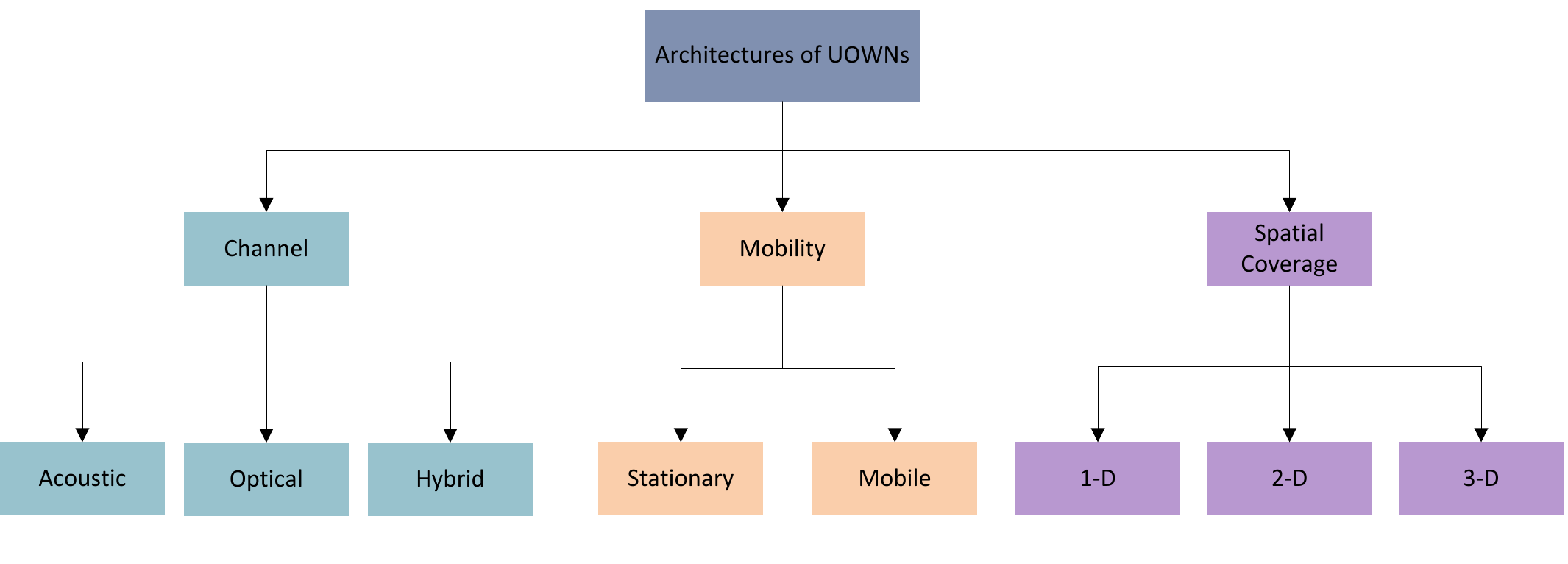}  
\caption{Potential architectures of UOWNs .\label{fig:architectures}}  
\end{center}  
\end{figure*}

\section{Potential Architectures of Underwater Optical Wireless Networks}
\label{sec:arch}
UOWNs can either operate in \textit{ad hoc} or \textit{infrastructure} modes: An ad hoc UOWN is a distributed type of wireless network which does not rely upon pre-installed network equipment. Hence, traffic requests are carried out by the participation of nodes along a routing path which is dynamically determined based on network connectivity and may necessitate self-configuration and self-organization skills because of the absence of a central control unit. Taking the potential connectivity challenges due to the directional light propagation with limited range, realizing a full ad-hoc UOWN is a non-trivial engineering task. On the other hand, infrastructure-based UOWNs may consist of omnidirectional optical access points (OAPs) or optical base stations (OBSs) each of which creates an underwater local area network (LAN) by serving and coordinating nodes in its vicinity or cell coverage area, respectively. 

 Fig.~\ref{fig:uwsn} shows a cellular infrastructure based three dimensional architecture where the underwater sensor nodes communicate with each other and with the underwater OBSs by using optical waves represented by orange links and dark green links respectively. The communication between the OBSs at same depth is represented by red colored optical links, i.e., horizontal haul (H-Haul) links, while the information from the OBSs which are at greater depth is relayed to the central OBS at the surface station by the OBSs at low depth, i.e., vertical haul (V-Haul) links drawn in blue color. It is also shown in Fig.~\ref{fig:uwsn} that the surface buoys can be operated on solar power thus improving the energy efficiency of the network. Furthermore, the submarines and autonomous underwater vehicles (AUVs) can also communicate with the OBSs by using the UOWC. Finally, the information gathered at the surface station can be transmitted to the onshore station or mobile station by using terrestrial RF networks.
Different possible architectures of UOWNs can be classified based on three principles, i.e., the spatial coverage, mobility of the sensor nodes, and channel. Based on these three principles, Fig.~\ref{fig:architectures} summarizes the possible architectures for UOWNs, which are discussed in detail below: 
\begin{itemize}

\item Stationary one-Dimensional UOWNs:
Static one-dimensional UOWNs refers to networks in which the optical sensor nodes form a line where each node attached either to the surface buoys or deployed on the seabed. Each optical sensor node in such stand-alone UOWNs process and transmit the sensed information directly to the surface station \cite{Hollinger2012}. The architecture of static one-dimensional UOWNs follows star topology, where the transmission between the surface station and optical sensor nodes is single hop \cite{Emad2015}.

\item Mobile one-Dimensional UOWNs:
Mobile one-dimensional UOWNs refers to networks in which the optical sensor nodes are deployed in an autonomous fashion. Each mobile optical sensor node process and transmit the sensed information directly to the surface station. The node in such stand-alone UOWNs is usually a floating buoy which senses the underwater environment and transmits back the information to the surface station or it can be a node deployed in the underwater environment for a specific period of time to sense the environment and floats back to the surface to transmit the sensed information. The architecture of static one-dimensional UOWNs also follows star topology, where the transmission between the surface station and optical sensor nodes is single hop \cite{Emad2015}.

\item Stationary two-Dimensional UOWNs:
In the stationary two-dimensional architecture of UOWNs, a group of static optical sensor nodes is deployed in underwater environment \cite{Guangjie2013}. Each group of nodes (i.e., cluster/cell) has a cluster head (i.e., OAP/OBS) which collects the sensed data and transmit it to the surface station. In two dimensional architecture, the sensor nodes communicate with the cluster head using horizontal communication link while the cluster head communicates with the surface station using vertical communication link. In two-dimensional UOWNs, it is assumed that all the sensor nodes are at the same depth. The network topology of this architecture depends on the application requirement and it can be a star, ring, cellular, or mesh topology.

\item Mobile two-Dimensional UOWNs:
In mobile two-dimensional UOWNs, a group of optical sensor nodes is able to float in the underwater environment. The two-dimensional mobile topology is more dynamic and challenging \cite{AKYILDIZ2005257}. In mobile two-dimensional UOWNs, the cluster head can be an AUV, which can move around in the network and collects the sensed information from different underwater optical sensors.

\item Stationary three-Dimensional UOWNs:
Generally, the three-dimensional UOWNs are used to sense and detect specific underwater phenomena which cannot be observed by seabed sensors \cite{Pompili2006, Teymorian2009}. In the stationary three-dimensional architecture of UOWNs, the depth of deployed optical sensor nodes is different, where each optical sensor node is floating at different depth \cite{Teymorian2009}. As the nodes are deployed at different depths the communication in these networks goes beyond two dimensions.  In this case, the three communication dimensions are given as: 1) Communication between nodes at different depth; 2) Communication between nodes and the cluster head (i.e., OAP/OBS); and 3) Communication between the cluster heads and the surface station.

\item Mobile three-Dimensional UOWNs:
Mobile three-dimensional UOWNs consists of underwater vehicles such as AUVs and remotely operated underwater vehicles (ROVs) which can move in different directions with different depths. Recently, AUVs and ROVs have been embedded into UOWNs to enhance the performance of typical underwater networks. In \cite{Vasilescu2005, Vasilescu2010} the authors have proposed a three-dimensional underwater surveillance, exploration, and monitoring system called autonomous modular optical underwater robot (AMOUR). Extra features such as localization, time division multiple access, and remote control have been added to AMOUR systems in \cite{Dunbabin2009}. The performance of AMOUR has been enhanced further in \cite{Dunbabin2005} by using cooperative AUVs.
\end{itemize}

\begin{table*}[htb!]
\centering
\caption{Comparison  of underwater wireless communication systems \cite{Camila2016}}
\label{Tableuwcsystems}
\begin{tabular}{|l|l|l|l|}
\hline
\hline
 Parameters              & RF  & Acoustic   & Optical  \\ \hline
 Transmission Distance   & 100 m             & Upto 20 Km     & 10-30 m      \\
 Attenuation             & Frequency and conductivity dependent & Distance and frequency dependent & Distance \\
 Speed                   & 2.255 $\times$ $10^{8}$ m/s &    1500 m/s         & 2.255 $\times$ $10^{8}$ m/s \\
 Transmit power          & Hundreds of Watts                & Few tens of Watts         & Few Watts          \\
 Cost                    & High              & High            & Low           \\
 Data rate               &  Upto 100 Mbps               &  	In Kbps             & Upto Gbps \\
 Antenna size            & 0.5 m             & 0.1  m & 0.1 m \\
 Latency                 & Moderate & High & Low\\
\hline
\hline        
\end{tabular}
\end{table*}

\section{Physical Layer: Essentials of UOWCs}
\label{sec:PHY}
The physical layer establishes the fundamentals of wireless networks and involves many essential communication functions including channel modeling and estimation, signal processing, modulation, coding, etc.  Compared to the higher layers, the physical layer of UOWNs is described well and studied more in-depth for both terrestrial optical wireless communications (TOWC) and UOWC systems.  In this section, we start with comparing the virtues and drawbacks of three main UWC systems: acoustic, radio frequency, and optical. Then, a detailed discussion of underwater propagation characteristics of optical waves is presented including absorption, scattering, turbulence, pointing, alignment, multipath fading, and delay spread. Thereafter, common modulator types and modulation techniques are addressed along with the underwater optical noise resources. 

\subsection{Waves Under the Sea: A Tour of the Underwater Communications}
\label{sec:waves}

The title of this subsection may evoke many unforgettable childhood memories since it is an adoption of the title for the renown adventure novel of Jules Verne,  \textit{``Twenty Thousand Leagues Under the Seas: A Tour of the Underwater World''}. As some parts of the underwater world are already full of mysteries, today's transportation and communication technologies still suffer from tremendous challenges of fulfilling underwater exploration and observation tasks. In particular, severe aquatic impairments pose a variety of obstacles to UWC systems depending on the nature of the following carrier waves:
\begin{figure}
\begin{center}  
\includegraphics[width=1\columnwidth]{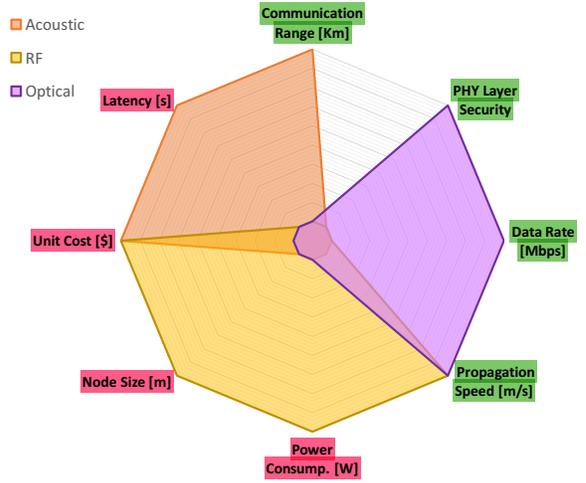}  
\caption{Comparison of acouistic, RF, and optical waves under different performance metrics which are highlighted with green and red colors if they favor for high and low values, respectively. \label{fig:waves_chart}}  
\end{center}  
\end{figure}

\subsubsection{Acoustic Waves}
\label{sec:acouistic_waves}

The involvement of human beings with acoustic waves in the oceans has been greatly motivated by the intellectual curiosity and necessity to response a possible threat. Today, UAWC systems are employed in almost every military and commercial applications of UWC \cite{Truax2001}. Accounting for hostile propagation characteristics of the seawater and enormous size of the still-vast oceans, the most prominent feature of the acoustic systems is their ability to reach very long distances up to tens of kilometers \cite{Sozer2000}. Nevertheless, UAWC systems cannot provide high quality of service due to the following innate restrictions: 1) The nominal propagation speed of the underwater acoustic signal is around 1500 m/s which yields latency in the order of seconds \cite{Pompili2009}. Hence, delay performance of acoustic systems are not desirable for real-time communication and control applications, 2) Operation bandwidth of underwater acoustic signals is between tens of Hertz to hundreds of kHz and achievable data rates of acoustic links are typically in the order of kbps, which is apparently not adequate to sustain the transmission of large data volumes \cite{AKYILDIZ2005257}, 3) Acoustic nodes are power hungry, expensive, and bulky \cite{Partan2006}. The cost per acoustic node makes creating a large scale underwater acoustic network economically demanding, energy inefficient which may necessitate battery replacement burden that can be quite a problematic task for nodes placed in the deep sea, and 4) Acoustic systems can also distress marine mammals such as dolphins and whales \cite{au1997}.

\subsubsection{Radio Frequency Waves}
\label{sec:RF_waves}

The exploitation of RF signals is especially considered to provide a smooth transition between terrestrial and underwater communication systems \cite{altgelt2005, Shamma2004}. Unlike acoustic waves, RF signals are more tolerant to turbulence and turbidity effects of the water, thus, can provide a faster propagation speed \cite{Shamma2004}. However, underwater RF communication is restricted to shallow waters and limited to the extremely low frequency band (i.e., 30 - 300 Hz) which yields a limited data rate even at very short communication ranges \cite{Frater2006}. Even if low-price terrestrial RF modules can be integrated into a Penny coin size, underwater RF nodes are costly, require huge antennas, and high transmission power is required to compensate for high antenna losses \cite{da2014, altgelt2005}.

\subsubsection{Optical Waves}
\label{sec:optical_waves}

In comparison to the acoustic and RF systems, UOWC can support high data rates on the order of Gbps over distances of tens of meters with very low delay performance thanks to the propagation velocity almost at the speed of light (i.e., $\approx$ 2.25 $\times 10^8$ m/s) \cite{Haltrin1999,Haltrin2002,Toublanc1996}. These two main advantages of optical waves can enable many real-time communication and control applications such as large-scale UWSNs and video-surveillance via AUVs. Furthermore, underwater optical wireless transceivers can be built in small sizes with low-cost and energy-conservative laser and photodiodes. Noting that optical communication generally takes place in a point-to-point fashion, it also provides an enhanced security as eavesdropping is much more difficult than in omnidirectional communications. 

Despite all these appealing virtues, there exist many challenges to implement UOWC systems in practice: Firstly, as it is the case for the free-space optical communication, misalignment of the optical transceivers can cause short-term disconnection which is generally a result of random movements of the sea surface \cite{Dong2013, Zhang2015}, depth depended variations and deep currents \cite{Johnson2013}, and oceanic turbulence \cite{Yi2015}. Secondly, even if the carrier wavelength of the light beam is chosen to be blue or green in order to mitigate the underwater attenuation effects \cite{absorbtion2017,Duntley:63,Gilbert1966}, light beam propagation still undergoes absorption, scattering, and thus multipath fading because of the interactions of water molecules and particulates with the photons \cite{Haltrin1999,Haltrin2002}. Such kind of impairments cause performance degradation and reduce the communication range significantly. 

Table \ref{Tableuwcsystems} compares these three technologies by tabulating the important state of the art system parameters. For the sake of a better visualization, we also draw a radar chart in Fig. \ref{fig:waves_chart} to highlight the potential of UOWC systems which obviously exhibit a good performance in terms of data rate, propagation speed, power consumption, latency, cost, and size. However, the main limitation is set by the short communication ranges which definitely entails range expansion via networking of optical nodes in order to operate in a large area of interest. Furthermore, misalignment of optical transceivers is one of the most challenging networking and control problems and necessitate precise alignment algorithms with inherited self-organization and self-configuration features to keep the nodes connected all the time. Therefore, it is of utmost importance to gain important insights into the UOWNs from a networking point of view including relaying, routing, deployment, localization, energy harvesting, mobility, network lifetime maximization, self-configuration, and self-organization, etc. Before proceeding to lower layers of UOWNs, however, we believe it is necessary to briefly discuss the physical layer aspects for the sake of completeness of the survey. Accordingly, the following subsections address the propagation characteristics, channel modeling, and modulation schemes of UOWC in some depth as they are building blocks of UOWNs.
     

\subsection{Underwater Propagation Characteristics of  Optical Waves    }
\label{sec:propagation}

Underwater communication channels exhibit quite different propagation characteristics varying with physio-chemical nature of oceans at different locations and depths. In particular, optical attributes of the aquatic medium is categorized based on inherent and apparent properties. While the inherent optical properties include absorption, scattering, and attenuation coefficients which heavily depend on the chemical composition of seawater \cite{zaneveld1995light}, apparent optical properties consist of radiance, irradiance, and reflectance factors which are determined by geometric parameters of light beams (e.g., diffusion and collimation) \cite{spinrad1994ocean}. In the remainder of this subsection, we cover these properties in more detail.  

\paragraph{Absorption \& Scattering} Absorption restricts the transmission range of an underwater optical wireless link by causing total propagation energy of an emitted light beam to continuously decrease. On the other hand, scattering spread the photons toward random directions such that some portion of them are not received by the receiver as it has a finite aperture size whereas the reception of some other portions may be delayed due to following different propagation paths. Thus, scattering leads to multi-path fading, time-jitter, and inter-symbol interference phenomena.  

Absorption and scattering coefficients can be formulated based on the proposed geometric model in \cite{mobley1994light}. This model considers a scenario where a volume of water $\Delta V$ with thickness $\Delta d$ is illuminated by a light beam with wavelength $\lambda$ and incident power $P_i$.  While a fraction of the incident power is absorbed by the water body $P_a=\alpha(\lambda) P_i$, another fraction $P_s=\beta(\lambda)  P_i$ is scattered due to the change of direction. The residual light power $P_t=\gamma(\lambda)  P_i$ continues to propagate on the transmitter trajectory. Fractions $\alpha(\lambda)$ and $\beta(\lambda)$ can be regarded as \textit{absorbance factor} and \textit{scatterance factor} respectively, and are related to $\gamma(\lambda)$  by $\alpha(\lambda) +\beta(\lambda) +\gamma(\lambda) =1$ as per the law of power conservation. Accordingly, the absorption and scattering coefficients are obtained by taking the limit as $\Delta d$ goes to zero \cite{mobley2010ocean}, i.e., $a(\lambda)= \lim_{\Delta d \to 0}\frac{\alpha(\lambda)}{\Delta d}$ and $b(\lambda)= \lim_{\Delta d \to 0}\frac{\beta(\lambda)}{\Delta d}$, respectively. 

\begin{figure}[t]
\begin{center}  
\includegraphics[width=0.8\columnwidth]{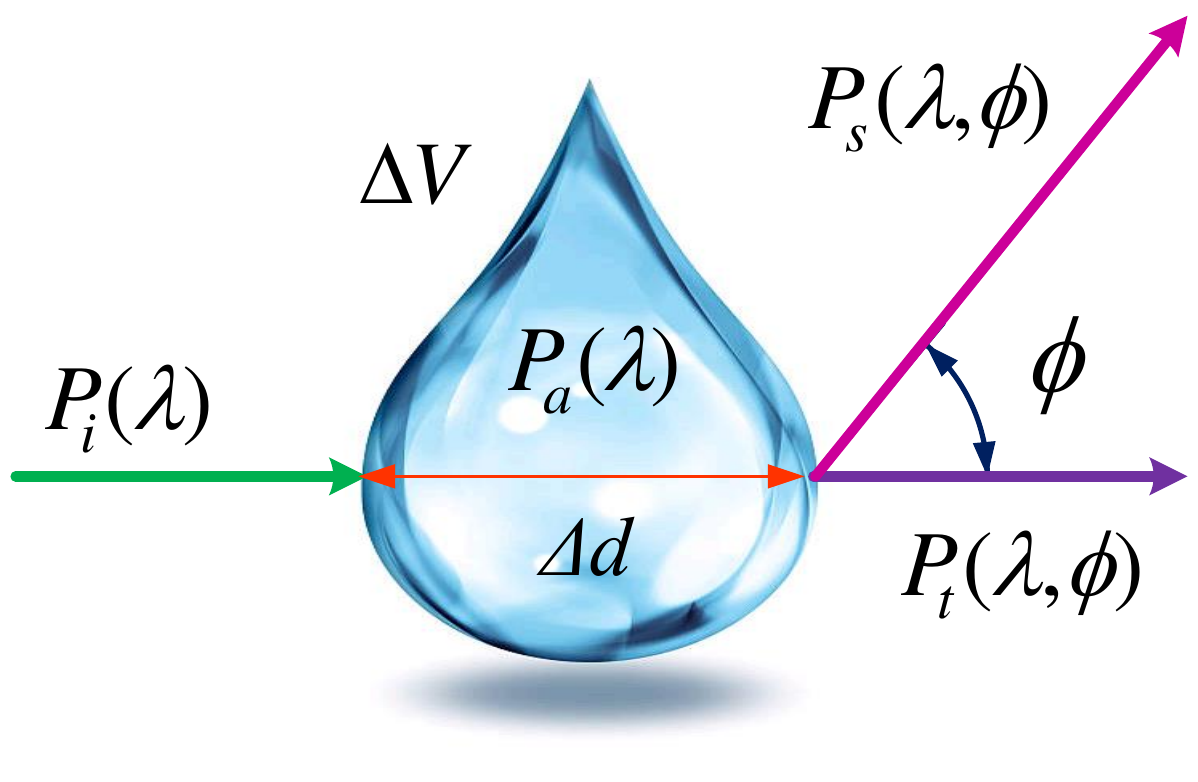}  
\caption{Geometric model for inherit optical properties \cite{mobley1994light}.}
\label{fig:geo_model}  
\end{center}  
\end{figure}

According to Jerlov \cite{Jerlov1976}, the absorption coefficient can also be modeled as a superposition of absorptions caused by pure seawater \cite{Smith1981}, colored dissolved organic materials (which are highly and less absorptive to blue \cite{Annick1981} and yellow \cite{Breves2000} wavelengths, respectively), photosynthesising of chlorophyll in phytoplankton \cite{Haltrin1999Chlorophyll}, and detritus (including organic and inorganic particles) \cite{hansell2002biogeochemistry}. Similarly, scattering coefficient can be represented as a summation of scattering effects resulting from pure sea water, phytoplankton \cite{Apel1988}, and detritus \cite{Kishino1986}. Scattering mostly depends on the density of the particulate matters rather than the wavelength as in the absorption.

The overall underwater attenuation effects are referred to as \textit{extinction coefficient} and expressed as the sum of the absorption and scattering coefficients, i.e., 
\begin{equation}
c(\lambda)=a(\lambda)+b(\lambda), \label{eq:exc}
\end{equation}
which heavily depends on water types and depths. 

Based on their influence on inherent optical properties, oceanic water types can be classified as follows \cite{Apel1988}: 
\begin{itemize}
\item \textit{ Pure sea water}: Pure sea water consists of pure water molecules (H$_2$O) and dissolved salts (NaCl, MgCl$_2$, Na$_2$SO$_4$, KCl, etc.), whose absorption effect sum mainly determines the total absorption in the pure sea water. As scattering coefficient of the pure sea water is negligible \cite{Smith1981}, light beam propagates in a straight line with very limited dispersion. 2)  3) 
\item  \textit{ Coastal ocean water}: Coastal ocean water are highly concentrated due to the dissolved particles, thus, display more severe absorption and scattering effects.
\item \textit{ Turbid harbor water}: Turbid harbor water shows the most hostile absorption and scattering levels as it has the highest concentration of suspended and dissolved particles.
\end{itemize}
Water depths are conceptualized by dividing oceans into various vertical zones based on the presence or absence of sunlight \cite{Costello2010}: The layer near the sea surface is called \textit{photic zone} which goes deep till which the sunlight penetrates. The uppermost stratum of the photic zone (0-200 m) is referred to as \textit{euphotic zone} and has sufficient light to support photosynthesis. The lower layer (200-1000 m) is known as \textit{dysphotic zone} (a.k.a twilight zone) which cannot support the adequate light for the photosynthesis. The water depths below the photic zone is called as \textit{aphotic zone} which is an abyssal region of pitch darkness. Noting that the average depth of the ocean is around 4.3 km, the photic zone covers only a thin layer but still poses the greatest biomass of all oceans. Extending from the sea surface to the bottom, the chlorophyll variation curve is observed to follow a skewed Gaussian profile\cite{Johnson2014}. Accordingly, the attenuation coefficient has shown to start from $0.05 \: m^{-1}$ and reaches the peak record $0.1 \: m^{-1}$ around 100 m depth, which starts decreasing for deeper waters \cite{Johnson2013}.  

\paragraph{Oceanic turbulence} 
Oceanic turbulence is defined as the rapid variations in the refraction index due to fluctuations in the aquatic medium parameters such as pressure, density, salinity, temperature, etc. \cite{Johnson2014}. This phenomenon provokes inconstant light intensity reception that is referred to as \textit{scintillation} and yields significant performance degradation.

\paragraph{Pointing \& Alignment}
Pointing and alignment are critical engineering tasks to maintain a constant reliable link between the optical transceivers. The pointing errors and misalignment are generally considered to be a result of the \textit{bore-sight} and \textit{jitter} \cite{Yang2014}. The bore-sight is defined as a fixed displacement between the transmitter trajectory (i.e., beam center) and center of the receiver aperture which may be caused by the inaccurate receiver location information. On the other hand, the jitter is random dislocations between the light-beam and aperture center as a result of oceanic turbulence \cite{Yi2015}, depth depended variations and deep currents \cite{Johnson2013}, and random movements of the sea surface \cite{Dong2013, Zhang2015}. Even though bore-sight can be mitigated by effective pointing and precise location information, jitter is still a problem as random nature of the oceanic environment cannot be controlled. We should note that as the scattering effects become more significant (i.e., coastal and turbid waters), tight pointing and alignment requirements are relaxed due to the high dispersion of the light-beam \cite{Sanchez2002}.

\paragraph{Multipath Fading and Delay Spread}
Due to the scattering and reflection effects, some portions of the emitted light-beam may follow different propagation paths with various traveling distance and reaches to the receiver aperture at different time instants, which yields time dispersion (i.e., delay spread) and inter-symbol interference (ISI). Unlike the UAWC where delay spread and ISI is quite considerable due to very long distances and low propagation velocity, these phenomena have not received much attention as a result of the high signal speed and limited communication ranges of UOWC. Multipath fading can be more significant in shallow waters because of the reflections from the sea surface, seabed, and obstacles in the vicinity. Impacts of spatial diversity on ISI was investigated in \cite{Jamali2017} where high data rates were observed to suffer more from ISI phenomenon.  In order to quantify the time spread, authors of  \cite{Hanson2008} have investigated the impacts of system design parameters such as divergence angle and receiver aperture size. In \cite{Jaruwatanadilok2008}, time spread analysis deduces that ISI is significant at 50 m for a polarized light beam with 1 Gbps data rate. However, a Monte-Carlo simulation based channel characterization concludes that time spread is negligible over short distances \cite{Gabriel2013}.

\subsection{Underwater Optical Wireless Channel Modeling}
In this section different UOWC channel models are briefly discussed. Firstly, the underwater optical wireless attenuation, absorption, and scattering models are presented which includes Beer-Lambert law, volume scattering function, radiative transfer equation, and Monte-Carlo methods. Secondly, the oceanic turbulence models are discussed for UOWC channels and finally, the models for pointing and misalignment are presented.
\label{sec:channel_modeling}

\subsubsection{Underwater Optical Attenuation Models} 
\label{sec:attenuation_model}

\paragraph{Beer-Lambert Law}
The simplest and thus most widely used model to describe the UOWC channel attenuation is Beert-Lambert Law which expresses the received signal power at the receiver as 
\begin{equation}
\label{eq:beert}
P_r(\lambda,d)=P_t e^{-c(\lambda) d},
\end{equation}
where $P_t$ is the transmission power of transmitter, $c(\lambda)$ is the extinction coefficient given in \eqref{eq:exc} and $d$ is the Euclidean distance between the transceivers. As previously discussed in detail, $c(\lambda)$ changes for different water types and depths \cite{Smart2005,Giles2005}. For practical values of $c(\lambda)$, we refer interested readers to the works in  \cite{Johnson2013, mobley2010ocean, mobley1994light, Apel1988, ALLDREDGE1988,Johnson2014, Cochenour2008, Hanson2008}. Assuming a perfect pointing between the transceivers, Beer-Lambert Law presumes all scattered photons are lost by ignoring the multipath arrival of the scattered photons. To overcome this deficiency, more sophisticated models were proposed, which are introduced in the following subsections. 
\paragraph{Volume Scattering Function} 
Volume scattering function (VSF) can be interpreted as the scattered intensity per unit incident irradiance per unit volume of water and expressed as \cite{mobley2010ocean}
\begin{equation}
\label{eq:VSF}
\vartheta(\lambda,\phi)= \lim_{\Delta d \to 0} \lim_{\Delta \omega \to 0} \frac{P_s(\lambda, \phi)}{\Delta d \Delta \omega},
\end{equation}
where $P_s(\lambda, \phi)$ is the power of scattered light beam into a solid angle which is centered on $\phi$ as shown in Fig. \ref{fig:geo_model}. Hence, scattering coefficient can be obtained by integrating the VSF over all directions, i.e., $b(\lambda)=\int \vartheta(\lambda,\phi) d\omega$. Furthermore, scattering phase function (SPF) can be expressed by normalizing the VSF by the scattering coefficient \cite{mobley2010ocean}, i.e., $\tilde{\vartheta}(\lambda,\phi)=\frac{\vartheta(\lambda,\phi)}{b(\lambda)}$ which is commonly represented by Henyey-Greenstein function \cite{Gabriel2013,Toublanc1996,Haltrin1999}.

\paragraph{Radiative Transfer Equation}
Even though VSF is an important inherent optical property to characterize the scattering effects, it is not easy to measure in practice \cite{Tan2013} and not suitable for a large number of photons as it only considers the scattering of a single photon \cite{Johnson2013}. To mitigate these drawbacks, radiative transfer equation (RTE) was proposed as an alternative and it can describe the energy conservation of a light beam passing through a steady medium \cite{HULST1980}. RTE is expressed as  \cite{Li2015, Arnon2012}
\begin{align}
\label{eq:RTE}
\nonumber \vec{r} \: \nabla L(\lambda,\vec{r},\vec{\ell})&=-c L(\lambda,\vec{r},\vec{\ell}) +\int_{2 \pi} \vartheta(\lambda,\vec{r},\vec{r} \: ' )L(\lambda,\vec{r},\vec{\ell}) d\vec{r} \: '\\
&+ E(\lambda,\vec{r},\vec{\ell})
\end{align} 
where $ \vec{r}$ is the direction vector, $\nabla$ is the divergence operator, $L(\lambda,\vec{r},\vec{\ell})$ represents the optical radiance at position $\vec{\ell}$ towards direction $ \vec{r}$, $\vartheta(\lambda,\vec{r},\vec{r}' )$ is the VSF, and $E(\lambda,\vec{r},\vec{\ell})$ denotes the source radiance.  By taking light polarization and multiple scattering into consideration, an analytic solution was developed in \cite{Jaruwatanadilok2008} by using Stokes vector. Another analytical solution was devised in \cite{Cochenour2008,Cochenour2008Spatial} where derivation was simplified by small angle approximation. Since it is very hard to find an exact analytical solution of RTE \cite{Arnon2012} which is generally obtained by making assumptions and simplifications \cite{Potter2013}, numerical solutions of RTE have gained more attention.
\paragraph{Monte-Carlo Methods}
Monte-Carlo simulation is a probabilistic numerical solver which mimics the underwater light propagation by emitting and tracking the large amount of photons \cite{Gabriel2013}. It has gained popularity due to its desirable features of accurate results, easy programming,  and high flexibility \cite{WANG1995}. However, it sill suffers from time complexity, efficiency, and statistical errors \cite{Lerner1982}. A robust Monte-Carlo based model was designed in \cite{leathers2004monte} by the U.S. Naval Research Laboratory. Recent research efforts on characterizing the UOWC channels by solving the RTE with Monte-Carlo simulations can be found in \cite{Gabriel2011underwater, Gabriel2013, Li2012, Li20121}.

\subsubsection{Oceanic Turbulence Modeling} 
\label{sec:turbulence_model}

Although UOWC channel modeling studies are mostly concentrated on obtaining a precise characterization of the absorption and scattering effects, the impacts of oceanic turbulence on the system performance has not received the attention it deserves. As physical mechanisms of atmospheric and oceanic turbulence share some similar features, several oceanic turbulence modeling studies employed traditional free-space optical (FSO) turbulence models. For example, the classical spectrum model of Kolmogorov was adopted for UOWC channels in \cite{Hanson2010}. Inspired by \cite{Hanson2010}, a generic channel model was proposed in \cite{Liu2015} by considering absorption, scattering, and turbulence which directly applies the well-known lognormal turbulence model, i.e., 
\begin{equation}
f_I(I)=\frac{1}{I \sqrt{2 \pi \sigma}} \exp \left( - \frac{ \left(\ln( I) - \mu\right)^2}{2 \sigma}\right),
\end{equation}
where $I$ is the received light intensity, $\mu$ is the mean logarithmic light intensity, and $\sigma$ is the scintillation index.

Impacts of oceanic turbulence and depth on the underwater imaging were analyzed in \cite{Hou2009, Woods2011}. Adaptive optics were proposed in \cite{HOLOHAN1997} to mitigate the negative effects of turbulence for UWOC and underwater imaging. In \cite{Nikishov2000}, authors have derived the power spectrum of refractive index fluctuations in turbulent sea water. Gaussian light-beam propagation in turbulent sea water was studied in \cite{KOROTKOVA2011,cox2012simulation,Ata2014}. In the weak oceanic turbulence case, an aperture averaging method was analyzed and shown to improve system performance by reducing the scintillation index \cite{Yi2015}. In \cite{Tang2013}, the average speed of moving oceanic turbulence has been shown to have a major impact on the temporal correlation of the irradiance whereas the link distance has minor effects. Using the Rytov method, scintillation indices of different optical waves was evaluated in turbulent aquatic medium \cite{Ata2014Scintillations}.  

\subsubsection{Pointing Errors and Misalignment Modeling} 
\label{sec:misalignment_model}

Neglecting the pointing errors caused by jitter, misalignment is modeled using the following beam spread function (BSF) \cite{Cochenour2008, Tang2012}
\begin{align}
\label{eq:BSF}
\nonumber & \mathrm{BSF}(\lambda,d,r)= P_r(\lambda,d) E(d,r) +\int_0^\infty P_r (\lambda,d) E(d,x) \\ & \times
 \left[ \exp \left (\int_0^d b(\lambda)  \tilde{\vartheta}\left( x(d-y)\right) dy  \right)-1 \right] J_0(yr)y dy,
\end{align} 
where $E(d,r)$ and $E(d,x)$ are the irradiance distributions of the laser source in spatial coordinates and spatial frequency domain, respectively; $d$ is the distance between transceivers; $r$ is the distance between the center points of aperture and the received light-beam; $\tilde{\vartheta}(\cdot)$ is the SPF. Using this model, the authors evaluated the BER performance of UOWC under misalignment condition. In \cite{Sanchez2002}, pointing error performance was investigated as a function of BSF under different water types. Effects of random movements of the sea surface on the jitter of transceivers were studied in \cite{Dong2013} where PDF of sea surface movements are considered as a two-dimensional Gaussian distribution. Impacts of transmitter parameters such as divergence and elevation angles were also analyzed and simulated using Monte-Carlo method \cite{Zhang2015}. In \cite{Gabriel2013}, misalignment of point-to-point (P2P) communication was studied by using Monte-Carlo simulations and verified with water tank experiments. Given sufficiently large transmission power, numerical results showed that a small misalignment does not yield a significant performance loss for any water type \cite{Cochenour2013}.

\subsection{Optical Wireless Modulation Techniques}
\label{sec:mod_techniques}

Optical wireless modulation schemes can be categorized into two main class: \textit{intensity modulation} (IM) (a.k.a. non-coherent modulation) and \textit{coherent modulation} (CM), which can be implemented either by a \textit{direct} or an \textit{external} modulator. 

\subsubsection{Modulator Types}
\label{sec:mod_types}

Direct modulators use the light source current by switching the light-source ON and OFF to transmit ``$1$" and ``$0$", respectively. Even though it has very low complexity and price, direct modulators are limited by communication range and achievable data rates due to the chirping effect. In external modulators, on the other hand, the light source is kept always on to transmit a continuous light-beam whose intensity or phase is modulated by an external device to pass or block the light-source to transmit the desired message. External modulators can provide very high data rates and long link ranges thanks to their switching speed and constant transmission power. However, they are not efficient in terms of power, cost, and complexity.

\subsubsection{Intensity Modulation}
\label{sec:IM}

The IM is carried out by modulating the intensity of the light source by a direct or external modulator. If the receiver demodulates the received light using a direct detector (DD), the overall system is referred to as IM/DD modulation. IM/DD is the most prominent modulation scheme due to its low cost and simplicity, as there is no need for the phase information. In what follows, we present and compare common IM schemes from the spectrum, power, cost, and monetary efficiency perspectives.

\paragraph{ON/OFF Keying (OOK)} 
The simplest form of the IM modulation is the OOK scheme where the ``$1$" and ``$0$" are represented by the presence and absence of light. OOK employs return-to-zero (RZ) or non-return-to-zero (NRZ) pulse formats. The NRZ format occupies the entire bit duration to represent ``1" while the RZ format only occupies part of the bit duration. The performance of OOK severely degrades with the channel variations, thus, a dynamic threshold mechanism can improve the overall performance by updating the detection threshold according to the channel state estimation \cite{Khalighi2014}. The low power consumption, bandwidth efficiency, and  simplicity makes OOK a popular and practical scheme which is theoretically and experimentally studied for UOWC in \cite{Akhoundi2015, Jamali2017, Ghassemlooy2018}.

\paragraph{Pulse Position Modulation (PPM)} 
PPM is one of the most widely used techniques which modulates each of $M$ transmitted bits as a pulse within $2^M$ time slots whose position corresponds to the message sent. PPM provides higher power and spectral efficiency in return for a more complicated transceiver. Even though it does not need a dynamic threshold mechanism, tight synchronization requirements cause significant performance loss due to jitter effects. Conventional PPM was improved by its variants such as differential PPM \cite{Shiu1999}, digital pulse interval PPM \cite{Ghassemlooy1998}, differential amplitude PPM and multilevel digital pulse interval modulation \cite{Ghassemlooy2006}. Analytic and experimental studies on the PPM can be found for UOWC in \cite{Sari1998, Chen2006, Meihong2009, Anguita2010, Anguita2010Optical, Hagem2012, He2012, Swathi2014}.

\paragraph{Pulse Width Modulation (PWM)} 
In an $M$-ary  PWM, pulses only appear in the first $M$ time slots where $M$ is equal to the decimal of the transmitted bits.  PWM reduces the peak transmission power by spreading the total power to $M$ time slots, which yields higher average power with the increase in $M$ \cite{Fan2007}. PWM is especially advantageous with its spectral efficiency and immunity to ISI effects \cite{Khalighi2014}.

\paragraph{Digital Pulse Interval Modulation (DPIM)} 
In DPIM, ``ON" pulses are followed by ``OFF" time slots whose number is equal to the decimal value of transmitted data symbol \cite{Gabriel2012Investigation}. Unlike PPM and PWM,  DPIM is an asynchronous modulation technique which can also support variable symbol lengths. Even if it provides higher power and spectral efficiency, the DPIM suffers from error propagation during the demodulation process. We refer interested readers to \cite{Doniec2009,Doniec2010,Donice2010II} for applications of DPIM in UOWC. 


\subsubsection{Coherent Modulation}
\label{sec:coherent_mod}

To the contrary of IM schemes, the CM employs both amplitude and phase information to encode the desired message. At the receiver side, a local oscillator converts the optical carrier down to baseband or RF intermediate frequency which is referred to as \textit{homodyne} and \textit{heterodyne} detection, respectively. The CM can provide higher receiver sensitivity, spectral efficiency, and resistance to background noise, but with extra cost and complexity. We refer interested readers to \cite{Cochenour2007, Cox2009II, Meihong2009, Dong2013II} and references therein for CM schemes including, phase shift keying and polarization shift keying. 

\subsubsection{Receiver Noise Sources}
\label{sec:noise_src}

The optical receiver is affected by many noise sources including the photo-diode (PD) dark current, transmitter noise, shot noise, thermal noise, and background noise \cite{gagliardi1976optical}. Noting that PD dark current is negligible in practice \cite{Khalighi2014}, the transmitted light is affected by the transmitter noise caused by the fluctuations of the light intensity. Transmitter noise is generally modeled by laser relative noise \cite{saleh1991fundamentals} which is shown to have a minor effect on the receiver performance \cite{Xu2011}. Thermal noise is generally modeled as a zero-mean Gaussian random process which results from the behavior of electronic circuitry, especially the load resistor. Shot noise (a.k.a. quantum noise) is modeled as a Poisson process which originate from random fluctuations of the PD current. If the received number of photons is large, the Poisson process can be approximated by a Gaussian process for both PIN diode based receivers \cite{saleh1991fundamentals} and avalanche PD based receivers \cite{Davidson1988}.

The background noise is highly dependent upon water types, depths, and optical carrier wavelength. In the euphotic zone, the solar interference can be regarded as the main contributor of the background noise, whose variation is given by \cite{Giles2005}
\begin{equation}
\label{eq:sol}
\sigma_{sol}=A (\pi \Phi)^2 \Delta   T_f L,
\end{equation}
where $A$ is the aperture size, $\Phi$ is the receiver's field of view (FoV), $ \Delta$ is the optical filter bandwidth, $T_f$ is the transmissivity, $L=\frac{E R L_{f} e^{-k h}}{\pi}$ is the solar radiance $[W/m^2]$, $E$ is the down-welling irradiance $[W/m^2]$, $R$ is the reflectance of the down-welling irradiance, $L_f$ is the diectional dependence of radiance, $k$ is the diffuse attenuation coefficient, and $h$ is the water depth. For deeper regions, another contributor is the bioluminescence (a.k.a. blackbody radiation) which is generally focused on blue-green wavelengths and given as \cite{Giles2005}
\begin{equation}
\label{eq:bb}
\sigma_{bb}=\frac{2 \hbar c^2 a    A (\pi \Phi)^2 \Delta T_a T_f}{\lambda^5 \left(\exp \left( \frac{\hbar c}{\lambda k T} \right)-1 \right) },
\end{equation}
where $c=2.25257 \times 10^8$ is the speed of light in the aquatic medium, $\hbar$ is the Plank's constant, $\kappa$ is the Boltzmann constant, $T_a=e^{-\tau_o}$ is the transmission in water, $T$ is the symbol duration, $a = 0.5$ is the radiant absorption factor, and $\lambda$ is the optical carrier wavelength. Accordingly, the overall background noise can be represented as a summation of solar and blackbody interference which is given as \cite{Jaruwatanadilok2008}, $\sigma_{bg}=\sigma_{sol}+\sigma_{bb}$.


\section{Data Link Layer: Link Configurations  and Multiple Access Schemes}
\label{sec:Data_Link}
Data link layer is the protocol layer that convey data between the neighbor network entities (i.e., single-hop or multi-hop connections) and may provide functions to detect and correct possible physical layer errors. Regardless of users ultimate destination, data link layer undertakes the task of arbitrating among the users, who compete for the same network resources such as time, frequency, space, wavelength, etc.,  in order to prevent frame collisions and specify protocols to detect and recover from such collisions. 

Accordingly, this section first compares wide-beam short-range and narrow-beam long-range transmission schemes and call attention to the fundamental tradeoff between divergence angle (i.e., coverage span) and communication range (or the received power for a given range). Then, the power budget of three main UOWC link configurations is presented: line of sight (LoS), non line of sight (NLoS), and retro-reflective links. Even though the contents of first two subsections are not solely related to the data link layer, covering them in this section is especially important to provide valuable insights into the cross-layer optimization of the first two layers. Following the error and data rate performance for UOWNs, potential multiple access schemes are addressed including time-division multiple access (TDMA), frequency-division multiple access (FDMA), code-division multiple access (CDMA), non-orthogonal multiple access (NOMA), wavelength-division multiple access (WDMA), and space-division multiple access (SDMA).

\subsection{Narrow Beam vs. Wide Beam Light Sources}
\label{sec:narrow_wide}

Depending upon the divergence angle specifications, light sources can be classified into two broad categories: wide-beam and narrow beam sources such as light emitting diodes (LEDs) and laser diodes, respectively. Let us first cover these two transmitter types in the realm of TOWC systems as they teach valuable lessons for UOWC applications. The visible light communication (VLC) operates on LEDs to combine their two main advantages: energy efficient indoor/outdoor illumination \cite{Cole2014} and high-speed data delivery \cite{Grobe2013}, which is already being commercialized by many startups, e.g., Light Fidelity (Li-Fi) \cite{Haas2016}. Since the VLC targets to serve multiple users concurrently, ongoing research efforts mostly concentrated on efficient resource sharing and multiple access schemes \cite{Marshoud2016}. However, FSO communication focuses more on long-range and high data rate outdoor TOWC applications such as wireless X-hauling \cite{Jaber2016}. Unlike the VLC, pointing, acquisitions, and tracking (PAT) functionality plays an important role for FSO communication systems to maintain a continuous system performance \cite{Hsien2004} since they are employed for P2P long-range outdoor links. 

\begin{figure}
    \centering
    \begin{subfigure}[b]{0.5\textwidth}
\includegraphics[width=1\columnwidth]{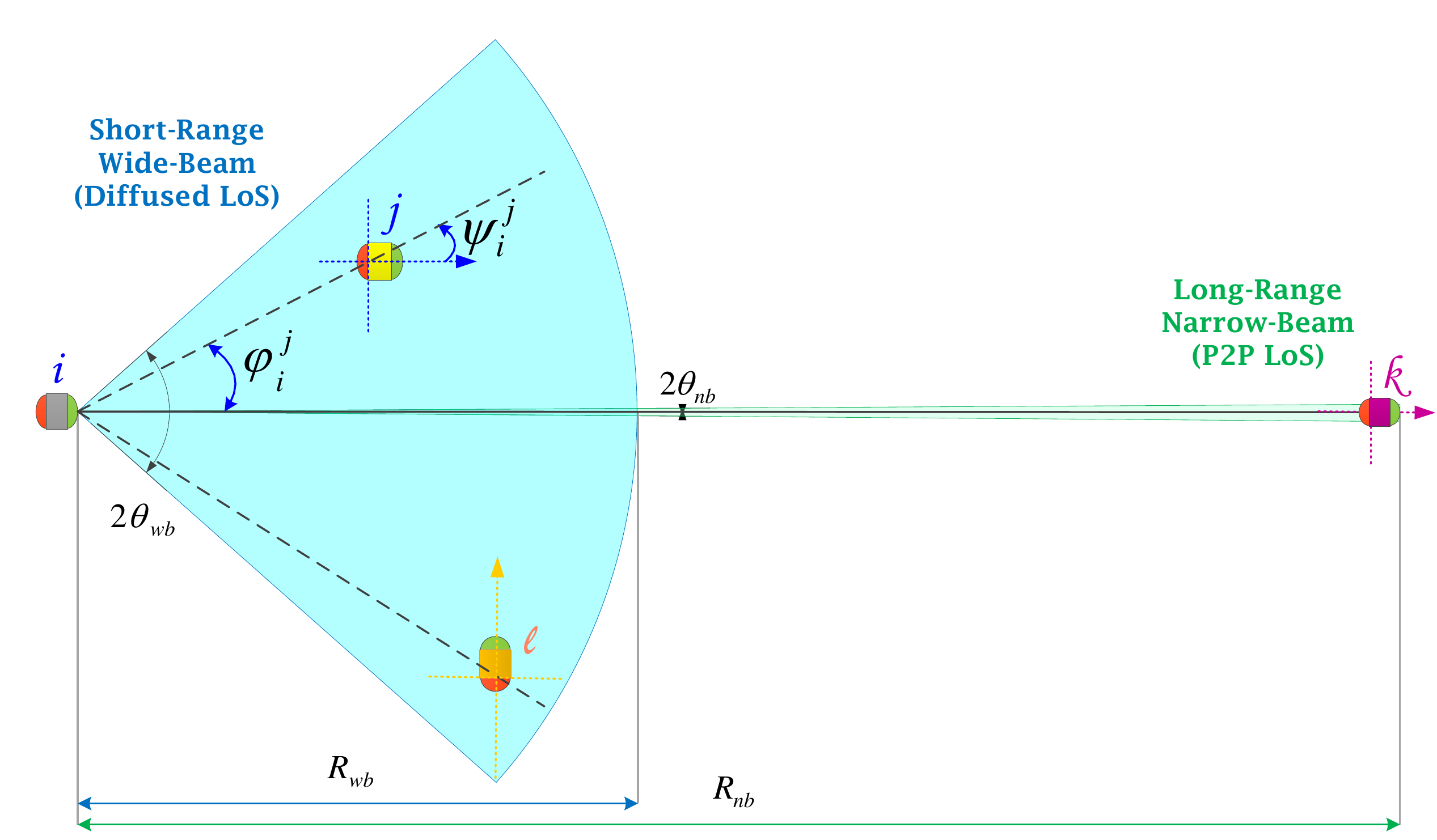}  
\caption{LOS}
\label{fig:LOS}  
    \end{subfigure}

    \begin{subfigure}[b]{0.5\textwidth}
\includegraphics[width=1\columnwidth]{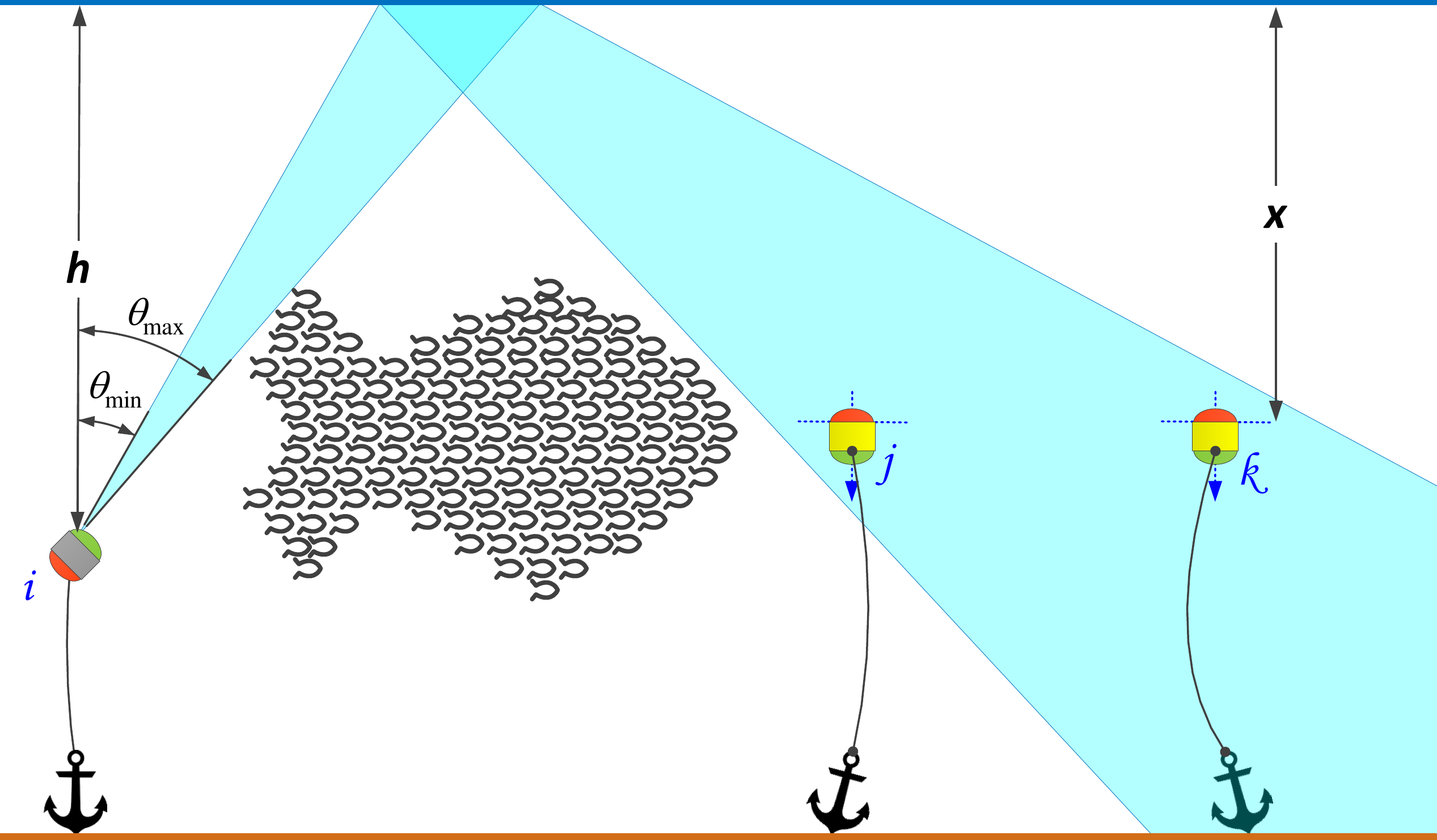}  
\caption{NLoS (Reflective)}
\label{fig:NLoS}  
    \end{subfigure}

    \begin{subfigure}[b]{0.5\textwidth}
\includegraphics[width=1\columnwidth]{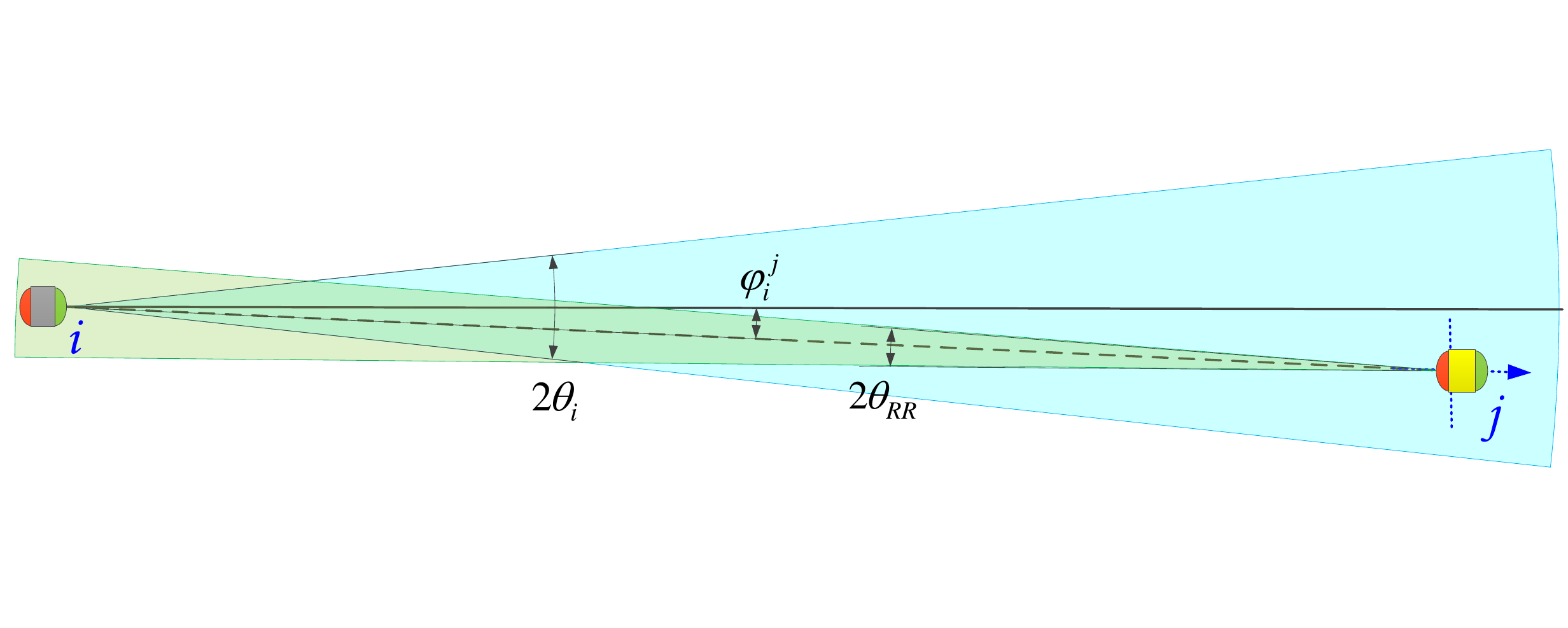}  
\caption{Retro Reflective}
\label{fig:RR}  
    \end{subfigure}
    \caption{Underwater optical link configurations: a) LoS, b) NLoS (Reflective), and c) Retro Reflective. }
\label{fig:links}
\end{figure}

 Atmospheric link losses are generally dominated by a beam spreading factor, $d^{-2}$, where $d$ is the communication distance. In the aquatic medium, however, extinction loss, $e^{-c(\lambda)d}$, of nearly collimated light beams (e.g., lasers) dominates the beam spreading factor. On the other hand, beam spreading factor $d^{-2}$ is the primary source of loss in the link budget calculations of light sources with broad divergence angles (e.g., LEDs). Hence, wide-beam light sources can communicate with nearby receivers scanned in a broad angle circular sector while narrow-beam light sources can reach distant receivers within a tight circular sector, as shown in Fig. \ref{fig:LOS}. In other words, there is a fundamental tradeoff between divergence angle (i.e., spanned coverage area) and transmission range (or received power for a fixed range). It must also be noted that even if the transmitter has a very tight divergence angle, the receivers can observe a slightly diffused light beam because of the aquatic medium, which is more significant in water types with severe scattering nature, e.g., turbid water.

 \begin{figure}
\begin{center}  
\includegraphics[width=1\columnwidth]{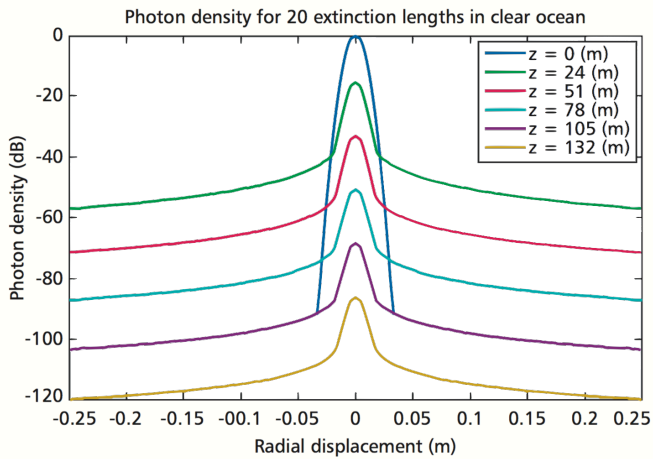}  
\caption{Photon density vs. radial displacement and propagation distance over 20 extinction lengths in clear ocean conditions \cite{Fletcher2015}.}
\label{fig:narrow1}  
\end{center}  
\end{figure} 

 \begin{figure}
\begin{center}  
\includegraphics[width=1\columnwidth]{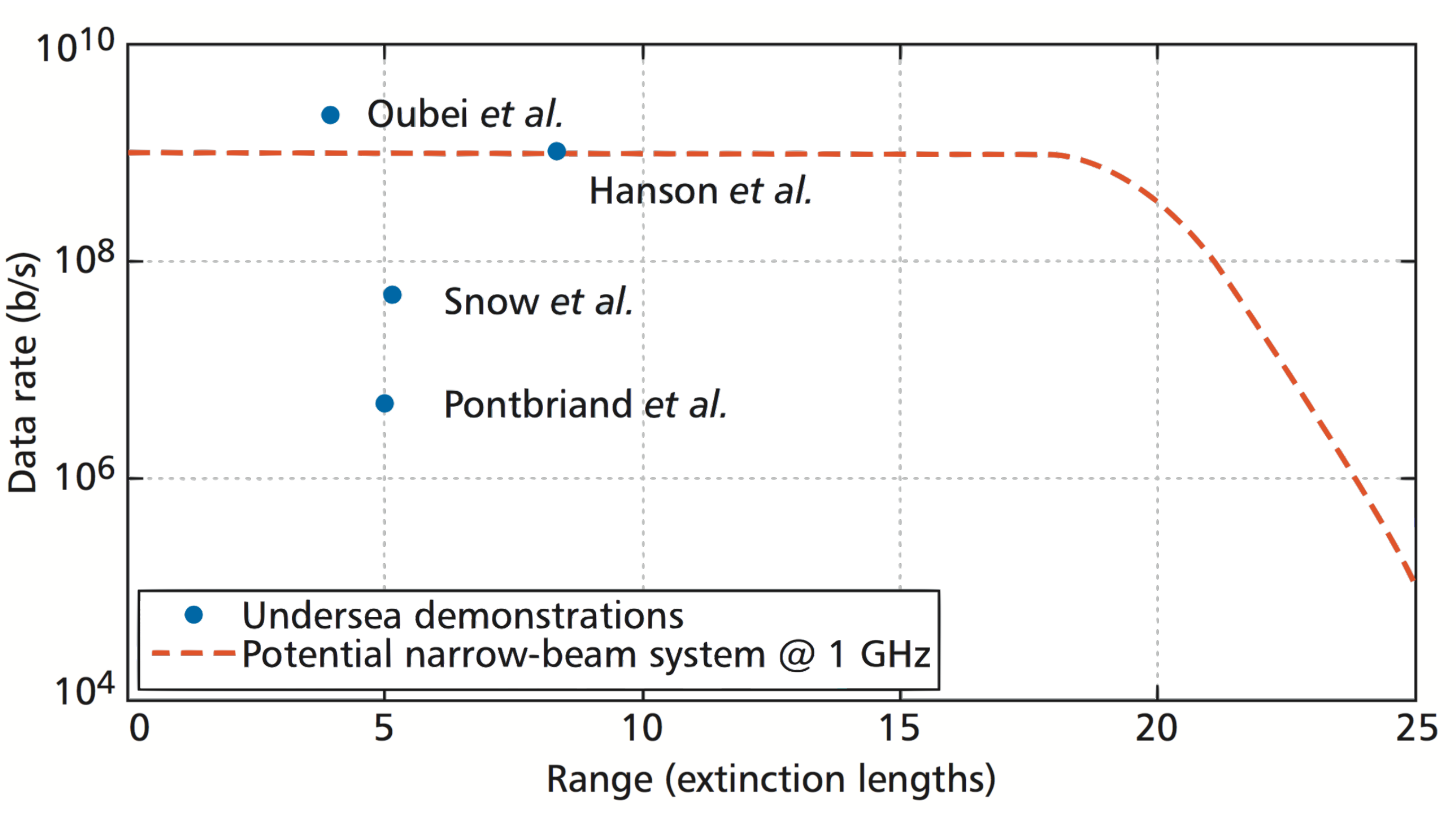}  
\caption{Comparison of underwater system demonstrations and potential narrow-beam system \cite{Fletcher2015}.}
\label{fig:narrow2}  
\end{center}  
\end{figure} 
 
Accordingly, narrow-beam light sources have the following advantages \cite{Fletcher2015}: (i) Higher power reception and longer communication ranges; (ii) reduced time spread due to the relatively high ratio of ``ballistic" photons which propagates without scattering. Monte Carlo simulations show that 90$\%$ of photons arrive within 10 ns and  2 ns for a wide-beam and narrow-beam transmissions, respectively.  The arrival time can even be reduced to 90 ps if the narrow-beam transmission is received by a receiver with 0.1 mrad FoV; and (iii) improved spectral and spatial filtering options are available since the receiver FoV can be reduced significantly due to the limited light diffusion at the receiver. Albeit these advantages, narrow-beam transmission requires accurate PAT mechanisms which are addressed in Section \ref{sec:relaying}.  

For a Gaussian light beam with 1 cm radius beam waist, Fig. \ref{fig:narrow1} shows the photon density [dB] vs. radial displacement [m]  and propagation distance [m] over 20 extinction lengths in clear ocean conditions. Fletcher et. al. considers a narrow-beam laser communication system over 20 extinction length (around 132 m for an extinction length of 6.6 m) with 100 mW transmission and 2 cm aperture size \cite{Fletcher2015}.  16-ary PPM with 1/2-rate forward error correction (FEC) achieves 1 Gbps capacity at a wavelength of 515 nm where the attenuation loss is 87 dB and noiseless sensitivity is  2.9 b/photon. For comparison purposes, the same set up was also considered for a wide-beam transmitter which achieves only 3.5 kbps. Underwater demonstrations in \cite{Oubei2015, Hanson2008, Snow1992, Pontbriand2008} were compared with the considered potential narrow-beam system in Fig. \ref{fig:narrow2} where the authors have assumed sensitive receivers, FEC, PAT, and photon-efficient modulation techniques. 

\subsection{Aquatic Optical Link Configurations}
\label{sec:link_config}

In this section, we consider three main link configurations for UOWNs: 1) LoS Links, 2) NLoS Links, and 3) Retro-Reflective Links.
\subsubsection{LoS Links} 
\label{sec:LoS}

LoS communication is the most straightforward form of optical links where transceivers communicate over an unobscured link which can either happen in a \textit{diffused} or a P2P fashion as illustrated in Fig. \ref{fig:LOS}. Even implementing the P2P LoS links for stationary transceivers is a trivial task in clear ocean, it may require sophisticated PAT mechanisms to keep transceivers bore-sighted in the case of mobility.

For a generic optical transmitter node $i$ and receiver node $j$, propagation loss factor is given based on Beer Lambert's Law as  \cite{Arnon:09}
\begin{equation}
\label{eq:ploss}
L_{ij}(\lambda, d_{ij})=\exp \left \{ - c(\lambda) d_{ij} \right \},
\end{equation}
where $d_{ij}$ is the Euclidian distance between the transceivers and $\varphi_i^j$ is the angle between the receiver plane and the transmitter trajectory. Likewise, geometric gain (a.k.a. telescope gain) of the LoS link is given as \cite{Shlomi2010}
\begin{equation}
\label{eq:tel_LoS}
G_{ij}^{LoS}= \begin{cases} \frac{A_j}{d_{ij}^2} \frac{\cos(\varphi_i^j)}{2 \pi  [1-\cos(\theta_i)]} 
, -\pi/2 \leq \varphi_i^j \leq \pi/2 \\
0, \hfill\text{otherwise}\end{cases},
\end{equation}
where $A_j$ is the receiver aperture area of node $j$ and $\theta_i$ is the beam divergence angle of transmitter node $i$. In order to concentrate the transmitted energy on receiver aperture, the divergence angle of laser-diodes are generally designed to be a few milliradians or less \cite{Juarez2006} whereas typical LEDs can have divergence angles less than $140$ milliradians to diffuse light to wide angles \cite{Dianhong2016}. Accordingly, received power can be formulated as a product of transmission power, transceivers' efficiency, telescope gain, and path loss factor, i.e., 
 \begin{equation}
 \label{eq:Pr_LoS}
 P_r^j=P_t^i \eta_t^i \eta_j^r G_{ij}^{LoS} \chi(\psi_i^j) L_{ij} \left(c(\lambda),  \frac{ d_{ij}}{\cos (\varphi_i^j)}\right),
 \end{equation}
where $P_t^i $ is the transmission power, $\eta_t^i$ and $\eta_j^r$ are transmitter and receiver efficiency, respectively; $\chi(\psi_i^j)$ is the concentrator gain \cite{Kahn97wireless}, which is defined for non-imaging concentrators as \cite{Ning1987}
\begin{equation}
\label{eq:conc}
\chi(\psi_i^j)=\begin{cases} \frac{n^2}{\sin^2(\Psi_j)}, 0 \leq \psi_i^j \leq \Psi_j \\
0, \hfill \psi_i^j> \Psi_j \end{cases},
\end{equation}
\begin{figure*}
\begin{equation}
\label{eq:telNLOS}
 G_{ij}^{NLoS}=\begin{cases}
 \frac{A_j cos(\varphi_i^j)}{2 A_{ann}}  \left( \left[\frac{\tan(\theta_t-\varphi_i^j)}{\tan(\theta_t+\varphi_i^j)}\right]^2 \left[\frac{\sin(\theta_t-\varphi_i^j)}{\sin(\theta_t+\varphi_i^j)}\right]^2    \right)  &, \theta_{\min} \leq \varphi_i^j \leq \theta_c \\
 \frac{A_j  cos(\varphi_i^j)}{2 A_{ann}}  &,  \theta_c  \leq \varphi_i^j \leq \theta_{\max} \\
 0 &, \text{otherwise} 
\end{cases}
\end{equation}
\hrule
\end{figure*}
 $\psi_i^j$ is the angle of incidence w.r.t. the receiver axis, $\Psi_j$ is the concentrator FoV which can be $\pi/2$ and down to $\pi/6$ for the hemisphere and parabolic concentrators, respectively;  and $n$ is the internal refractive index. Notice that the receiver gain increases as the FoV decreases. Hemispherical lens are common nonimaging concentrators \cite{Savicki1994} which can achieve  $\Psi_j \approx \pi/2$ and $\chi(\varphi_i^j) \approx n^2$ over its entire FoV. The compound parabolic concentrator \cite{Ning1987} is another type of nonimaging concentrators and can obtain a much higher gain in return for a narrower FoV, which is especially more desirable for P2P-LoS links. Since it is easy to implement, most of the experimental studies considered LoS links under different water characteristics and modulation schemes using a variety of transmitter hardware \cite{Cochenour2007, Cox2009II, Baiden2009paving, Oubei2015, Hanson2008, Snow1992, Pontbriand2008}.

%
%

\subsubsection{NLoS Links}
\label{sec:NLoS}

LoS links may  not always be available due to the obstructions within the underwater topology, PAT errors, mobility, and random orientations of the transceivers, etc. In such cases, a diffused light beam which is reflected over sea surface (or alternatively a mirror located in an appropriate location) can be beneficial to facilitate a point-to-multipoint (P2M) (a.k.a. multicasting) transmission to reach obscured receivers, as depicted in Fig. \ref{fig:NLoS}. Assuming that the transceivers are oriented vertically upward, the transmitted light beam is characterized by inner and outer angles $\theta_{\min}$ and $\theta_{\max}$, respectively. As per the Fresnel's law, propagating light is partially refracted and partially reflected at interfaces between the mediums with different refractive indices. Therefore,  the light beam transmitted from depth $h$ is partially reflected from the sea surface and illuminate an annular surface $A_{ann}$ at depth $x$ with equal power density. $A_{ann}$ is given by
\begin{equation}
\label{eq:ann}
A_{ann}=2 \pi (h+x)^2 \left[ \cos (\theta_{\min})-\cos (\theta_{\max}) \right],
\end{equation}
which defines an annular area taken from a sphere of radius $h + x$  \cite{Arnon:09}.  Assuming that sea surface is modeled as smooth (i.e., incident angle is equal to the perpendicular angle between the receiver plane and the transmitter-receiver trajectory, i.e., $\varphi_i^j$), the telescope gain of the NLoS links is given in (\ref{eq:telNLOS}) where $\theta_t$ is the angle of transmission, $\theta_c \triangleq \sin^{-1} \left( \frac{n_A}{n_W}\right)$ is the critical angle (i.e., the angle of incidence above that the total internal reflection (TIR) occurs), $n_A$ is the refraction index of air, and $n_W$ is the refraction index of water. Accordingly, received power at node $j$ is expressed as follows
\begin{equation}
\label{eq:NLoS}
P_r^j=P_t^i \eta_t^i \eta_j^r G_{ij}^{NLoS} \chi(\psi_i^j) L_{ij} \left(c(\lambda),   \frac{h+x}{\cos (\varphi_i^j)} \right).
\end{equation}

LoS and NLoS links have been compared by Jasman et. al. in \cite{Jasman2013} where they have demonstrated that 100 MHz bandwidth availability of LoS links is reduced to 20 MHz in case of NLoS even in clear water conditions. Indeed, such a reduction is not a surprise due to the reflection losses at the sea surface and diffusion of the reflected light beam. Furthermore, multi-scattering effect of NLoS links was addressed in \cite{Kedar2006} and \cite{Kedar2007}.

\subsubsection{Retro-Reflective Links}
\label{sec:Retro}

Similar to backscatter communication in RF systems, retro-reflective communication consists of a light source and a reflector. While the light source could be a sophisticated system with high transmission power, the reflector behaves as an interrogator as it lacks the ability to fulfill transceiver operations due to its simple architecture with low power availability. Therefore, the continuous light beam emitted from the source is modulated and reflected back to the receiver. Retro-reflective communications can be considered in two cases \cite{Kaushal2016underwater}: photon limited case and contrast limited case which take place in clear and turbid water, respectively. In the former case, absorption is the dominant effect which reduces the number of photons received by the reflector. Furthermore, the accuracy of PAT mechanisms at both sides plays a significant role in receiving enough information-bearing photons. In the latter case, scattering is the dominant factor which mainly determines retro-reflective link range and capacity. Contrast limitation is especially important for underwater imaging applications as a reduction in photon quantity directly reduces the image contrast, which can be considerably improved by exploiting polarization discrimination \cite{Mullen2009, Brandon2009}. If the receiver has enough power resource, the reflector can even amplify the modulated light beam in order to achieve a better performance both in photon and contrast limited scenarios \cite{celik2018modeling}.

Based on the geometric gain of LoS links in \eqref{eq:tel_LoS}, telescope gain of the retro-reflective links is expressed as \cite{Shlomi2010}
\begin{equation}
\label{eq:RRgain}
G_{ji}^{RR}= \begin{cases} \frac{A_j}{d_{ij}^2} \frac{\cos(\varphi_i^j)}{2 \pi  [1-\cos(\theta_i)]} 
 \frac{A_{RR} \cos(\varphi_j^i)}{\pi \left[ d_{ij} \tan(\theta_{RR})\right]^2}
, -\pi/2 \leq \varphi_j^i \leq \pi/2 \\
0, \hfill\text{otherwise}\end{cases}
\end{equation}
where $A_{RR}$ is the aperture area of the reflector, $\theta_{RR}$ is the divergence angle of the reflector, and $\varphi_i^j$ is the angle between receiver trajectory of the source and the reflector trajectory. Accordingly, reflected light beam is received back by the source node $i$ as follows
\begin{equation}
\label{eq:RR}
P_r^i=P_t^i \eta_t^i \eta_i^r \eta_j^{RR} G_{ij}^{RR} \chi(\psi_j^i) L_{ij} \left(c(\lambda), \frac{2 d_{ij}}{\cos (\varphi_i^j)} \right),
\end{equation}
where $\eta_j^{RR}$ is the retroreflector efficiency.

 \begin{figure}
\begin{center}  
\includegraphics[width=1\columnwidth]{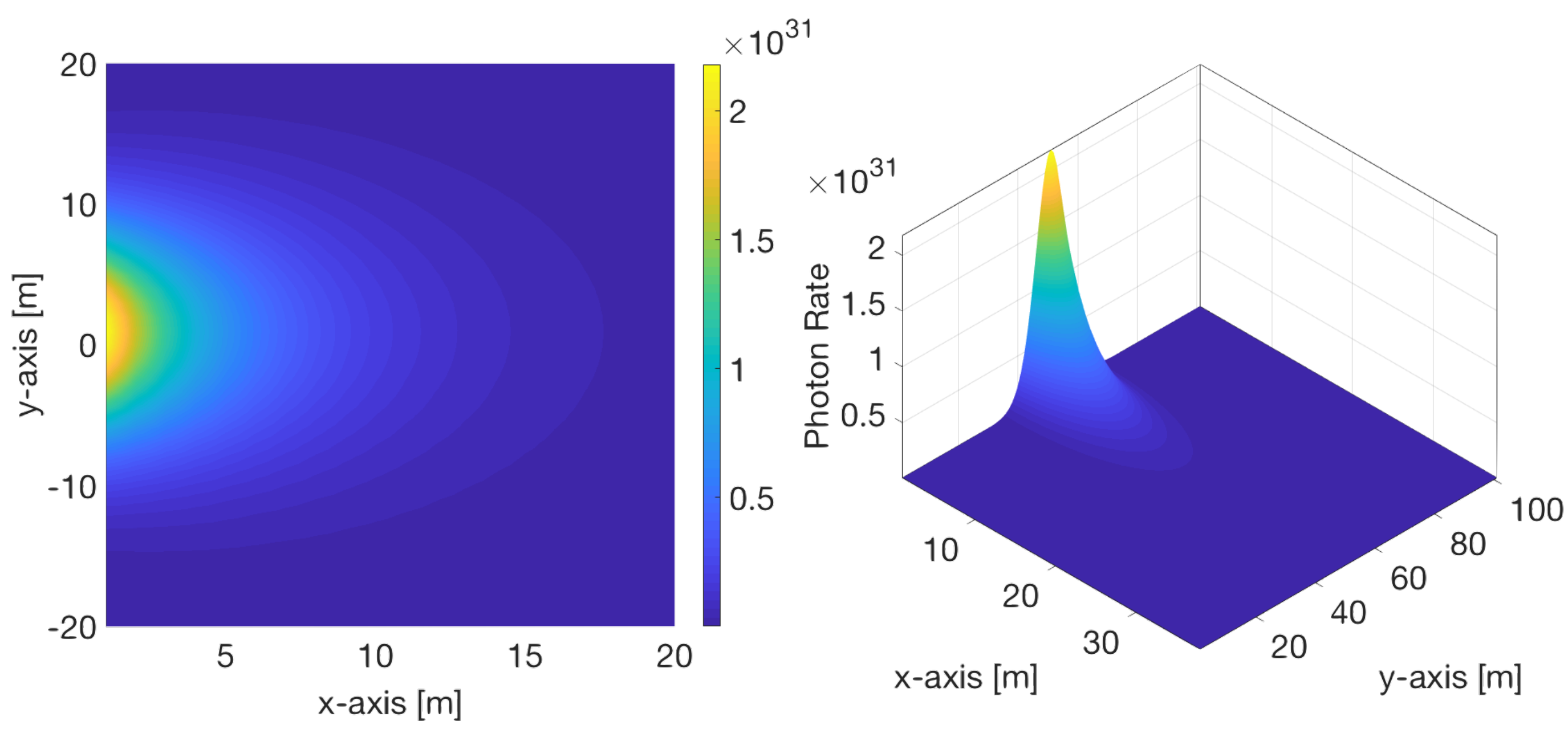}  
\caption{Demonstration of photon arrival rate for an LoS link \cite{celik2018modeling}.}
\label{fig:photon}  
\end{center}  
\end{figure}

\subsection{Error and Data Rate Performance}
\label{sec:BER}

Before proceeding to the medium access schemes, it is important to quantify the error and data rate performance of these link configurations based on a common and straightforward detection technique. Therefore, the authors in \cite{eraerds2007sipm} considered IM/DD OOK with silicon photo-multipliers (SiPMs) based photon counter detectors. Photon arrivals are generally assumed to be a Poisson distributed function, therefore, the photon arrival rate within a slot duration $T$ is given by
\begin{equation}
\label{eq:photons}
p_i^j=\frac{P_j^r \eta_D^j}{T R_i^j  \hslash c},
\end{equation}
where $P_j^r$ is the received power, $\eta_D^j$ is the detector efficiency, $R_i^j$ is the data rate, $\hslash $ is the Plank constant, and $c$ is the speed of light. Fig. \ref{fig:photon} shows the photon arrival rate w.r.t. xy-coordinates for a sensor fixed at origin and pointing in positive x-axis direction while the receiver is located at different location and its receiver directed to the origin. Assuming a large number of photon reception, then according to the central limit theorem, Poisson distributed photon arrivals can be approximated by a Gaussian distribution and the bit error rate (BER) is given by
\begin{equation}
\label{eq:BER}
BER_i^j=\frac{1}{2}\mathrm{erfc} \left[ \sqrt{\frac{T}{2}} \left( \sqrt{p_{ij}^1} - \sqrt{p_{ij}^0}\right) \right],
\end{equation}
where $\mathrm{erfc}(\cdot)$ is the complementary error function, $p_{i,j}^0=p_{bg}+p_{dc}$ and $p_{i,j}^1=p_i^j+p_{i,j}^0$ are the photon arrival rates when binary 1 and binary 0 are transmitted, respectively; $p_{bg}$ and $p_{dc}$ are the background illumination noise and additive noise due to dark counts, respectively. For a given BER $\overline{BER}_i^j$, achievable data rate can be obtained from \eqref{eq:BER} as 
\begin{equation}
\label{eq:rate}
R_{i}^j=\frac{P_i^j \eta_D^j \lambda}{T \hslash c \left[ \left( \mathrm{erfc}^{-1} \left(2 \overline{BER}_i^j \right) \sqrt{\frac{2}{T}} + \sqrt{p_{ij}^0} \right)^2- p_{ij}^0\right]}.
\end{equation}
Since hard decision forward error correction (HD-FEC) can successfully identify and correct all bit errors below an FEC-BER threshold, one can set $\overline{BER}_i^j \leq 3.8 \times 10^{-4}$ as recommended by the International Telecommunication Union Standardization Sector (ITU-T) \cite{recommendation2004forward}.


\subsection{Multiple Access Schemes}
\label{sec:multi_access}

\begin{figure}
\begin{center}  
\includegraphics[width=1\columnwidth]{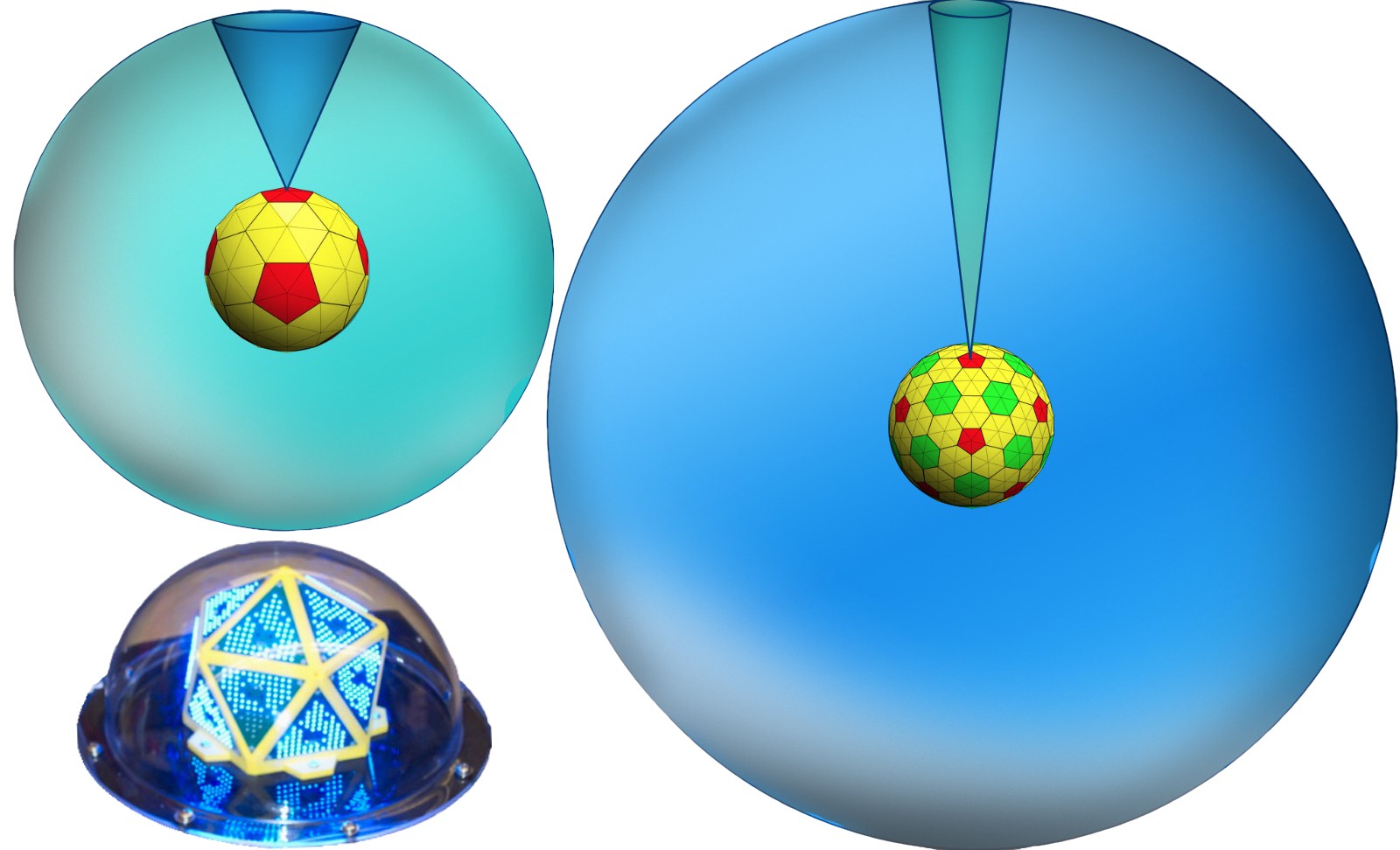}  
\caption{Illustration of optical polyhedron transceivers (SOTs) and its implementation  \cite{Baiden2007,Baiden2009paving}.}
\label{fig:ball}  
\end{center}  
\end{figure}

For infrastructure based UOWNs, many researchers conceptualized omnidirectional OAPs/OBSs by designing them in multi-faceted spherical shape which has single or multiple transceivers at each face \cite{Baiden2007, Baiden2009paving, Bilgi2008, Farr2005}. Therefore, underwater OBSs can be designed as geodesic polyhedra as shown in Fig. \ref{fig:ball} along with its implementation \cite{Baiden2007, Baiden2009paving}. Geodesic polyhedra approximates spheres with triangles and can be a good solution against underwater pressure as the geodesic domes are known to withstand heavy structure loads by distributing the structural stress over its rigid triangular building blocks \cite{ingber1998architecture}. Notice that as the number of faces (pentagonal/hexagonal shapes in Fig. \ref{fig:ball}) increases, it is possible to employ narrower transmitter divergence and receiver FoV angles which naturally yields longer transmission range and higher receiver gain (Fig. \ref{fig:ball}), respectively. In addition to their spatial reuse and angular diversity advantages \cite{Bilgi2008}, OBSs can also provide flexibility as each LED on a face can be exploited to serve for fulfilling a specific task. 

In OBS based cellular UOWNs, there exists two main interference scenarios: intercell interference (ICI) and intracell interference which is also referred to as multiple access interference (MAI).  While the former happens when a user receives signals from other users using the same network resources within the adjacent cells, the latter occurs when a user observes interfering signals from users sharing the same cell resources. Compared to VLC systems, intercell interference expected to be at low levels due to the severe aquatic channel impairments and can even be further reduced by intelligent OBS deployment strategies. Nonetheless, intracell interference still stays as a first and foremost research challenge for both downlink (DL) and uplink (UL) transmission. Hence, in addition to efficient resource allocation strategies, OBSs necessitate multicarrier transmission schemes and multiple access protocols to serve several users simultaneously. 

\begin{figure}
\begin{center}  
\includegraphics[width=1\columnwidth]{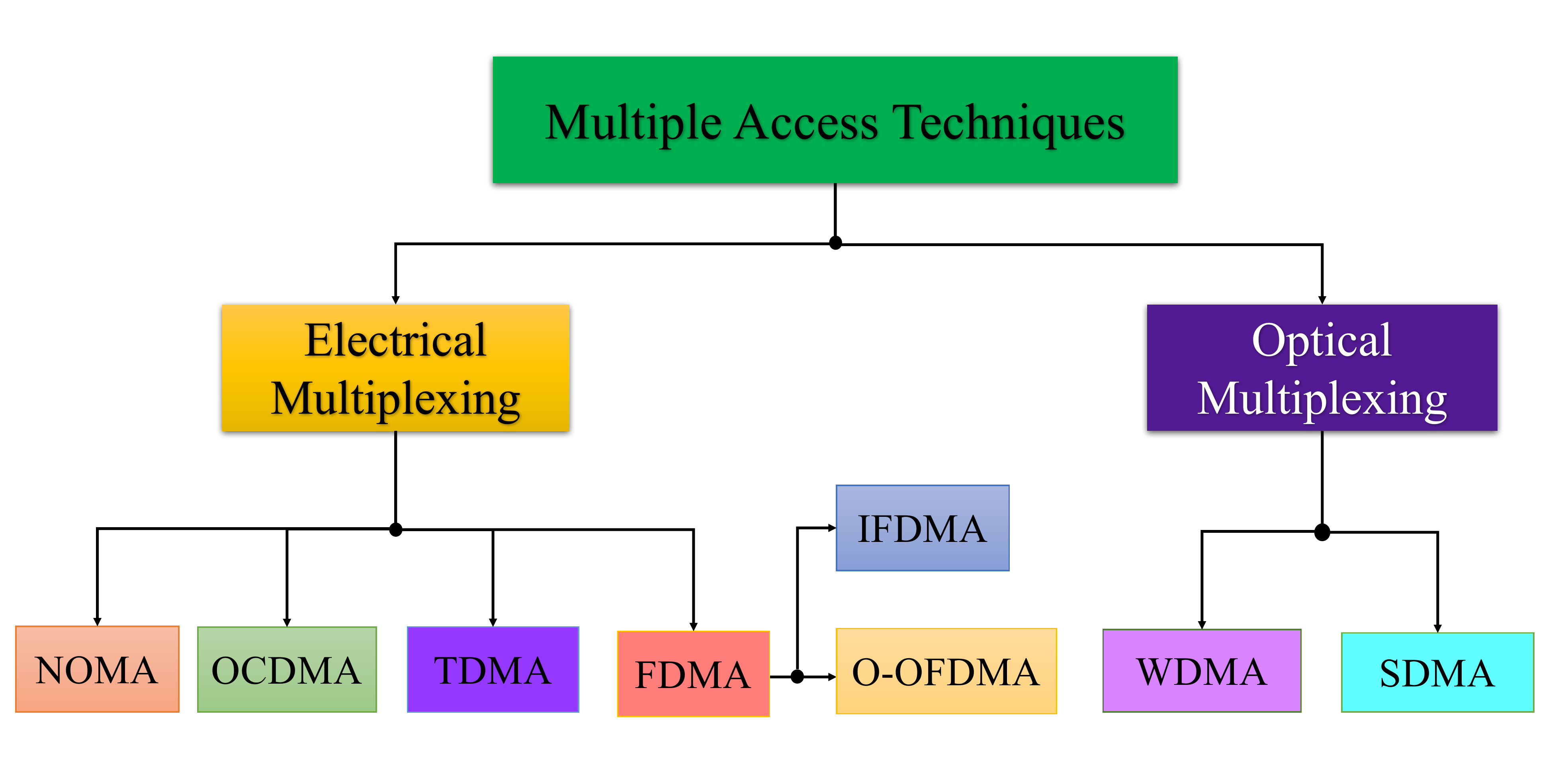}  
\caption{Classification of multiple access schemes.}
\label{fig:MAC}  
\end{center}  
\end{figure}

As diagramed in Fig. \ref{fig:MAC}, multiple access schemes can be categorized into electrical and optical multiplexing subcategories. Electrical multiplexing schemes consist of time division multiple access (TDMA), frequency-division multiple access (FDMA), code-division multiple access (CDMA), and non-orthogonal multiple access (NOMA) whereas optical multiplexing schemes contain wavelength-division multiple access (WDMA) and space-division multiple access (SDMA). To the best of authors' knowledge, there is no research efforts on UOWN multiple access schemes excluding the optical CDMA \cite{Akhoundi2015, Jamali2016, Akhoundi2016cellular}. In what follows, we present multicarrier transmission techniques along with corresponding multiple access schemes for UOWNs.

\subsubsection{Time Division Multiple Access}
\label{sec:TDMA}

TDMA is a synchronous channel access scheme where non-overlapping time slots are assigned to different users as per the requested QoS levels. Hence, TDMA does not allow nodes to transmit simultaneously and independently. In UAWC systems, TDMA provides a limited bandwidth efficiency because low propagation speed requires long time guards to prevent packet collisions of the adjacent time slots \cite{Sozer2000}, which may not be the case for UOWC systems thanks to low propagation delays. TDMA can support high energy efficiency in return for reduced capacity per user \cite{Kahn97wireless}. Nevertheless, TDMA requires efficient scheduling techniques in order to overcome the MAI. A potential scheduling scheme could be based on users rather than LEDs embedded on OBSs as they can be much larger than the number of users. Even though TDMA has not attracted the attention for UOWNs yet, it can be motivated by research efforts on TDMA based VLC systems: As a potential solution, each LED is orthogonally allocated to a time slot in \cite{Nadeem2015} and a block encoding TDM is exploited in \cite{Hou2015} where one LED from each LED group is allowed to transmit. In \cite{Wang2014}, TDMA is considered for UL transmission where each user has certain time slots to transmit such that identity of the transmitting users can be recognized as per the scheduling policy.   

\subsubsection{Frequency Division Multiple Access}
\label{sec:FDMA}

FDMA scheme permits multiple users to transmit momentarily over non-overlapping frequencies/subcarriers within a cell area. Noting that FDMA is not suitable for acoustic systems due to the limited bandwidth availability \cite{AKYILDIZ2005257}, it offers high spectral efficiency and robustness again intersymbol interference (ISI) \cite{Armstrong2012} for optical wireless communications. However, it lacks energy efficiency which deteriorates with the increasing number of subcarriers \cite{Elgala2011}. Orthogonal FDMA (OFDMA) and interleaved FDMA (IFDMA) are two well-known schemes studied extensively for OWCs \cite{Marshoud2016}.

OFDMA allocates each user with several time slots and frequency blocks which spans a number of orthogonal frequency division multiplexing (OFDM) subcarriers. Because of the real and unipolar valued signal requirements of the IM, conventional OFDM cannot be directly applied to optical OFDM (O-OFDM) systems. In return for losing half of the bandwidth, reality constraint can be satisfied by applying Hermitian symmetry on inverse fast Fourier transform inputs. Positivity of the signals can be achieved either by direct current biased optical OFDM (DCO-OFDM) \cite{Sanya2015} or asymmetrically clipped optical OFDM (ACO-OFDM) \cite{Li2012}. The former adds a DC bias before transmission which may cause overheating and high signal distortion. At the expense of BER performance degradation and increased complexity \cite{Tan2016}, several peak-to-average power ratio (PAPR) reduction techniques were proposed to overcome these problems \cite{Zhang2014, Yu2014}. In order to obtain unipolarity, ACO-OFDM technique clips the signal at zero level \cite{Vucic2012} and transmits only the positive part of the signal. Even if it is more energy efficient than the DCO-OFDM, bandwidth utilization is quite low because of using only half of the subcarriers for data transmission. Optical OFDMA (O-OFDMA) was proposed in \cite{Dang2012} which has a lower decoding complexity and power efficiency in comparison with O-OFDM based interleave division multiple access (IDMA). In \cite{Dinc2015}, the authors have considered two handover schemes for users within the intersection area of two optical transmitters: In the first scheme, the user combines the signal of both transmitters, while in  the second scheme the each transmitter use a dedicated band for the user. IFDMA was proposed in \cite{Lin2015} to mitigate the high PAPR effects of O-OFDMA where it was shown that IFDMA have lower computational complexity than O-OFDMA and it reduces the synchronization errors.

\subsubsection{Code Division Multiple Access}
\label{sec:CDMA}

\begin{figure}
\begin{center}  
\includegraphics[width=1\columnwidth]{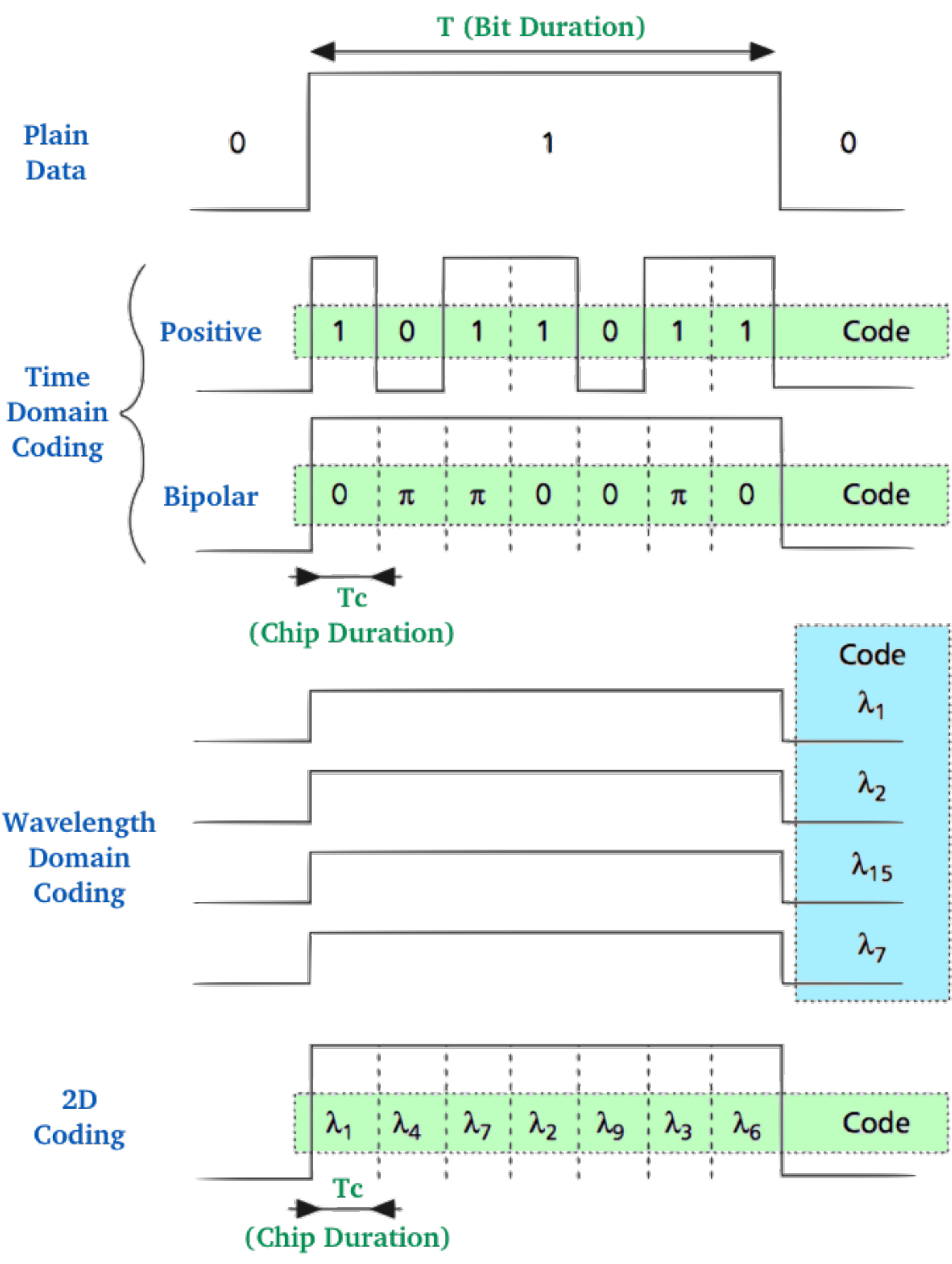}  
\caption{Illustration of OCDMA dimensions \cite{Fouli2007}.}
\label{fig:OCDMA}  
\end{center}  
\end{figure}

Optical code division multiplexing (OCDM) is a multiplexing scheme where communication channels are distinguished by optical orthogonal codes in addition to  time and wavelength \cite{Jazayerifar2006, Ghaffari2008}. As shown in Fig. \ref{fig:OCDMA}, the data stream is multiplied by a code sequence either in the time domain, wavelength domain, or even as a combination of both (i.e., 2D coding) \cite{Fouli2007}. In the time domain, a bit duration is divided into smaller time slots which are called chips. Bipolar time-encoding is a coherent technique that manipulates the phase of the optical signal and needs phase accuracy. As an alternative, positive time encoding is non-coherent which manipulates the power of the optical signal without requiring any phase information \cite{Stok2002}. On the other hand, a wavelength-encoded signal consists of a unique subset of wavelengths in order to form the code. Finally, 2D coding combines both time spreading and wavelength assignment such that a data stream is constituted as successive chips of different wavelengths. In the receiver side, decoding is performed by applying the reverse operations of the encoding. 

Accordingly, optical CDMA (OCDMA) employs OCDM technique to mediate multiple asynchronous nodes in sharing common network resources. Thanks to its high spectral efficiency, distributive, and asynchronous nature; OCDMA has received much attention to be employed in UOWNs \cite{Akhoundi2015, Jamali2016, Akhoundi2016cellular}. In \cite{Akhoundi2015}, the authors have addressed the structures, principles, and performance analysis of OCDMA based cellular UOWNs where OBSs are connected to a central optical network controller. In \cite{Jamali2016}, the performance of relay-assisted OCDMA networks was characterized by the turbulent channels. Finally, potential and challenges (e.g., mobility, cell edge coverage, blockage avoidance, power control, etc.) of OCDMA networks were presented in \cite{Akhoundi2016cellular}. 

\subsubsection{Non-Orthogonal Multiple Access}
\label{sec:NOMA}

 NOMA is also referred to as power domain multiple access where user signals are superposed in a way that each signal is allocated to a distinct power level depending upon the channel conditions. While NOMA allocates more power to users with bad channels conditions compared to those with good channel condition. Employing successive interference cancellation, the user allocated with high powers can cancel the interference of the users with the low power allocation. Thus, all users can occupy the available entire time-frequency resources and increase overall system performance significantly \cite{Marshoud2016}. Even though NOMA has attracted attention for VLC systems \cite{Marshoud2016NOMA, Marshoud2017}, there is no study targeting NOMA for UOWCs. 

\subsubsection{Wavelength Division Multiple Access}
\label{sec:WDMA}

WDMA facilitate the multi-user access harnessing the wavelength division multiplexing (WDM)  such that each user has a dedicated wavelength along with an optical tunable reception filter in order to operate on assigned wavelength. WDM multiplexes a number of optical signals at different wavelengths (i.e., color) into a single one. Coarse and dense WDM are two standard types which are named based on the available number of channels and their spacing. Even if WDMA reduces the signal processing complexity to a grate extent, it may significantly increase the hardware complexity and cost \cite{Tsonev2014}.  Since underwater operational wavelength is different from TOWCs, it is necessary to standardize the WDM channels and their spacing for blue-green wavelengths. It is also important to develop efficient wavelength assignment policy as the nodes in UOWNs can observe different channel conditions at different wavelengths because of varying water types and depths. 
\subsubsection{Space Division Multiple Access}
\label{sec:SDMA}

SDMA harnesses the spatial distribution of the users and directivity of the light beam propagation to permit parallel transmission on the same network resources which can either be in time, frequency/wavelength, or code domains. In  \cite{Chen2015}, random grouping and optimal grouping approaches were proposed for an SDMA based VLC system and obtained results have shown that SDMA can offer 10 times higher throughput than the conventional TDMA scheme. Notice that SDMA is a potential technique to be employed for underwater OBSs as they can benefit from both spatial and angular diversity.

\section{Network Layer: Relaying Techniques and Routing Protocols}\label{sec:Network}
Due to the communication range limitations of UOWCs, relay-assisted UOWC is a key enabler technique to realize UOWNs by expanding coverage area, extending the communication range, enhancing energy efficiency, providing cooperative diversity,  and improving the end-to-end system performance \cite{Khalighi2014}. However, the full benefit of relay-assisted UOWCs can be obtained by effective routing algorithms taking the underwater propagation characteristics of light beams into account. Therefore, this section first covers \textit{serial relaying} and \textit{parallel relaying} techniques using \textit{decode-and-forward }(DF), \textit{amplify-and-forward} (AF) methods, and \textit{ Bit-detect-and-forward }(BDF). Thereafter, potential routing protocols for UOWNs are surveyed including location-based routing, source-based routing, hop-by-hop routing, cross-layer routing, clustered routing, and reinforcement learning based routing.

 \begin{figure*}
\begin{center}  
\includegraphics[width=1.75 \columnwidth]{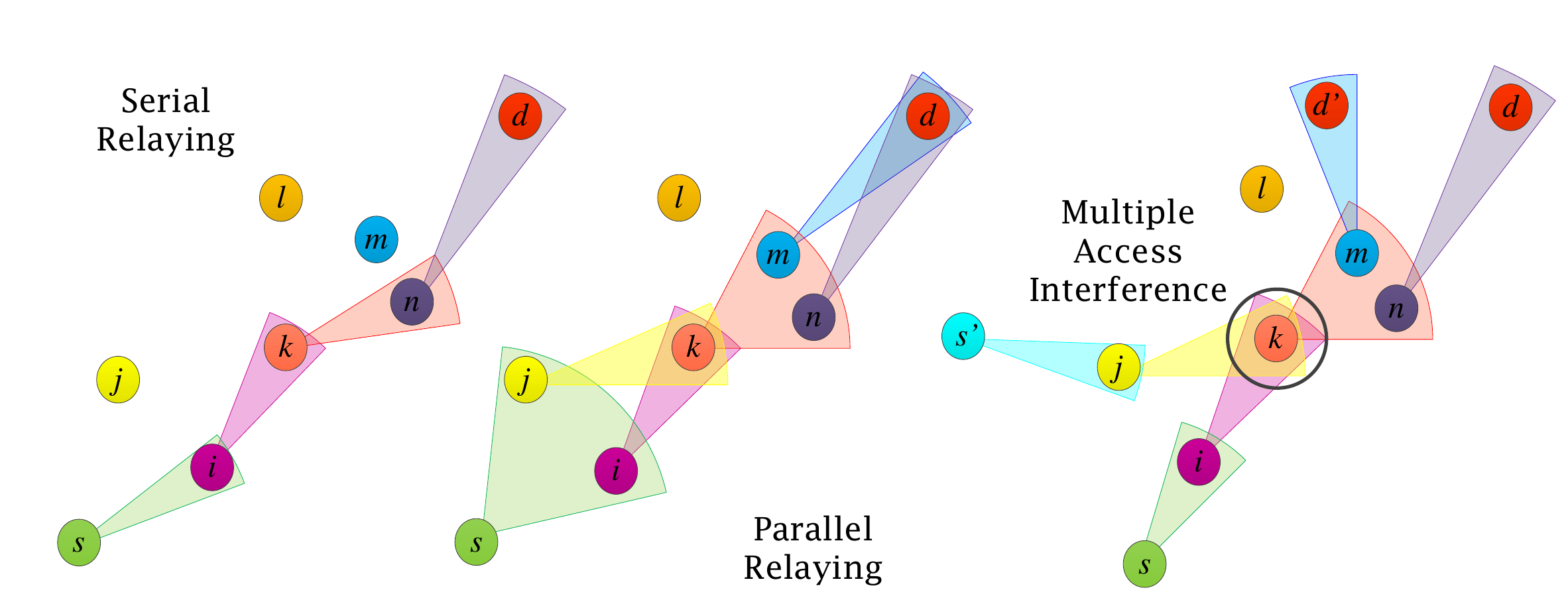}  
\caption{Illustration of serial and parallel relaying techniques between a source and destination pair along with an MAI interference scenario.}
\label{fig:relaying}  
\end{center}  
\end{figure*}

\subsection{Relaying Techniques}
\label{sec:relaying}

As depicted in Fig \ref{fig:relaying}, relaying can be implemented by involving either a single node or multiple nodes at each hop, which are referred to as \textit{serial} and \textit{parallel} transmission, respectively. 

\subsubsection{Serial Relaying and PAT Mechanisms}
\label{sec:serial}
Serial transmission (a.k.a. multihop transmission) employs the relaying nodes in a serial configuration along a certain routing path \cite{celik2018modeling, Jamali2016, Jamali2017}, which is especially beneficial to extend the communication range and expand the cell coverage in ad hoc and cellular UOWNs. In \cite{Jamali2016}, authors exploited the serial relaying to expand the coverage area of OCDMA based UOWNs. They evaluated the end-to-end performance of the proposed relay-assisted OCDMA network under absorption, scattering, and turbulence effects. In \cite{Jamali2017}, end-to-end BER performance of a multi-hop transmission was analytically evaluated by using single-hop BER expression as a building block. Authors in \cite{Jamali2016, Jamali2017} have applied Gauss Hermite quadrature formula and derived the closed-form BER solution under the lognormal fading channel.  In \cite{Jamali2017}, end-to-end BER performance is obtained by assuming that each hop experience the same error of probability, which may not be the case in reality. Therefore, an end-to-end BER performance analysis was considered in \cite{celik2018modeling} where we have distinguished the error probability of each transmission hop. 

The key point in multi-hop transmission is to employ narrow-beam light sources in order to concentrate the received signal power at the receiver aperture area. Although narrow-beam transmission significantly enhances the system performance at each hop, it requires highly directional beams and rapid PAT mechanisms which accounts for beam wander and jitters due to aquatic turbulence and random motion patterns (roll, pitch, and yaw) of the transceiver platforms \cite{Juarez2006}. Furthermore, the precision of the localization algorithms is quite decisive for positioning the complementary node in its FoV during the acquisition \cite{Ho2004}. Lastly, a fast closed-loop tracking and wavefront control is necessary to sustain a constant link \cite{Nikulin2006}. To the best of our knowledge, there is no study addressing the PAT mechanisms for UOWNs yet. When the location accuracy is low, pointing errors and misalignment could be mitigated by ensuring a certain diffusion area (proportional to the localization error) rather than directly pointing to the estimated receiver location. Therefore, it is essential to develop robust and adaptive divergence and power control schemes \cite{LoPresti2006}.

\subsubsection{Parallel Relaying and Relay Selection Protocols}
\label{sec:parallel}

Parallel transmission (a.k.a. cooperative transmission) is an alternative relay-assisted transmission scheme and basically built upon the idea that the source node may be overheard by a number of neighboring nodes which can act cooperatively to relay traffic request of the source node. In other words, a set of transmitting nodes (probably each with a single optical transmitter) jointly process and transmit the traffic request by creating a virtual antenna array \cite{Uysal2009}. This cooperation naturally increases the degree of diversity and provides opportunities to mitigate multipath fading effects. Even though parallel relaying has received quite an attention in TOWCs (please see \cite{Khalighi2014} and references therein), there is no UOWC work addressing the virtue and benefits of the cooperative relaying. 

Relay selection is an interesting research topic for cooperative relaying schemes because involving all the neighbor in transmission may always not yield the desired results \cite{Chatzidiamantis2013}. This is mainly because of the fundamental tradeoff between the divergence angle and received transmission power (or the communication range for a fixed power reception). In Fig.~\ref{fig:relaying}, for instance, relay node $\ell$ does not participate in relaying as it does not provide a better performance than involving relay nodes $m$ and $n$ only. MAI raises another issue when a relay node is incorporated with relaying to convey two different data streams as shown in Fig. \ref{fig:relaying} where node $k$ is not able to serve data streams $s \rightarrow d$ and $s' \rightarrow d'$ at the same time unless it employs an efficient multiple access scheme. Notice that node $k$ constitutes the bottleneck of these two data streams and such critical nodes mainly determine the overall network performance. It is important to develop adaptive divergence and power control schemes \cite{LoPresti2006} for employing efficient relay selection strategies in order to sustain and improve the network performance. 

\subsubsection{Traffic Forwarding Methods}
\label{sec:forwarding}
Inspired by the methods in the well-known TOWC parts, several signaling strategies can be employed for UOWCs:
\paragraph{DF Relaying}
In DF, the received optical signal at each hop is converted into electrical signal, then decoded, and finally re-encoded before retransmission for the next hop. Although DF greatly improves performance as it limits background noise propagation, it may introduce significant power consumption and encoding/decoding delay to the system \cite{celik2018modeling}.

\paragraph{AF Relaying}
AF is conventionally realized by executing optical-electrical-optical (OEO) conversion at each node, amplifying the received signal electrically, and then retransmitting the amplified signal for the next hop. However, actual merits of AF relaying over the DF counterpart emerges only if OEO conversion is eliminated. Alternatively, all-optical AF relaying process received signal in the optical domain and requires only low-speed and low-power electronic circuitry to adjust the amplifier gain \cite{Kazemlou2011}. The main drawback of the AF transmission is the propagation of noise added at each node, which is amplified and accumulated through the path \cite{celik2018modeling}.

\paragraph{BDF Relaying}
Different from the DF method, the relay node detects each transmitted bit of the source and forwards it to the next relay without applying any error correction \cite{Karimi2009}.

\begin{figure}
\begin{center}  
\includegraphics[width=1\columnwidth]{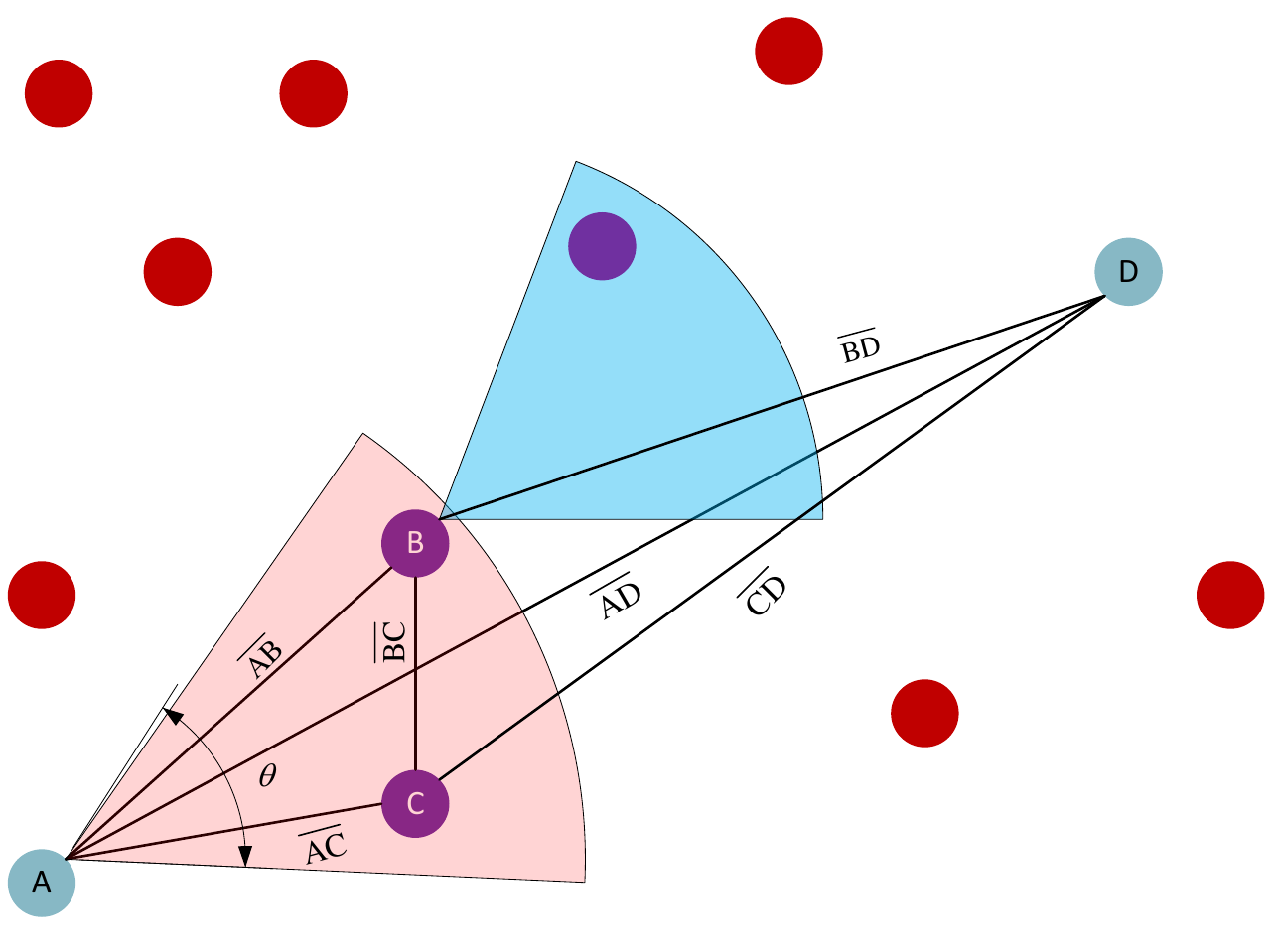}  
\caption{Illustration of focus beam routing protocol \cite{Jornet2008}.}
\label{fig:fbr}  
\end{center}  
\end{figure}

\subsection{Underwater Routing Techniques}
\label{sec:routing}

Routing holds a significant place in order to keep the UOWNs connected by discovering and maintaining the transmission routes. The physical layer issues for UOWNs are well studied in the recent past but the research on network layer issues such as routing is still in its infancy. A number of routing protocols for underwater acoustic wireless networks have been highlighted in \cite{AYAZ2011, Ning2016, Ahmed2017, Lu2017, Sahana2018} some of which can be well adapted for UOWNs. The key point in adapting the existing routing protocols is that designers should take the angular sector shaped coverage region of optical nodes along with the fundamental tradeoff between the angle and radius of this sector. In what follows, we highlight some of the routing algorithms proposed for underwater acoustic networks, which can also be adopted to apply for UOWNs:
\subsubsection{Location based routing}
The location information of underwater sensors is used in location-based routing strategy to discover the best route from the source to the destination node. In location-based routing, every node should be aware of its location, the target area, and neighbors' locations. The data is forwarded in accordance with the location information. AUV based routing protocols were proposed in \cite{Hirai2012,Tanigawa2015} which integrate localization and routing. An energy efficient and reliable routing protocol was introduced in \cite{Fu2013}, where the transmission from the source node starts with local flooding and then an adaptive mechanism is established to find the optimal route with minimum energy consumption. Directional flooding protocol were proposed in \cite{Hwang2008,Ahmed2017r} where the source node knows its own location, the sink location, and the location of its neighbors. The flooding region in \cite{Hwang2008,Ahmed2017r} was defined by the link qualities among the neighbors. The flooding phenomena can burden the network therefore, in \cite{Jornet2008} the authors have proposed a routing protocol based on focused beam. It is assumed in \cite{Jornet2008} that every node knows its location and location of the destination node, where the decision about the next hop is made at each intermediate node. Focus beam routing is a good candidate for UOWNs due to its directive nature from source to destination. Fig.~\ref{fig:fbr} shows the data forwarding scheme used in focus beam routing, where node ``A" is the sender node and node ``D" is the destination node, the intermediate nodes are selected based on the cone angle $\theta$ (which can be considered as twice of the divergence angle). Nodes which lies within the cone angle $\pm \theta/2$ of  the sender node, are selected as relay nodes for forwarding the data.

\begin{figure}
    \centering
    \begin{subfigure}[b]{0.4\textwidth}
\includegraphics[width=1 \columnwidth]{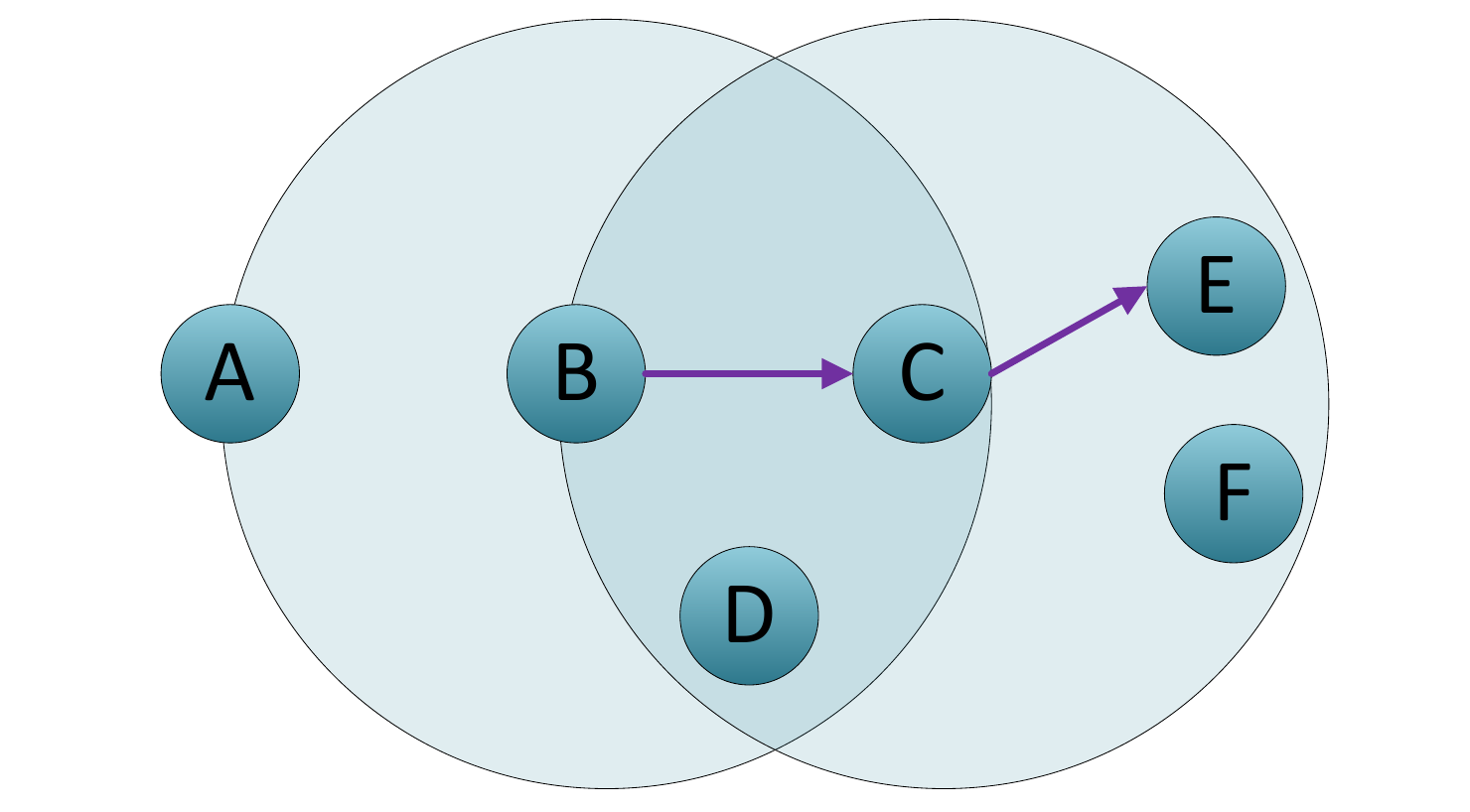}
    \caption{Multihop omnidirectional optical links.}
    \label{fig:omni}
    \end{subfigure}

    \begin{subfigure}[b]{0.4 \textwidth}
\includegraphics[width=1 \columnwidth]{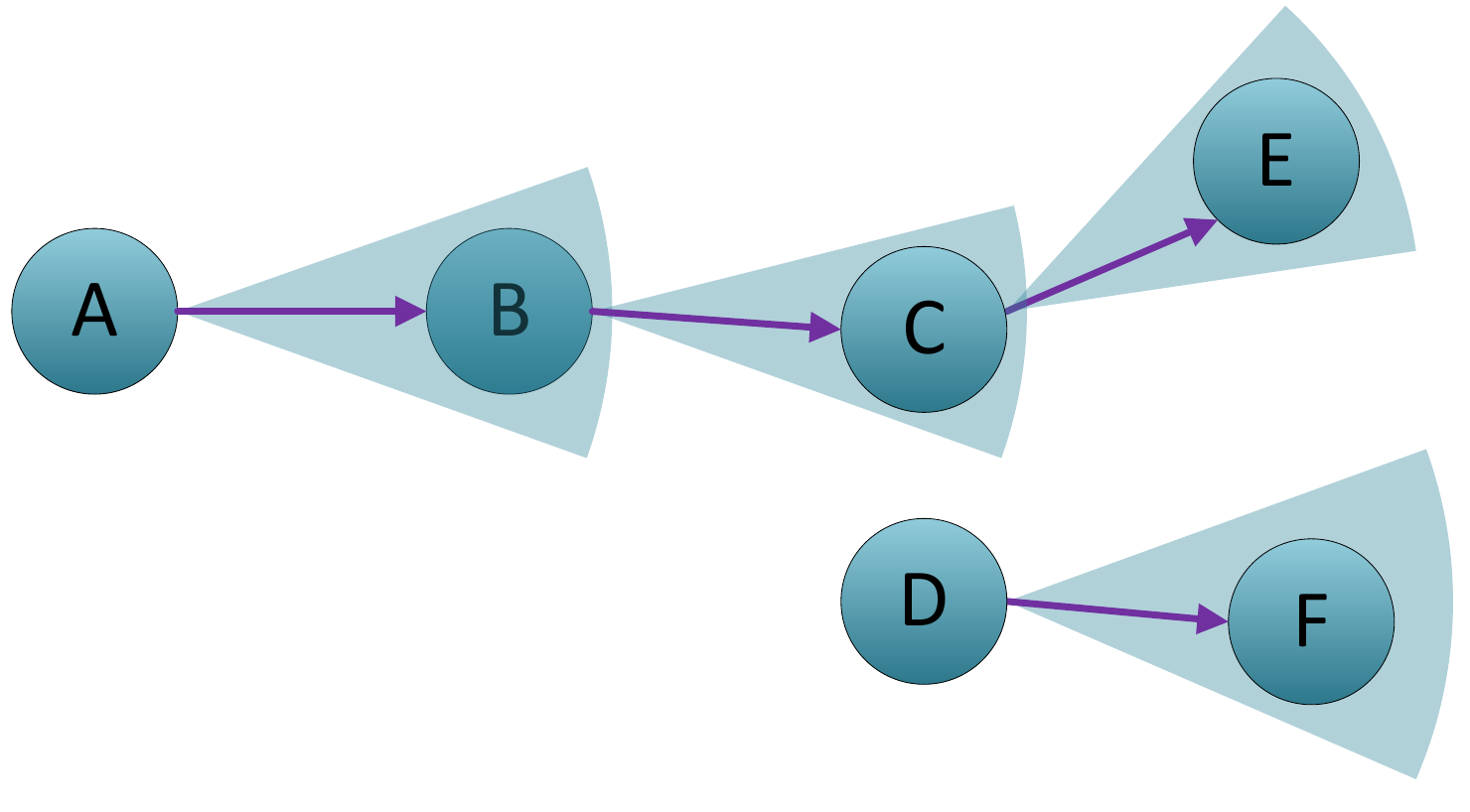}
      \caption{Multihop directional optical links.}
      \label{fig:directive}
    \end{subfigure}

    \begin{subfigure}[b]{0.4 \textwidth}
\includegraphics[width=1 \columnwidth]{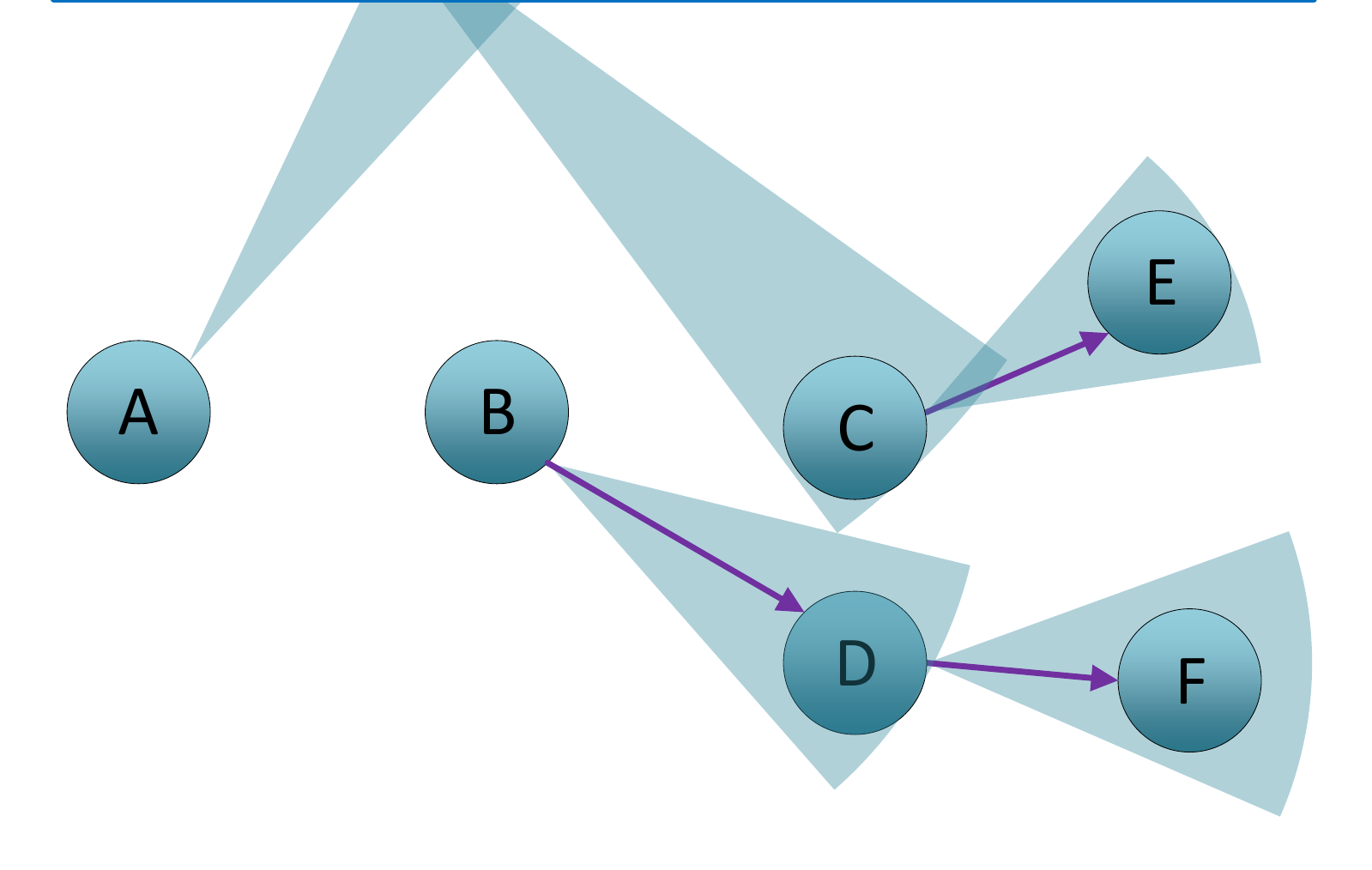}
      \caption{Multihop directional optical links with reflection.}
      \label{fig:directiveref} 
    \end{subfigure}
    \caption{Multihop underwater routing protocols: a) Omnidirectional, b) LoS directive, and c) NLoS directive (Reflective) \cite{Emokpae2012}.}
\label{fig:gorilla}
\end{figure}

A geographical reflection enabled routing protocol was introduced in \cite{Emokpae2012} which tries to find the stable route between the source node and the destination node. Directional antennas were used in \cite{Emokpae2012} to consider both LoS and NLoS links between the neighbor nodes. Fig.~\ref{fig:gorilla} shows the different scenarios for the proposed routing protocol in \cite{Emokpae2012}, where it can be seen from Fig.~\ref{fig:directive} and  Fig.~\ref{fig:directiveref} that the directive and NLoS communication can help in simultaneous transmissions respectively, thus improving the throughput of the network. The proposed routing protocol in \cite{Emokpae2012} was designed for acoustic underwater networks which can also be well adopted for UOWNs. Sector based routing protocols were designed in \cite{Chirdchoo2009, Zhang2013r} with the location prediction of destination. The network topology in \cite{Chirdchoo2009, Zhang2013r} is fully mobile where each node moves along a pre-defined route. Comparative study of location-based routing protocols for underwater acoustic networks was carried out in \cite{Khalid2017}.  In all of the location-based routing protocols, it is assumed that the underwater sensor nodes find its location by using GPS or by using underwater local positioning systems. However, GPS cannot work in the underwater environment  and the underwater local positioning techniques have large localization error due to the hostile underwater environment.

\subsubsection{Source based routing}
A simple and energy efficient source based routing protocol was introduced in \cite{Wei2014}. The protocol in \cite{Wei2014} selects the route with minimum transmission delay from source to the sink node. Once the route is defined, the nodes along the route can also transmit the data to the sink node. The average end to end delay, average energy consumption, and packet delivery ratio of the proposed protocol in \cite{Wei2014} outperforms other traditional routing protocols. Another source based routing protocol for small size UAWC networks was proposed in \cite{Kim2007} where each node just share information with its single-hop neighbor nodes and find a minimum cost path from source to the destination. Source-based routing protocols are good for reducing the energy consumption of routing protocols.
\subsubsection{Hop-by-hop routing}
In hop-by-hop routing, the intermediate nodes (or relay nodes) selects the next hop by itself. Hop-by-hop routing provides flexibility and scalability to the network but the route selection may always not be optimal. Channel aware hop-by-hop routing protocols were introduced in \cite{Chen2014, Chen2017}  where the speed of acoustic waves in different depth were taken into consideration for the relay nodes to reduce the end to end transmission delay. In \cite{Zhang2013r}, the authors have proposed a hop-by-hop routing protocol based on beam-width and direction of the intermediate nodes. Adaptive depth based routing protocols were introduced in \cite{Chen2012, Coutinho2013} which takes into account the speed of acoustic waves at different depth levels, depth of sink node, and distance to sink node.  A MIMO-OFDM based routing protocol was introduced in \cite{Kuo2012} to take the advantage of multiplexing and diversity gain adaptively. The proposed cross-layer design in \cite{Kuo2012} adapts itself to the noise and interference for underwater acoustic channels and selects a suitable transmission mode for the subcarriers. An energy efficient and network topology aware greedy routing protocol was proposed in \cite{Wu2010} which assigns adaptive weights to the highly connected nodes. For underwater delay tolerant networks a redundancy-based routing protocol was designed in \cite{Guo2008} which adopts a tree-based forwarding method to replicate packets.

\subsubsection{Cross Layer routing}
The cross-layer routing protocols take the information available from different layers into account and provide a solution to several networking issues such as scheduling, defining routing policy, and power control. Cross-layer routing protocols can also select the next hop for transmission by considering the transmission delay, distance to sink, channel conditions, and buffer size of the candidate node. Cross-layer strategy increases the overall network performance and minimizes the energy consumption of the network. Cross-layer protocols for the 3D underwater environment were investigated in \cite{Pompili2006r, Goetz2012} which utilizes the channel efficiently and sets the optimal packet size for transmission. Multipath power control routing protocols were proposed in \cite{Zhou2011, Min2012} which combine multipath routing and power control at the sink node. Channel aware cross-layer routing protocol are also investigated in \cite{Basagni2012} which exploits the link quality for the relay selection.
\subsubsection{Clustered routing}
Cluster based routing is especially suited for infrastructure based UOWNs as shown in Fig. \ref{fig:uwsn}. In cluster based routing, the network is divided into a number of clusters/cells based on the geographical location of the nodes. Once the network is divided into clusters, the cluster head (i.e., OAP/OBS) is selected for each cluster by using any cluster head selection strategy. The cluster head is used as a gateway to communicate between the clusters and to the sink node. A location-based routing protocol was introduced in \cite{Anupama2008} which divides the network into clusters and the data from the nodes are gathered by the cluster heads. A distributed clustering based protocol was proposed in \cite{Domingo2007} where the communication between the cluster head and the sensor node was single hop. Location unaware cluster based multihop routing protocol was proposed in \cite{Wang2011} where the sensor nodes do not know their location and location of the cluster head. The interested readers are referred to \cite{Sandeep2017} where a number of cluster based routing protocols are highlighted for underwater wireless sensor networks.
\subsubsection{Reinforcement learning based routing}
The routing protocols based on reinforcement learning uses Q-learning method for the network states and adapts itself to the topology change. The node analyzes its remaining energy and energy of its neighbor nodes, applies a reinforcement function, and then selects the optimal node to forward the data \cite{ Tiansi2013}. The routing problem in \cite{ Tiansi2013} is fully distributed and formulated as a Markov decision process where the state space consists of all the nodes. A machine learning based routing protocol was proposed in \cite{Hu2010}  which is energy efficient and improves the lifetime of the network. A layer structured based routing protocol was introduced in \cite{Hu2012} for hybrid acoustic and optical architecture where the upper layer cluster heads supervise the routing in lower layer by using the Q-learning function.

\begin{figure*}
    \centering
    \begin{subfigure}[b]{0.48\textwidth}
\includegraphics[width=0.99 \columnwidth]{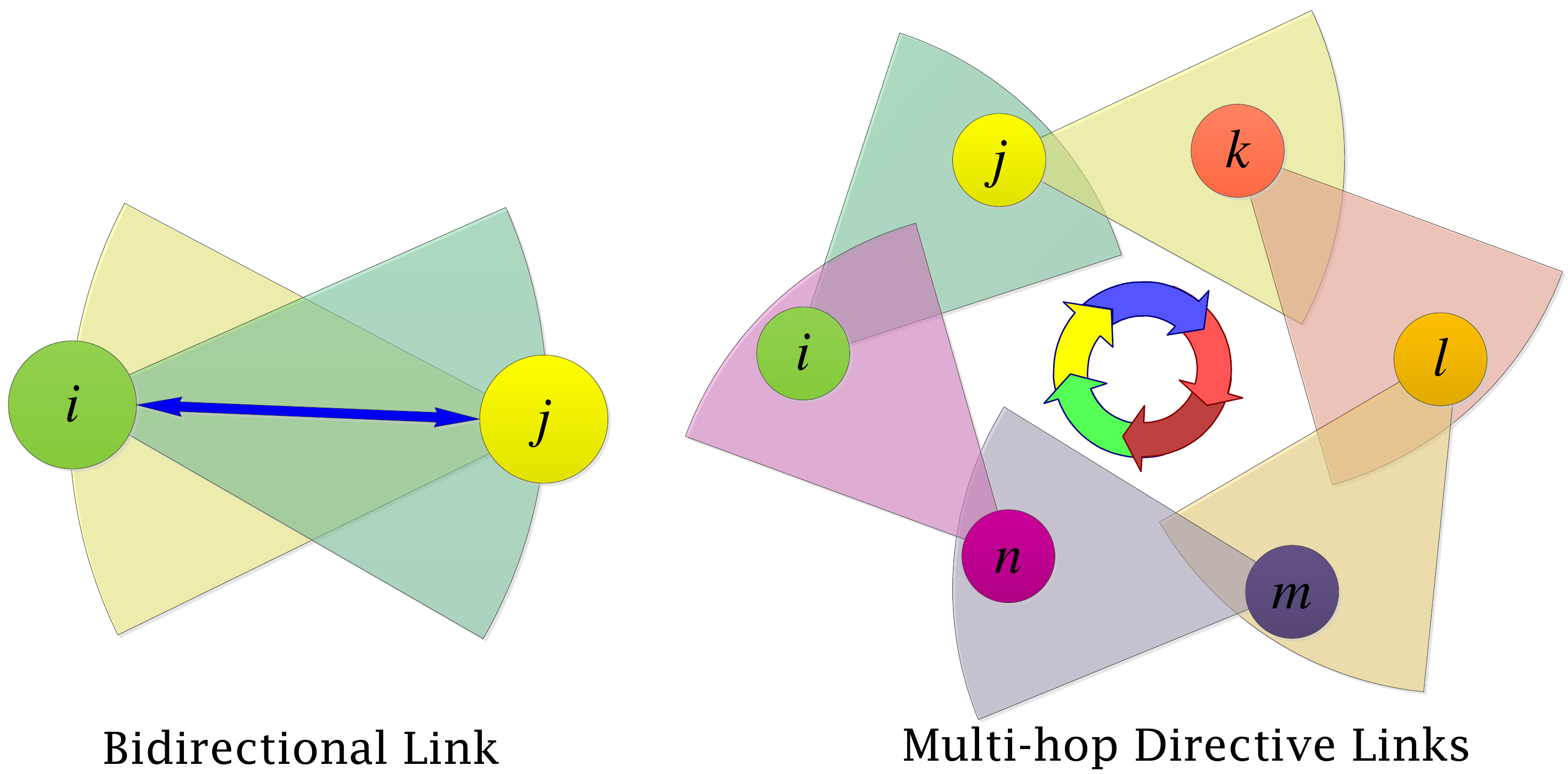}  
\caption{Connection types in random sector directed graphs.}
 \label{fig:connecttype}
    \end{subfigure}
    \begin{subfigure}[b]{0.48\textwidth}
\includegraphics[width=1\columnwidth]{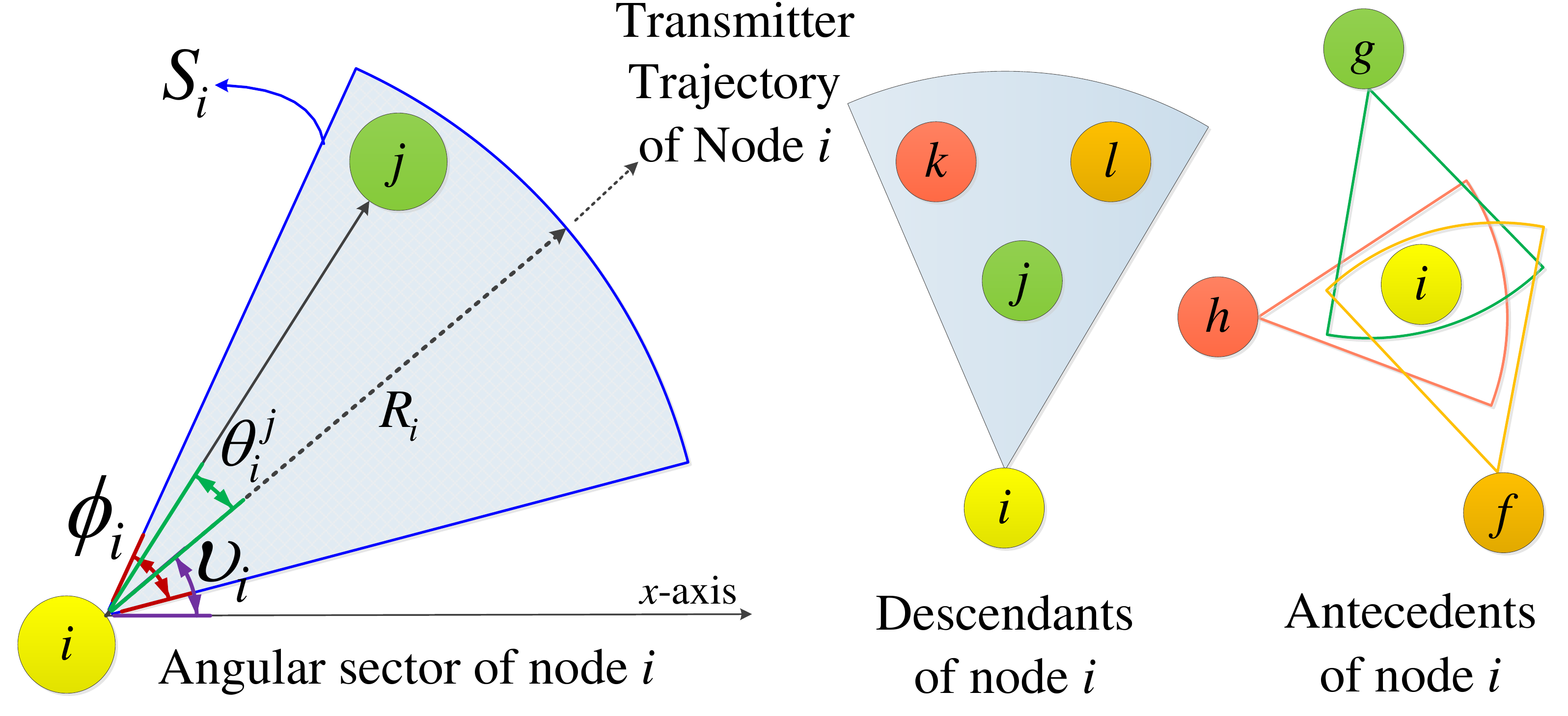}  
\caption{Connectivity parameters, descendants, and antecedents of $n_i$.}  
\label{fig:connectdemo}
    \end{subfigure}
\caption{Depiction of connection types, connection parameters, descendants, and antecedents for random sector directed graphs.}
\label{fig:connect}
\end{figure*}

\section{Transport Layer: Connectivity, Reliability, and Flow/Congestion Control}
\label{sec:transport}
Unlike the first two lower layers, transport layer of UOWNs is still in a primitive stage and remains totally unexplored. Therefore, this section discusses the fundamental challenges for developing an efficient transport layer including connectivity, reliability, flow control, and congestion control aspects of UOWNs. 

\subsection{Connectivity analysis of UOWNs}
Connectivity of UOWNs is the most critical component of transport layer as other network functions heavily depend on a connected network assumption. It is also used as a metric for different performance parameters such as survivability, robustness, and fault tolerance \cite{Dall2002}. Connectivity is measured by number of links in the network and a network is referred to be as connected if there exists at least one connecting path between any two nodes in the network. In strongly connected networks bidirectional links exist between any pair of nodes while in a directed network the links are usually unidirectional until and unless both nodes are in the beam scanning angle of each other \cite{Erdos1960}. The problem of network connectivity is addressed in \cite{Betstetter2004} for omnidirectional networks such that no node is obscured for RF wireless sensor networks. The exact closed-form analytical expression of connectivity in multihop wireless networks for physical layer parameters still remains as an open research problem. The connectivity parameters of UOWNs depends on the transmission range of optical sensor nodes, number of optical sensor nodes, number of descendants and antecedents, node orientation, and the beam width (see Fig. \ref{fig:connectdemo}).

Range limitation of UWOC can be augmented with multi-hop UOWNs where nodes can share information for long distances through intermediate nodes. Indeed, multi-hop cooperative communications have been extensively studied for RF networks \cite{Uysal2009},  underwater acoustic networks \cite{Liao2016}, and TOWNs \cite{alquwsiee2015}.  Due to the omnidirectional communication capability of RF and acoustic signals, wireless sensor networks are traditionally modeled as geometric random graphs \cite{Gupta1998} where two sensor nodes $n_i$ and $n_j$ are generally assumed to establish a bidirectional communication link (i.e., $n_i \leftrightarrows n_j$). On the contrary, such a model is not suitable for UOWNs because a node can only reach to the nodes within a certain beam scanning angle around their transmission trajectory, that is, optical wireless nodes are connected via unidirectional links. Directed communication networks are generally modeled by random scaled sector graphs \cite{Wu2014} where a unidirectional communication link from node $n_i$ to $n_j$ (i.e., $n_i \rightarrow n_j$) is established if and only if $n_j$ is positioned within the beam scanning angle of $n_i$. Notice that a directed reverse path is possible (i.e., $n_j \rightarrow n_i$) if  $n_i$ is in the beam-width of $n_j$  or through other multi-hop path as illustrated in Fig.~\ref{fig:connecttype}. A connectivity framework for multihop UOWNs was discussed in \cite{Vavoulas2014} where the authors have assumed bidirectional links between every pair of optical sensor nodes. In \cite{Nasir2018icc}, we have analyzed the connectivity of  UOWNs by using random sector graphs where we have considered unidirectional links between underwater optical sensor nodes.

%

\begin{figure*}
    \centering
    \begin{subfigure}[b]{0.48\textwidth}
    \includegraphics[width=0.99 \columnwidth]{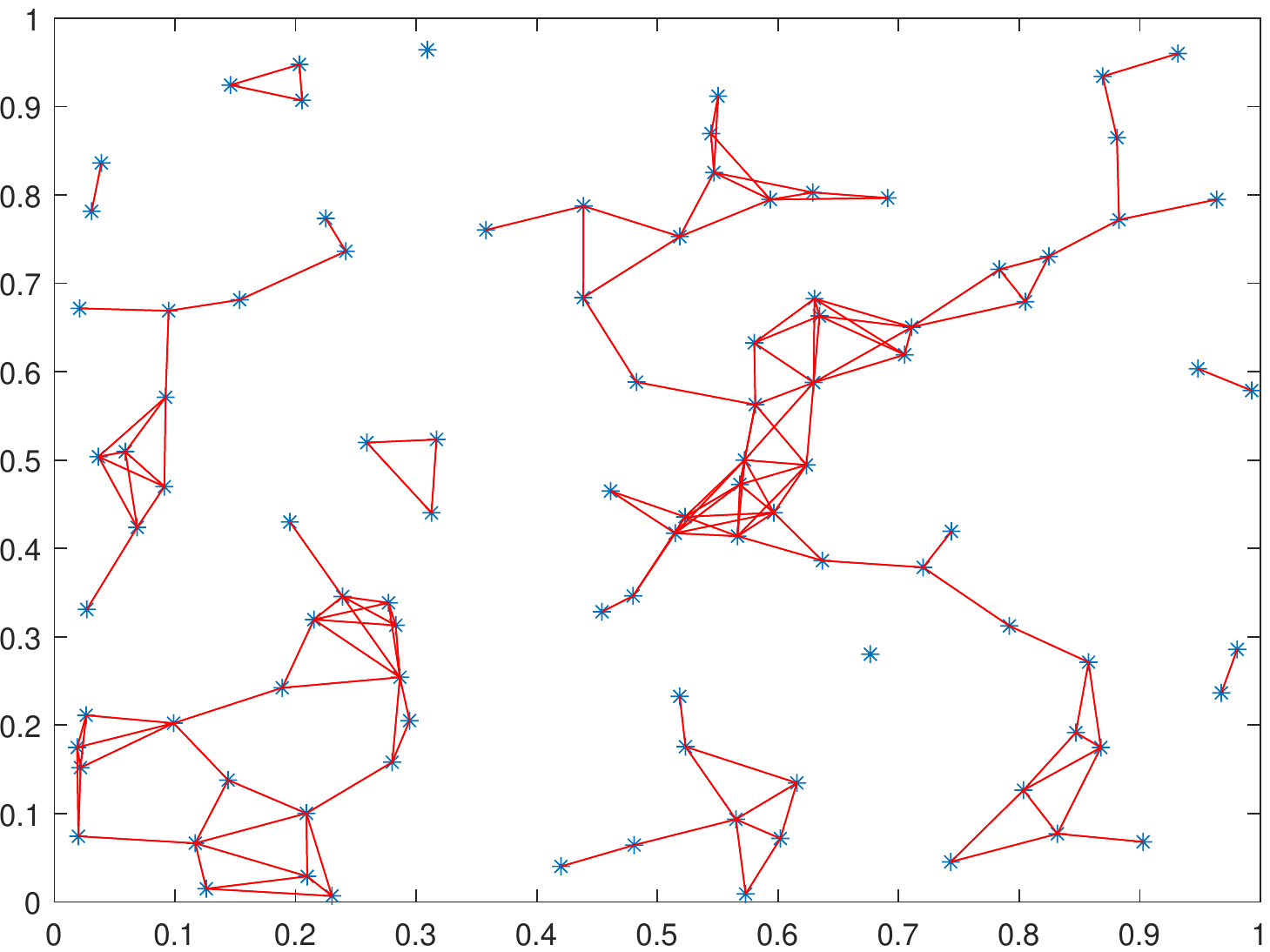}
    \caption{ $m=100$, $\phi = \frac{\pi}{3}$, and $R=0.2 m$.}
    \label{fig:pibythree}
    \end{subfigure}
    \begin{subfigure}[b]{0.48\textwidth}
      \includegraphics[width=0.99 \columnwidth]{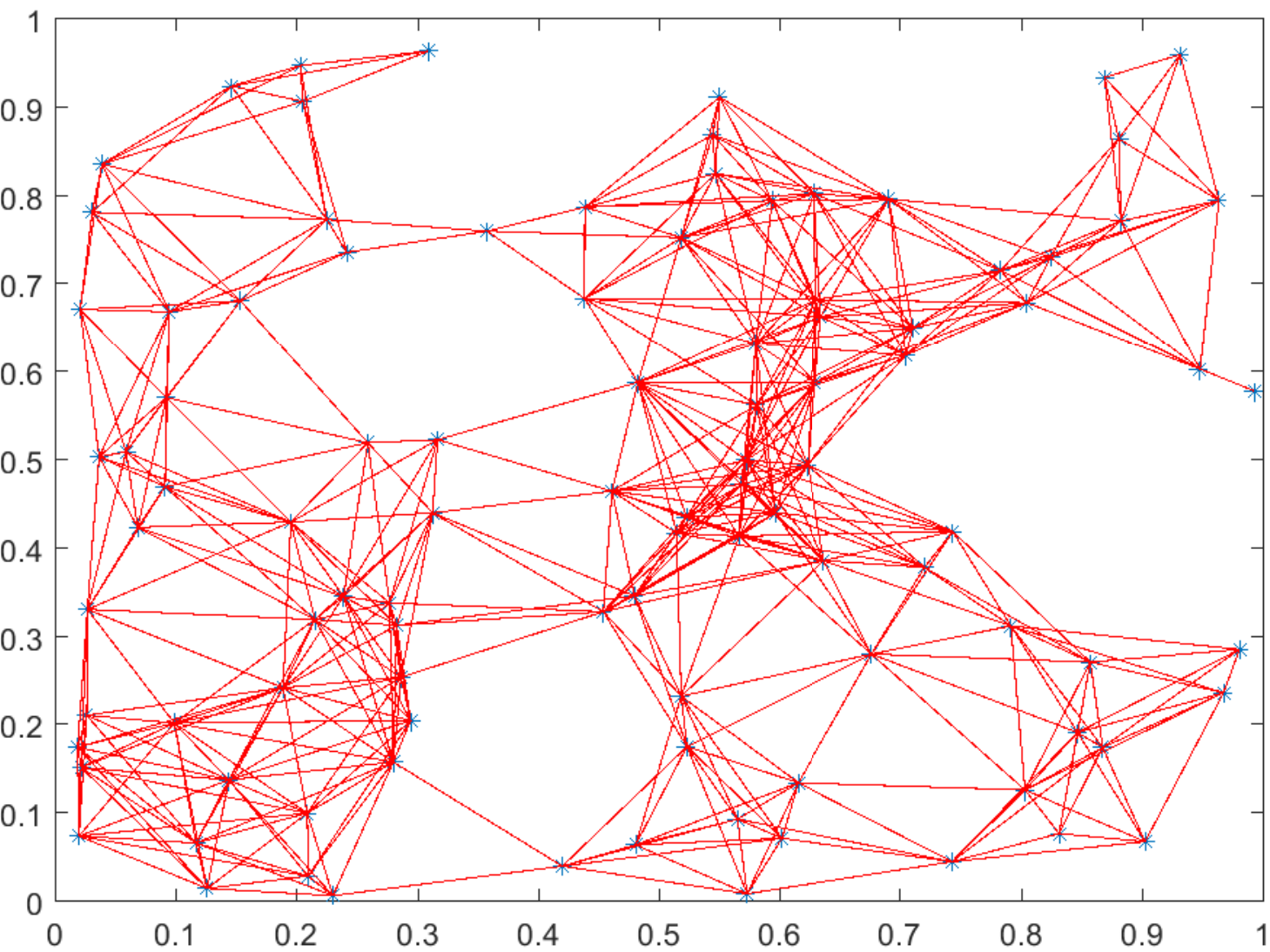}
      \caption{$m=100$, $\phi = 2\pi$, and $R=0.2 m$. }
      \label{fig:twopi}
    \end{subfigure}
\caption{Illustration of random directed graph scenarios for different parameters.}
\label{fig:random_graph}
\end{figure*}

%

\begin{table*}[htb!]
\centering
\caption{Comparison  of connectivity analysis of wireless communication systems}
\label{Tableconnectivity}
\begin{tabular}{|l|l|l|l|}
\hline
\hline
 Literature              & Channel model        & Link type   & Graph model  \\ \hline 
\cite{Gupta1999, Gupta2000, Ozgur2007, ONAT2008, Ammari2009, Ammarii2009, Coon2015, Yang2017}                      & Terrestrial RF       & Bidirectional  & Random graphs      \\
 \cite{Ravelomanana2004, Akkaya2009, Pompili2006, SENEL2015, Wang2017}                      & Underwater Acoustic  & Bidirectional & Random graphs \\ 
\cite{Diaz2003, Kundur2006, Kundur2007, Xie2011,  Rong2010, Ferrero2014}                      & Terrestrial Optical  & Bidirectional/Unidirectional & Random graphs/Random sector graphs\\
\cite{Vavoulas2014, Kaushal2016underwater, Nasir2018limited, Nasir2018icc}                      & Underwater Optical   & Bidirectional/Unidirectional         & Random sector graphs          \\
\hline
\hline        
\end{tabular}
\end{table*}

In order to define a random sector graph consider that the total number of optical nodes are $m$, the scanning sector (coverage area) of $n_i, 1\leq i \leq m$, is defined as a tuple of random orientation $\zeta_i$, scanning angle $\phi_i$, communication range $R_i$, and sensor node coordinates $\mathbf{c}_i$, i.e., $\mathbf{S}_i=(\zeta_i,\phi_i, R_i, \mathbf{c}_i)$ which is illustrated in Fig. \ref{fig:connectdemo}. Accordingly, UOWNs can be defined as a random sector directed graph $\mathcal{G}(\mathbf{\mathcal{V}},\mathbf{\mathcal{E}})$  where $\mathbf{\mathcal{V}} = \{ \mathbf{c}_1, \ldots, \mathbf{c}_i, \ldots, \mathbf{c}_M \}$ represent the set of vertices and $\mathbf{\mathcal{E}} \in \{0,1\}^M$ is the set of links which is primarily characterized $\mathbf{S}=\mathbf{S}_1, \ldots, \mathbf{S}_i, \ldots, \mathbf{S}_M$. Notice that $\mathbf{\mathcal{E}}_{i,j}=1$ only if $n_i \rightarrow n_j$ holds. Random sector directed graphs and random geometric graphs are identical in case of $\phi = 2\pi$ \cite{Erdos1960, Martin1997, Dall2002, Wu2014}. Notice that two nodes $i$ and $j$ are connected when the distance between them is less than $R$ in random geometric graphs, however, the connectivity of random directed sector graphs also depends on the beam scanning angle and its orientation. Fig.~\ref{fig:pibythree} and Fig.~\ref{fig:twopi} shows two different random directed sector graphs with scanning angles of $\phi = \frac{\pi}{3}$ and $\phi = 2\pi$, respectively. It is obvious that increasing the scanning angle for each node from $\phi = \frac{\pi}{3}$ to $\phi = 2\pi$, increases the number of links in the graph. These asymmetric and directional characteristics of the random directed sector graphs require us to define descendant and antecedent neighbors for every node.  The descendants of node $n_i$ are defined as $\mathcal{D}_i \triangleq \{n_j | \: ~\forall~j:~\mathcal{E}_{i,j}=1\}$, i.e., the set of nodes who lies within the coverage region of $n_i$, antecedents of $n_i$ are defined as $\mathcal{A}_i \triangleq \{n_j | ~\forall~j:~\mathcal{E}_{j,i}=1 \}$  the set of nodes who can reach to $n_i$. 
In Fig.~\ref{fig:connectdemo}, the set of descendants and antecedents of $n_i$ are shown as $\{n_j,n_k,n_l\}$ and $\{n_g,n_h,n_f\}$, respectively.

\begin{figure*}
    \centering
    \begin{subfigure}[b]{0.48\textwidth}
\includegraphics[width=1\columnwidth]{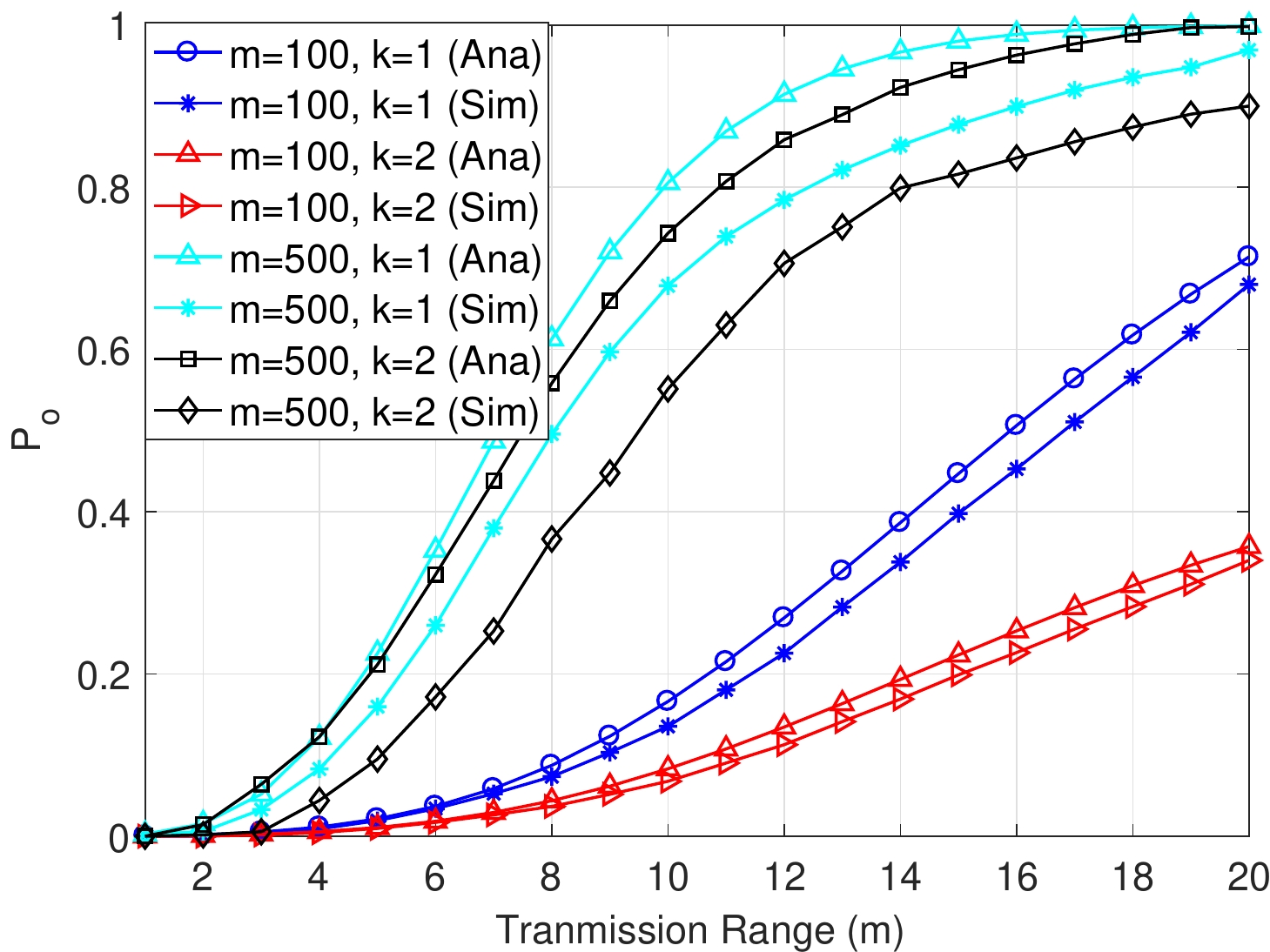}  
\caption{Probability of connectivity vs. transmission range for $\phi = \frac{2\pi}{9}$.\label{fig:twopibynine}}  
    \end{subfigure}
    \begin{subfigure}[b]{0.48\textwidth}
\includegraphics[width=1\columnwidth]{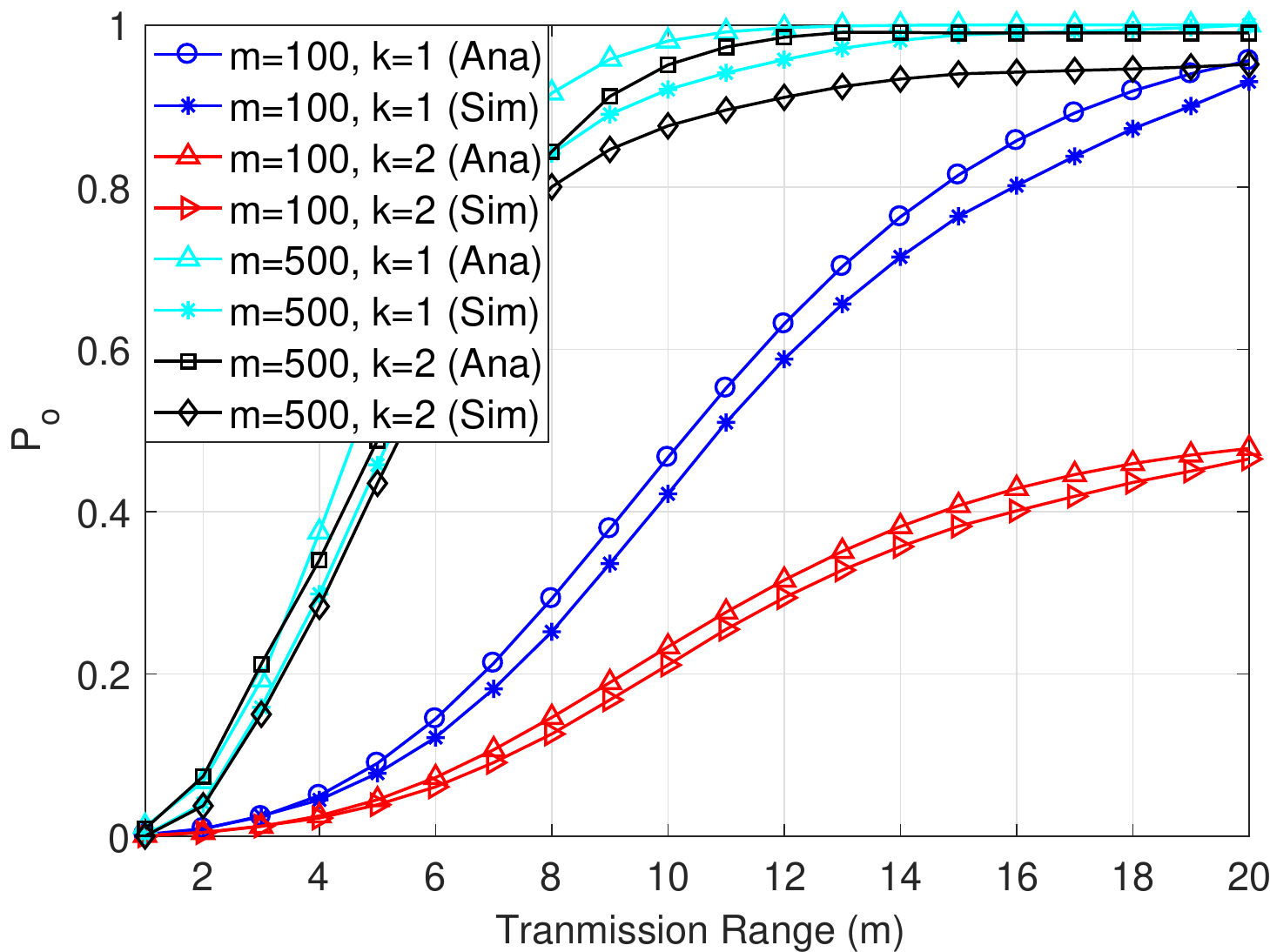}  
\caption{Probability of connectivity  vs. transmission range for $\phi = \frac{\pi}{2}$.\label{fig:pibytwo}}  
    \end{subfigure}
    
    \begin{subfigure}[b]{0.48\textwidth}
\includegraphics[width=1\columnwidth]{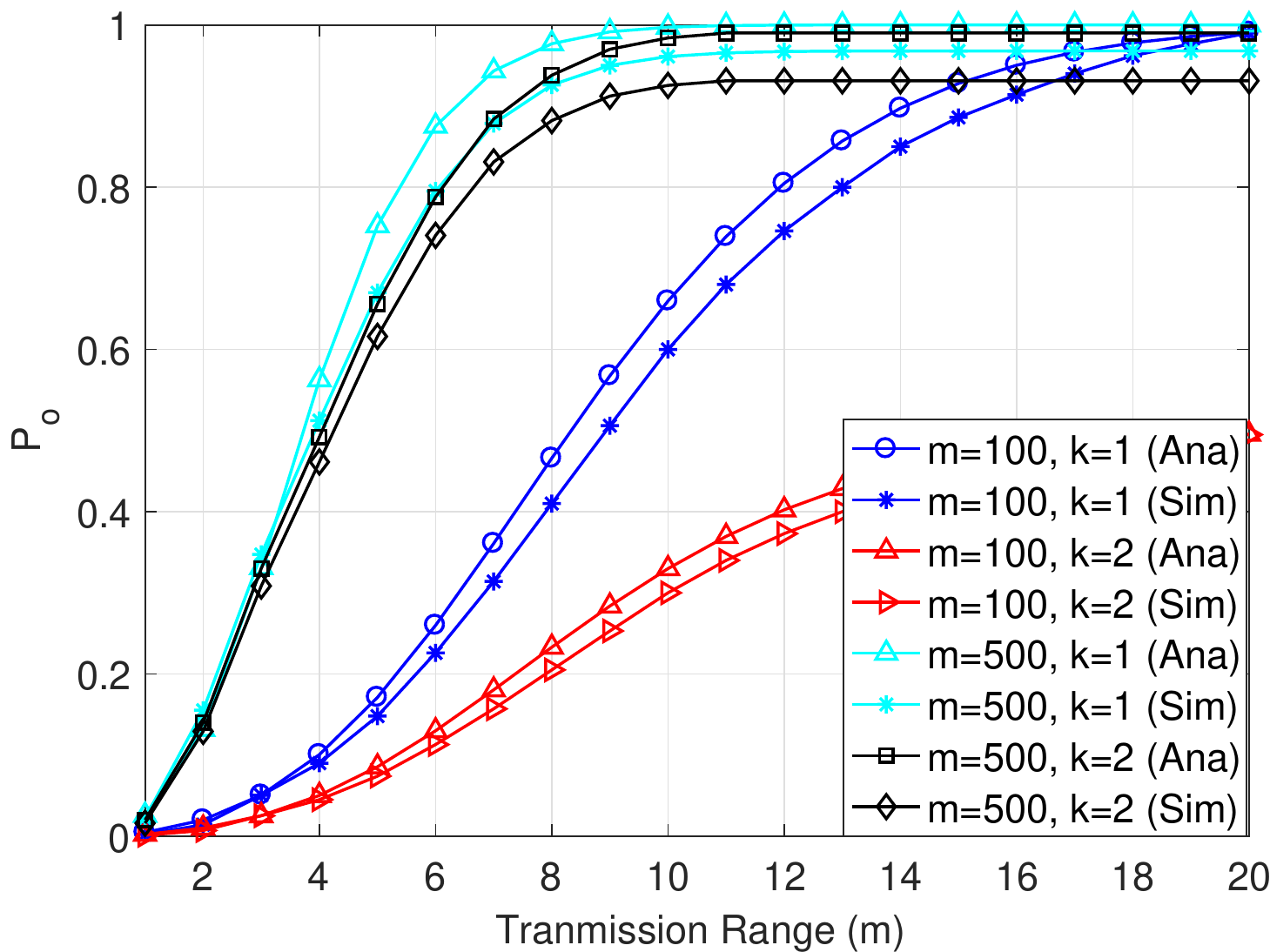}  
\caption{Probability of connectivity vs. transmission range for $\phi = \frac{3\pi}{4}$.\label{fig:threepibyfour}}  
    \end{subfigure}
    \begin{subfigure}[b]{0.48\textwidth}
\includegraphics[width=1\columnwidth]{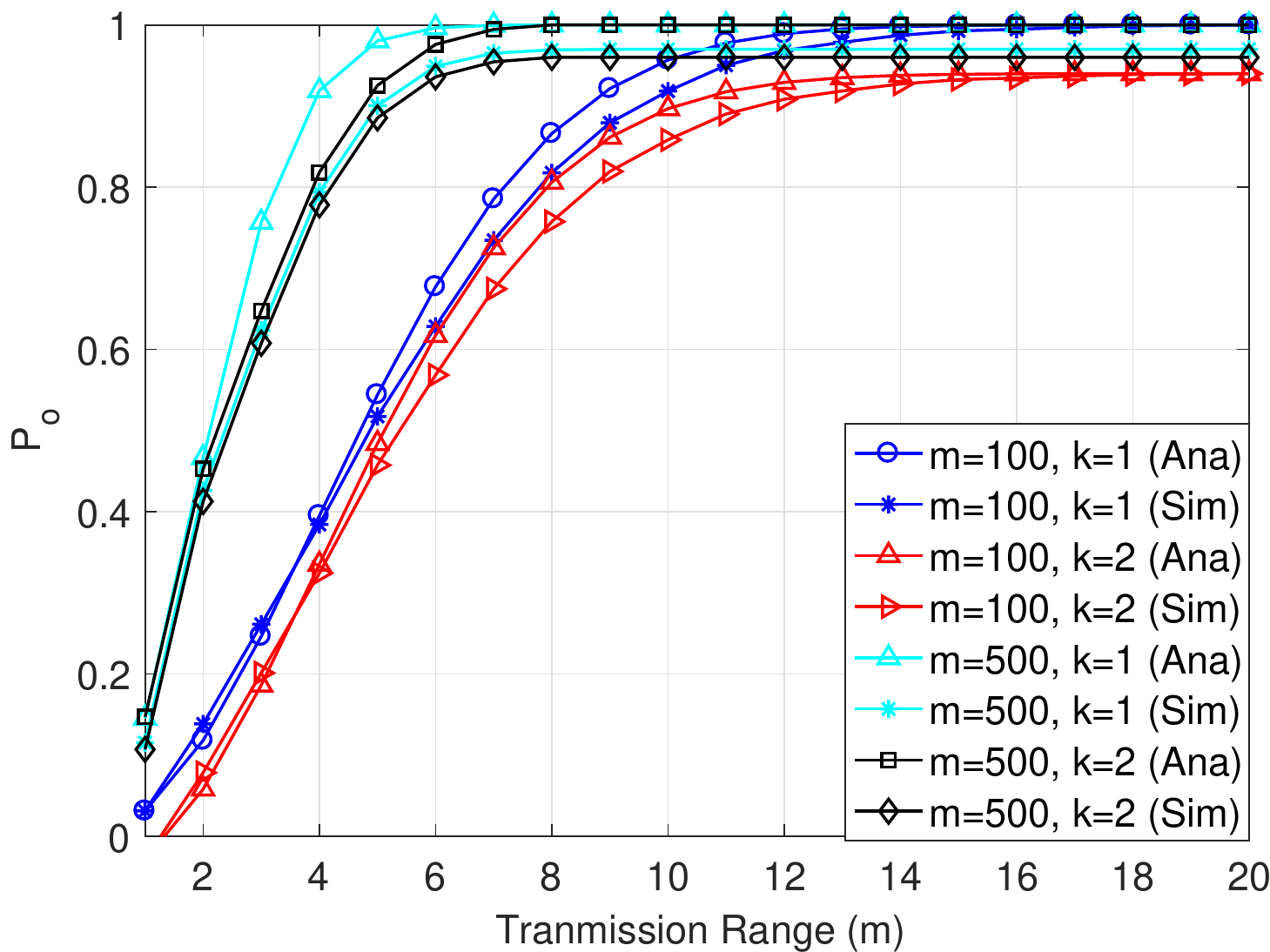}  
\caption{Probability of connectivity vs. transmission range for $\phi = 2\pi$.\label{fig:twopiii}}  
    \end{subfigure}    
\caption{Probability of connectivity vs. different transmission angles and ranges.}
\label{fig:prop_connect}
\end{figure*}

In order to find out the probability of a connected UOWN, we have considered networks of $m=100$ and $m=500$ optical sensor nodes randomly deployed in underwater $100~m \times 100~m$  square area respectively. The probability of a connected network is evaluated when each node is connected to at least one node ($k=1$) and when each node is connected to at least two other nodes ($k=2$). The transmission range $R$ varies from 1 to 20 meters and we set the beam scanning angles of the nodes with different widths of $\phi = \frac{2\pi}{9}, \frac{\pi}{2}, \frac{3\pi}{4}$, and $2\pi$ to see the impact of scanning angles on the probability of a connected network. Fig.~\ref{fig:twopibynine} - Fig.~\ref{fig:twopiii} shows that increase in the beam scanning angles, number of nodes, and transmission range results in high probability of a connected network. Table \ref{Tableconnectivity} summarizes the literature on connectivity analysis of different wireless networks.


\subsection{Reliability}
Packet losses may occur during the transport as a result of the hostile underwater channel impairments and network congestion. Hence, transport protocol can check the data corruptions by means of error correction codes and verify the correct receipt by the ACK/NACK messages to the source node. Considering the relation between a node and its  antecedents and descendants as described above, optical sensor nodes may always not be able to convey ACK/NACK messages to its antecedents. That is, operation of such a mechanism requires a fully connected network such that there is always another communication path to deliver ACK/NACK messages to the source node. Hence, it is essential to handle \textit{shadow zones} where temporary connectivity loss and high bit error rates occur \cite{Pompili2009}. Transmission control protocol (TCP) is the best-known connection-oriented transport layer protocol which assumes congestion as the only cause of packet loss and reduces the rate if packet losses occur. However, obstruction, pointing and misalignment events are quite common in UOWCs and an efficient UOWN transport layer protocol must distinguish between packet losses due to the congestion and channel impairments. Alternatively, user datagram protocol (UDP) is a connection-less transport layer protocol which may suite the UOWN better for very simple transmission applications. Rather than traditional end-to-end approaches, reliability can also be characterized in a hop-by-hop fashion. However, a hop-by-hop based reliability may not guarantee an end-to-end reliable network. Therefore, UOWNs paradigm necessitates novel transmission protocols which ensures the reliability by accounting for the underwater channel impairments and limited connectivity of UOWNs. 

\begin{figure*}[htb!]
\begin{center}  
\includegraphics[width=2\columnwidth]{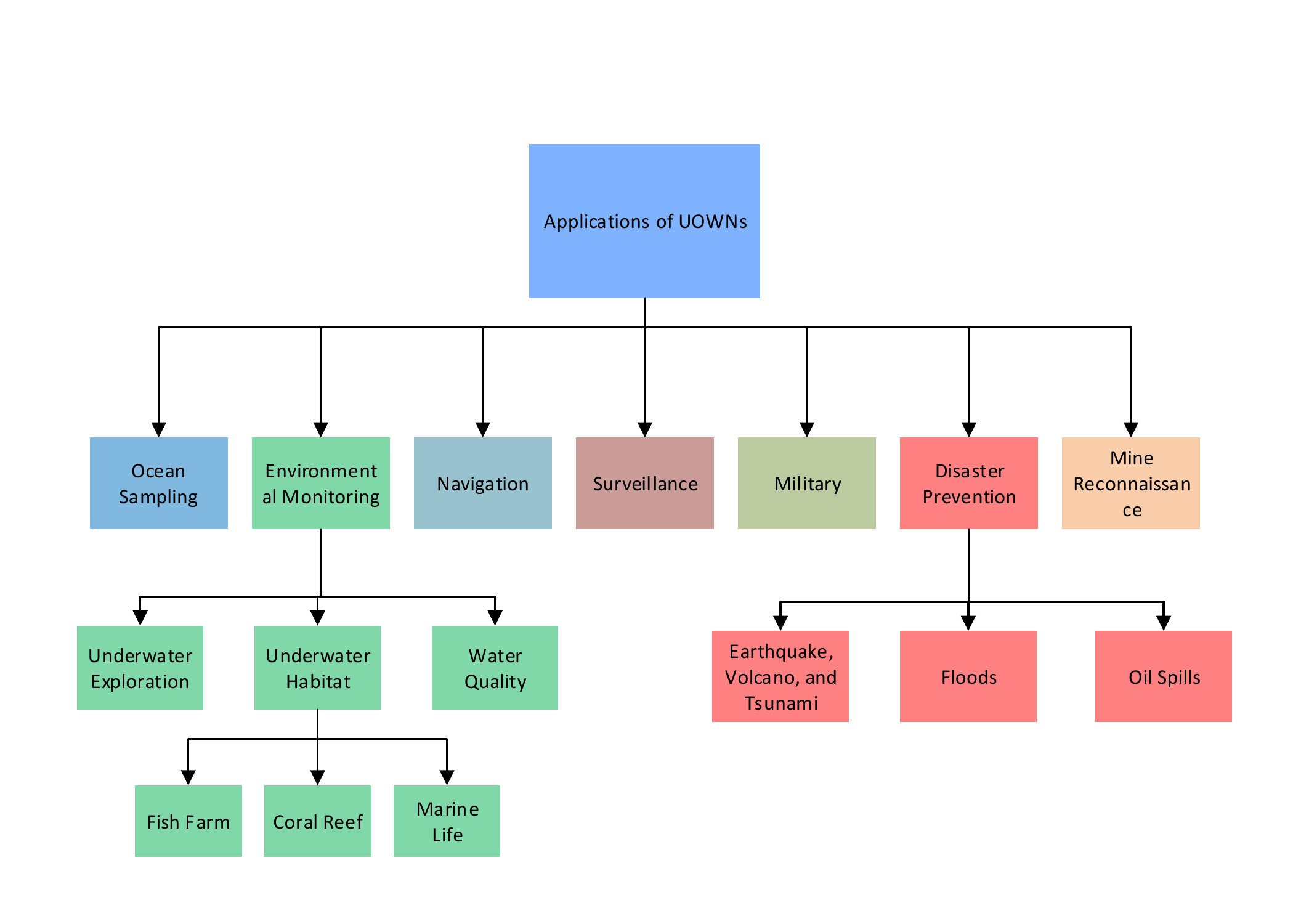}  
\caption{Classification of UOWNs applications.\label{fig:applications}}  
\end{center}  
\end{figure*}

\subsection{Congestion and Flow Control}
Congestion control is needed in order to avoid from being congested due to oversubscription of many traffic flows which may not be affordable by available network capacity whereas flow control is required to manage the sender's transmission rate in order to prevent buffer overrun at the receiver. Due to the window-based mechanism which relies upon the accurate round trip time (RTT), most of the TCP implementations are unsuited for underwater acoustic networks as they incur end-to-end delay with high mean and variance \cite{Pompili2009}. Even if UOWCs provide very high propagation speeds, a potential transport protocol still needs to take the link failures because of the dynamic topology changes of UOWNs into account. Alternatively, rate-based transport protocols do not depend on windows-based mechanism and can provide a flexible rate control, however, it requires feedback messages to dynamically adapt the transmission rate according to the packet losses. The rate-based scheme is not appropriate for UOWNs due to high mean and variance of feedback delay \cite{AKYILDIZ2005257} where some of the UOWN nodes may not receive any feedback messages if there is not a connecting path from the receiver to the transmitter in case of limited connectivity. Accordingly, proposed congestion and flow control mechanisms should account for such kind of specific challenges related to UOWNs. It is especially important to leverage critical information from lower layers to predict and handle shadow zones as connectivity can be regarded as the main delimiter of any potential transport layer protocol.

\section{Application Layer}
\label{sec:applications}
Even though one can count numerous applications for UOWNs, application layer protocol is a completely unexplored  area of research. The main purpose of a potential application layer protocol is multifold \cite{AKYILDIZ2005257}: 1) to provide a mediating language to query the entire UOWNs, 2) to advertise events and assign the tasks, and 3) to provide  efficient network management tools which can see and manipulate the hardware and software features of the lower layers. Having these functionalities in the hand, application layer protocols are needed to be customized according to the QoS requirements of target applications, which are summarized in Fig.~\ref{fig:applications} and surveyed below:

\subsubsection{Ocean Sampling}
Ocean sampling provides an embedded ocean research capability with the help of mobile and networked sensors. In a comprehensive form, ocean sampling combines observation tools with efficient modeling to reduce error in the state estimation for oceans. The network of underwater optical sensors and AUVs can perform different tasks such as synoptic adaptive sampling of three dimensional coastal of oceans, measuring the physical properties of oceans, ecosystem productivity, and other fundamental tests to understand the behavior and capabilities of ocean processes. Some of the well-known projects developed on ocean sampling are autonomous ocean sampling network \cite{aosn2017}, Bermuda bio-optics project \cite{bermuda2017}, Littoral ocean observing and prediction system \cite{YU2002, DICKEY2003}, coastal ocean dynamics \cite{PHILPOT2004}, and ocean research interactive observatory networks \cite{Orcutt2004}. 
\subsubsection{Environmental Monitoring}
Environmental monitoring applications of UOWNs are specifically related to monitor the physical underwater environment. Underwater environmental monitoring applications can further be categorized into three major categories i.e., monitoring of underwater exploration, monitoring of underwater habitat, and  monitoring of the water quality.
\begin{itemize}
\item Monitoring of Underwater Exploration: There are abundant resources such as oil and gas present in the underwater environment which is required to be explored. Water is covering most part of the Earth’s surface, the dry parts of the Earth are connected by underwater cables and pipelines. These underwater cables and pipelines provide some of the most basic necessities for instance gas pipelines, oil, and optical fiber. Therefore,  UOWNs can be used to monitor these underwater cables and pipelines, as well as UOWNs can be used to explore the underwater precious resources. In \cite{ Nishida2017}, the authors have developed an underwater monitoring system to find the manganese crust. AUVs were used to observe the manganese crust on the seabed at Katayama sea-mount and detailed three-dimensional seabed images were taken by the AUVs for investigation. An underwater monitoring system was proposed in \cite{ Srinivas2012} which combines ROVs and AUVs to discover the underwater mineral resources. Large and deep scale areas can be scanned using this system for oceans exploration. A deep ocean monitoring system was also proposed in \cite{Acar2006} for ocean exploration. The monitoring system was tested in coastal areas by deploying cameras in the ocean. Detailed literature on different ocean exploration monitoring systems and their possible architectures using underwater sensor networks were presented in \cite{AKYILDIZ2005257, Pompili2006, Heidemann2006, Tuna2017, Luo2018}.
\item Monitoring of Underwater Habitat: The study of underwater habitat is one of the most interesting and challenging fields of natural sciences. Underwater habitat monitoring includes fish farm monitoring, marine life monitoring, and study of underwater coral reef and plants. 
\begin{itemize}
\item Fish Farm Monitoring: Fish farming is considered to be a good economical resource, but it requires strict monitoring of the habitat conditions. For the monitoring purposes, underwater sensors are deployed to monitor the underwater habitat. In \cite{Lloret2011} and \cite{Lloret2015}, the authors have developed an underwater sensor network which was able to estimate the amount of residual food and waste in the underwater habitat. Similarly, an underwater monitoring system was developed in \cite{Yalcuk2015} to maintain the ecosystem for trout fish by sensing the water quality. A Zigbee based underwater sensor network has been proposed in \cite{LOPEZ2009} to monitor the properties of small fish farms. The underwater sensors were able to sense the temperature, pH, and NH4+ of underwater habitat and send the information to the central station using Zigbee.  An underwater monitoring system was developed in \cite{Nasser2012} for large lakes and fish farms. The proposed system was able to measure the depth of the farm and acidic level of the water. The depth was measured by using the optical sensors while acidic levels of water were measured by using the pH and oxygen sensor. An increased lifetime water quality monitoring system was proposed in \cite{Cario2017} for fish farms and tested at the farming cages located in Mediterranean sea, Italy. Recently a semi-automatic fish monitoring system has been developed in \cite{Kratzert2017} which monitors the migration of fish by using the videos taken by the AUVs.
\item Marine Life Monitoring: The applications of marine life monitoring include overseeing various species related to oceans. An underwater marine life monitoring system was developed in \cite{Acar2006} which was able to monitor the marine life in the underwater environment within its coverage area. An underwater sensor network was deployed in \cite{Alippi2011} to monitor the marine life at different levels. In the proposed system the sensor nodes were divided into small clusters and the cluster head directly sends the sensed information to the surface buoy. The proposed system was deployed in Queensland, Australia to monitor underwater temperature and luminosity. The authors in \cite{Trevathan2012} claimed a cost-effective monitoring system called smart environmental monitoring and analysis technology (SEMAT). SEMAT was used for marine life research and water quality monitoring. Another marine life monitoring system was developed in \cite{Pérez2011} and tested at Menor coast, Spain. An intelligent architecture and different protocols for marine life study have been studied in \cite{Shaowei2012}. An interesting seashell monitoring system has been deployed in Zhejiang, China \cite{Yang2009}.  A detailed survey on marine life research using underwater sensor networks is presented in \cite{Guobao2014}.
\item Coral Reef Monitoring: Coral reefs are one of the important and diverse underwater ecosystems which are built by the microorganisms. In \cite{Acar2006} the authors have developed an underwater sensor network for coral reef monitoring. In the proposed systems stationary sensor nodes were deployed at the target coral places. AUVs were used to deploy the sensors and collect the data from the sensors. The prototype of the proposed system was deployed at Okinawa, Japan. An intelligent surface buoy has been developed in \cite{Alippi2011}  to monitor the coral reef. The proposed system harvest energy from sea waves, thus improving the lifetime of the monitoring system. In \cite{Bainbridge2011} and \cite{Bainbridge2017}, underwater sensor networks have been deployed to monitor coral reef at northeastern Australia. The proposed system is still in operation from last two years and reliably monitoring the largest coral reef ecosystem on planet Earth.
\end{itemize}
\item Monitoring of Water Quality: Life has started from Water and it is an essential resource on Earth for living organisms. Therefore, it is very important to keep an eye on the quality of water. An application to monitor the quality of pool water for trout farms was developed in \cite{Yalcuk2015, Anthony2014}. Different parameters of the trout farm such as demand for oxygen, ammonium nitrogen, electrical conductivity, and pH, were monitored for 270 days. A low cost and effective water quality monitoring system was presented in \cite{Haroon2013} for lakes and ponds which consumes low power and reliable data transmission.  Similarly, an underwater sensor network has been developed in \cite{Gurkan2013} to monitor the quality of drinking water. Sensors were integrated with AUVs to collect the water samples and information was sent to the surface station for investigation. To monitor the quality of water in Indian rivers, a water quality monitoring system has been deployed in \cite{Menon2012}. In addition, pollution and wreckage detection techniques have been presented to monitor the quality of water \cite{Jenkins2008}. A comprehensive survey of water quality monitoring systems by using sensor networks is presented in \cite{AduManu2017}.
\end{itemize}

\subsubsection{Navigation}
Underwater habitat is extremely irregular and dark with growing depth. Therefore, navigation in such an environment is a challenging task and require assistance. The common assistive technologies used for navigating the ships, boat, and vessels on the surface of water cannot be used for underwater navigation due to the different medium for transmission. Underwater sensor networks are the promising technology to assist the navigation in the underwater environment. In \cite{ Waldmeyer2011} the authors have proposed an assistive navigation system using AUVs for underwater sensor networks. An on-demand underwater technique for locating the underwater sensors was proposed in \cite{ Carroll2014}. In short, many researchers developed assistive localization schemes for the underwater environment based on acoustic waves. Navigation and localization of underwater acoustic networks have been studied widely in the past and number of surveys are written on this subject \cite{Chandrasekhar2006, kantarci2011survey, TAN20111663, Tuna2017, Luo2018}.  Since the UOWC channel poses new challenges, the existing localization techniques used for terrestrial wireless networks and underwater acoustic networks are not directly applicable to UOWNs. Therefore, novel time of arrival (ToA) and received signal strength (RSS) based distributed localization schemes are developed in \cite{Akhoundi2017underwater} for UOWNs.  Recently, in \cite{Nasir2018limited, Saeed2017, Nasir2018twc} we have proposed RSS based centralized localization schemes for UOWNs.
\subsubsection{Surveillance}
Underwater surveillance is quite important, especially for intruder detection. Underwater sensor networks can be used for offshore and onshore surveillance. Onshore surveillance applications include battleships detection and logistics arrival. The warfare surveillance system (GLINT10 field trail) has been tested in \cite{ Kemna2011}. AUVs were used in complete autonomous fashion with signal processing capabilities to act and share the information about the underwater battlefield. The protection of offshore and onshore equipment's and infrastructures were investigated in \cite{ Caiti2013 } by using the underwater acoustic network. The proposed system was composed of four acoustic stationary nodes, one mobile node, and two AUVs. The proposed system was deployed in Norway and worked successfully for continuous five days. A novel layout for underwater surveillance has been proposed in  \cite{ CAYIRCI2006 } which consists of surface sensors. The surface sensors are then moved to specific depths to get maximum coverage. Data mining tools were used to classify and detect different objects such as submarines and divers.  Electromagnetic waves based different architectures for underwater surveillance were proposed in \cite{ Kumudu2016 }. For the interesting readers, we refer to \cite{ Ferri2017 }, which is the most recent survey presented on underwater surveillance systems .
\subsubsection{Mine Reconnaissance}
As the sensors are able to sense different physical parameters, it is rational that sensors can detect underwater mines. Detection of underwater mines can assist the ships for a safe voyage. An underwater mine detection system has been developed in \cite{ Kumar2004} by Naval warfare center, Florida which can detect the underwater clutters. An underwater mine detection systems have also been proposed in \cite{ Williams2010, Rao2009, Leonard2013, Rajeshwari2015, Yu2016} which considers image processing techniques to localize the mines. Four different types of AUVs and two different sonars were used in \cite{Khaledi2014} to find out the underwater mines. The proposed system was tested in the Chesapeake Bay, U.S. to detect different kind of underwater mines. A machine learning and deep learning approaches were used in \cite{ Barngrover2015, Denos2017} to classify the underwater mines from other objects.
\subsubsection{Military}
Underwater sensors are also deployed for a number of military applications. The underwater network consists of a number of high-resolution cameras, sonars, and metal detectors combined with AUVs to detect mines, secure submarines, and ports. Norwegian defense has developed an AUV suitable for underwater military applications. This AUV was named as HUGIN 1000 and it was able to work in different applications such as localization and classification of underwater objects, mine detection, route surveys, and environment assessment \cite{ Hagen2003}. U.S. military has recently launched a project of worth 37 million US dollars which involves the development of intelligent underwater surveillance sensors. This project is named as Ocean of things with two major purposes: building efficient low-cost underwater sensors and developing data processing techniques to get the useful information \cite{ Keller2017}.
\subsubsection{Disaster Prevention}
Natural disasters are imminent among which water-based disasters are deadly and cause enormous destruction. Due to the insufficient resources and challenging environment of oceans, development of disaster monitoring system for water-based natural disasters is still a significant challenge. Prediction of a natural disaster using a disaster monitoring system and taking preventive mechanism is very important. Underwater sensor networks offer broad range of disaster monitoring applications from predicting a disaster to the aftermath of a disaster. Water-based disasters include underwater volcanic eruptions, floods, underwater earthquakes, and oil spills.
\begin{itemize}
\item Earthquake, Volcano, and Tsunami: Monitoring of underwater earthquakes and volcanoes is very important otherwise it can lead to enormous destruction. In \cite{Kumar2012}, an early warning system was proposed by using underwater sensor networks for earthquakes and tsunamis. An efficient architecture of underwater sensors to detect tsunamis was presented in \cite{Casey2008}. The proposed system utilized seismic pressure to detect the tsunami and transmitted the information to the surface station. To the best of our knowledge, there is a lot of literature which address underwater natural disasters but very few practical underwater disaster monitoring systems exist. 
\item Floods: The destruction caused by floods and its increased frequency require to timely generate flood alerts. To generate timely flood alerts deployment of underwater and on the surface sensors can be used. A flood monitoring system was proposed in \cite{ Tyan2013} which consists of an AUV, sensors, and a remote station. The authors in \cite{Rafael2012} have proposed a  flood warning system based on the underwater sensor network. The proposed system was tested and implemented in 650 $\text{km}^2$ watershed in southern Spain. Similar flood monitoring systems have also been implemented in 15 different flood regions in Nigeria \cite{ Edward2013}. Recently, a cloud computing based flood monitoring system have been proposed in \cite{ Afzaal2016}. 
\item Oil Spills: Large underwater oil spills cause a serious biological impact on marine life. Therefore, to monitor these large oil spills, underwater sensor networks provide a good solution. In \cite{ Jenkins2008}, an efficient underwater sensor network was deployed to detect the ocean pollution. The thickness and location of the oil spill have been investigated in \cite{ Koulakezian2008, Denkilkian2008, Massaro2012} by using underwater sensors. Optical detection methods were used in \cite{ Oh2012, Pangilinan2016 } to detect underwater oil spills with the help of LEDs.
\end{itemize}

\begin{figure*}[htb!]
\begin{center}  
\includegraphics[width=2\columnwidth]{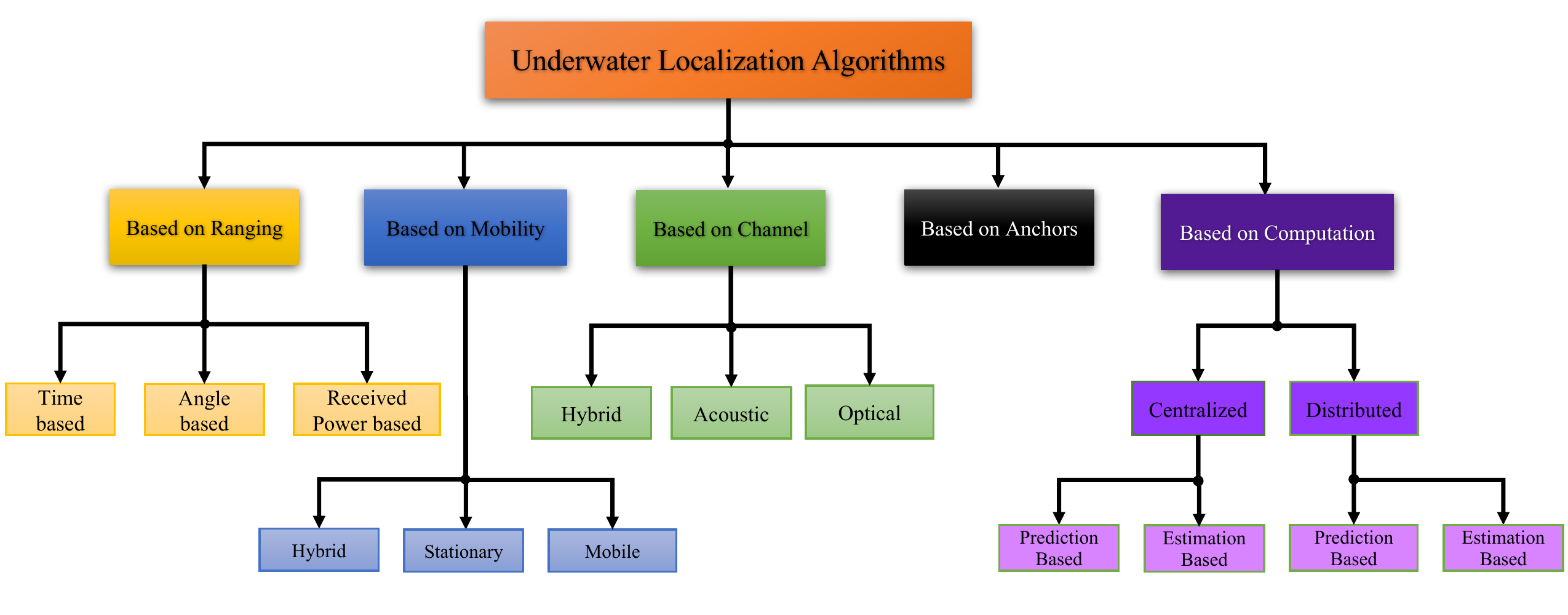}  
\caption{Taxanomy of underwater localization algorithms.\label{fig:taxanomy}}  
\end{center}  
\end{figure*}

\section{Localization in Underwater Optical Wireless Networks}
\label{sec:loc}
Numerous acoustic based localization techniques have been well investigated in the past since it is important for tagging the data, detection of an underwater object, tracking of underwater nodes, underwater environment monitoring, and surveillance. Nevertheless, due to the challenges discussed in previous sections for each layer of UOWNs, there is a dire need to develop  novel localization techniques for UOWNs. Therefore, this section provides the fundamental concepts of underwater localization, state of the art underwater localization systems, and development of localization techniques for UOWNs.
\subsection{Basics of Underwater Localization}
Localization of underwater sensors is an important part of UOWNs for many applications such as resource exploration, surveillance, underwater environment monitoring, and disaster prevention. The large propagation delay of acoustic channels and high attenuation of RF/optical channels pose significant challenges for underwater localization. The major challenges for underwater localization are
\begin{itemize}
\item Deployment of nodes: Most of the localization algorithms depends on the distribution of sensor nodes and the anchor nodes to form a network \cite{Kantarci2010, TAN20111663}. 
\item Mobility of the nodes: Due to the uncontrollable phenomena such as winds, turbulence, and current, the underwater sensor nodes inevitably drifts from its position. The location of anchor nodes on the surface buoys can be accurately measured by using GPS but the location of the underwater nodes cannot be precisely measured \cite{kantarci2011survey, TAN20111663}. 
\item Harsh underwater channel: Variations in the underwater wireless communication channel is very severe for all type of carrier waves. The effects of attenuation, absorption, reflection, scattering, and noise do not allow for accurate range measurements, thus reflecting large localization estimation error \cite{Tuna2017}.
\item Synchronization: As the GPS signals are not available in the underwater environment, it is hard to achieve the time synchronization between the sensor nodes. Thus, if the time of arrival based ranging is used, this miss-synchronization will lead to large localization error \cite{Diamant2013, Cario2016, crowell2017}.
\end{itemize}

A large number of underwater localization algorithms have been proposed in the past for underwater acoustic wireless communication systems. All of these localization algorithms consider different parameters of the network such as network topology, range measurement technique, energy requirement, and device capabilities.  In addition, the accuracy of localization algorithms also depends on many other factors which include propagation losses, number of anchor nodes, the location of anchor nodes, time synchronization, and scheduling \cite{Ramezani2015}. The underwater localization algorithms can be classified based on different parameters such as range-based/range free, anchor-based/anchor free, acoustic/optical, stationary/mobile, and centralized/distributed. Centralized and distributed algorithms can further be classified into estimation based algorithms and prediction based algorithms. The taxonomy of underwater localization schemes is shown in Fig.~\ref{fig:taxanomy}. Every underwater localization algorithm requires distance estimation between the nodes or between the node and anchors. The distance is estimated by using acoustic ranging or optical ranging for UWC systems.  
\subsubsection{Acoustic Ranging}
Underwater acoustic channels suffer from two kinds of major losses, i.e., attenuation loss and spreading loss \cite{rusworth85}. Attenuation loss is a result of scattering, diffraction, absorption, and leakage from ducts while spreading loss is a combination of cylindrical and spherical losses \cite{ETTER2001}. Acoustic ranging based localization algorithms can further be classified as time-based, received signal strength based, and angle based algorithms. Most of the underwater localization algorithms are time-based which includes time of arrival (ToA) and time difference of arrival techniques (TDoA).  
\begin{itemize}
\item ToA based acoustic ranging: ToA based ranging is one of the most popular ranging technique used for underwater distance estimation. Indeed, ToA is more preferred in underwater acoustic systems compare to terrestrial systems since the ToA technique for RF signals requires high-resolution stable clocks. But as the speed of sound waves is very slow compared to the speed of RF signal, ToA is best suited for underwater ranging \cite{Stojanovic2009}. In ToA based underwater acoustic localization systems the anchor nodes transmit the acoustic signals and each sensor nodes require to receive the acoustic signal from at least three anchor nodes to carry out trilateration and find its estimated position. In \cite{Kanaan2004, Emokpae2011} the authors have defined the objective function for ToA based underwater acoustic localization as
\begin{equation}
f(x,y)=\sqrt{{\sum}_{i=1}^{N}(\hat{d}_{i}-\sqrt{(x-X_{i})^{2}+(y-Y_{i})^{2}})^{2}}
\end{equation}
where $x$ and $y$ is the two-dimensional estimated location of a sensor node, $N$ is the total number of anchor nodes, $\hat{d}_i$ is the estimated distance to anchor node $i$, and $X_{i}~Y_{i}$ is the two-dimensional location of anchor node $i$. The underwater acoustic ToA distance estimation is mainly affected by dispersion of underwater acoustic channels, multipath fading, and Doppler spread \cite{Walree2013}.
\item TDoA based acoustic ranging: In ToA based ranging techniques the sensor nodes and anchor nodes must be synchronous while in TDoA based ranging the sensor node do not need to be synchronized to the anchor nodes, thus, mitigating the limitation of time synchronization requirement of ToA techniques \cite{Poursheikhali2015}. The authors in \cite{Cheng2008} have proposed a TDoA based underwater localization system called silent positioning which relies on the range differences collected from four different anchor nodes. The problem of three-dimensional underwater localization by using TDoA based acoustic ranging has been studied in \cite{Teymorian2009, Isbitiren2011, Choi2017}. The performance of TDoA based localization was investigated in \cite{Scheuing2008} for the reverberant underwater environment and multipath propagation channel. Localization of underwater acoustic networks has been investigated in \cite{Ioana2017} by using Hausdorff distances for TDoA ranging. Localization in shallow water using TDoA measurements was achieved in \cite{Kouzoundjian2017} by considering two hydrophones mounted under the boat. In \cite{Jung2017}, the authors have implemented an accurate and precise underwater acoustic localization system by using TDoA measurements with the help of a microphone and a speaker. Recently, a real-time acoustic ranging technique has been proposed in \cite{Kim2017} to improve the accuracy of TDoA measurements. The TDoA ranging can provide better accuracy compare to ToA ranging at the expense of more complexity and cost of the system \cite{Cheng2008}.
\item RSS based ranging: Due to the problems of attenuation and spreading loss for underwater acoustic channel, RSS is not well suited for underwater ranging. However, due to the simplicity and nice transmission behavior of RSS at certain depth \cite{Hosseini2011}, it was used in \cite{Xu12016} for underwater source localization. Source localization in an underwater  environment by using RSS was also studied in \cite{Zhang2016} by using Thorp's propagation model \cite{Thorp1967}.
\end{itemize}
\begin{figure*}[htb!]
\begin{center}  
\includegraphics[width=1.8\columnwidth]{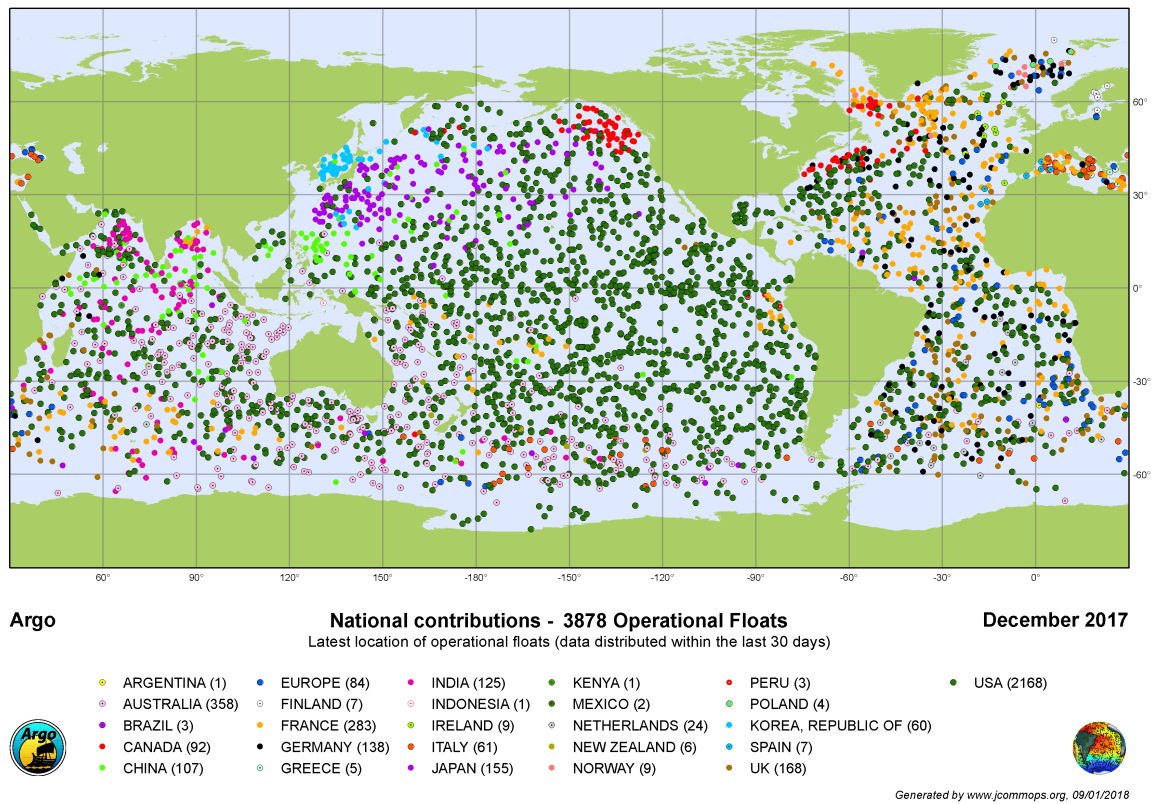}  
\caption{Distribution of floating sensors in Argo \cite{argo}.\label{fig:argo}}  
\end{center}  
\end{figure*}
\subsubsection{Optical Ranging}
Optical light passing through the aquatic medium suffers from widening and attenuation in angular, temporal, and spatial domains \cite{Akhoundi2017underwater}. In the literature, only ToA and RSS based localization techniques exist for UOWNs. In  \cite{Akhoundi2017underwater} the authors have proposed an underwater optical positioning system, where an OBS was considered as an anchor node which transmits optical signals. The sensor nodes receive the optical signals from multiple anchors and locate itself using simple linear least square solution. In \cite{Arnon:09}, an RSS based distance estimation technique is developed for UOWNs for a given data rate. The RSS based distance estimation strongly depends on different parameters such as characteristics of the underwater optical channel, divergence angle of the transmitter, field of view of the receiver, transmitted power, and trajectory angle. For a LoS link and achievable data rate $R_{i}^j$, the estimated distance $\hat{d}_{ij}$ between node $i$ and $j$ is obtained in \cite{Vavoulas2014} from \eqref{eq:Pr_LoS} as    
\begin{equation}\label{eq: estimateddistanceo}
\hat{d}_{ij}= \frac{2\cos\varphi_i^j}{c(\lambda)} W_0 \left(\frac{c(\lambda)}{2\cos\varphi_i^j\sqrt{\frac{2\pi T \hbar c R_{i}^j r_j (1-\cos\theta_i)}{\eta_t^j \lambda P_{t_i}\eta_t^i\eta_j A_j\cos\varphi_i^j}}}\right),
\end{equation}
where $T$ is the pulse duration and  $W_0(.)$ is the real part of Lambert $W$ function \cite{Corless1996}. Table \ref{Tableranging} summarizes the literature on different ranging techniques for underwater localization systems. 
\begin{table*}[htb!]
\centering
\caption{Comparison  of different ranging techniques for underwater localization.}
\label{Tableranging}
\begin{tabular}{|l|l|l|l|l|}
\hline
\hline
 Literature              & Channel model        & Ranging Technique   & Accuracy & Complexity \\ \hline
\cite{Stojanovic2009, Kanaan2004, Emokpae2011, Walree2013}                       & Acoustic     &       ToA & High & Moderate\\
\cite{Cheng2008, Poursheikhali2015, Teymorian2009, Isbitiren2011, Choi2017, Scheuing2008, Ioana2017,Kouzoundjian2017}                         & Acoustic     &      TDoA & High & High\\
 \cite{Hosseini2011, Xu12016, Zhang2016}                        & Acoustic    & RSS & Low & Low\\
  \cite{Akhoundi2017underwater}                       & Optical    & ToA & High & Moderate\\
   \cite{Akhoundi2017underwater}                       & Optical    & RSS & Low & Low\\
\hline
\hline        
\end{tabular}
\end{table*}
\subsection{State of the Art Underwater Monitoring and Localization Systems}
In the past various techniques have been used by the oceanographers for oceans exploration. The most common monitoring equipment include ocean floor sensors, floating sensors, surface buoys and surface stations. Sensed data from the sensors on the ocean floor is collected by the surface buoys. The surface buoys are fixed and they can send the collected data to the surface station using wired or wireless communications. In case of floating sensors, the sensors do not have fixed location and drift with ocean currents. Floating sensors are dynamic in nature and they can sense a reverberant underwater environment. At present, the largest ocean monitoring system is developed by Global Ocean Data Assimilation Experiment (GODAE)  and Oceanview called Argo \cite{argo, Gould2006}. Argo consists of 3800 free drifting floating sensors which measure the salinity, currents, and temperature of the ocean up to 2000 m of depth. The location of Argo float is determined using GPS once it is on the surface of the ocean and it also transmits the data to the onshore station using satellite communication. In Argo project, the floats do not interact with each other and work independently. Fig.~\ref{fig:argo} shows the current distribution of Argo floats in oceans all over the world. China has announced recently a similar project to Argo, to build underwater monitoring systems across the south and east China seas for intruder detection \cite{cnnreport}.
In 1980, the U.S. Navy has developed a large scale network of underwater devices called Seaweb \cite{Rice2008}. Seaweb consisted of AUVs, surface buoys, gliders, repeaters, and surface stations. Seaweb has used acoustic waves for underwater communication and RF waves for terrestrial communication.

Acoustic localization systems for underwater monitoring utilizes two different approaches, namely long baseline (LBL) and short baseline (SBL) \cite{collin2000}. In the LBL approach, the acoustic transponders are installed in the underwater operation area. Sensor nodes that are in the coverage of these acoustic transponders respond by using a certain ranging method and localize itself either by using triangulation or trilateration \cite{Foley2002}. In the SBL approach, the surface station follows the underwater sensor nodes and transmits short-range acoustic signals for the sensor nodes to localize itself. The SBL underwater positioning systems have been used by Woods Hole Oceanographic Institution to find out the location of a deep underwater ROV \cite{BALLARD1993}. In addition to the LBL and SBL approaches, there exists an underwater localization system called GPS intelligent buoy (GIB). The GIB is a commercial system in which the surface buoy acts as a relay between the surface station and the seabed sensors. GIB collects the distance estimation from the sensors to itself and sends it to a central station where the central station finds the global view of all the seabed sensors. GIB systems have numerous applications which include weapon testing and training \cite{Kayser2005}, tracking AUV's \cite{Chen2013},  global view of the network \cite{Chen2013}, and intruder detection \cite{Zhou2010}. Table \ref{Tablelocsystems} summarizes some of the well known commercial underwater localization systems.
\begin{table*}[htb!]
\centering
\caption{Comparison  of different commercial underwater localization systems.}
\label{Tablelocsystems}
\begin{tabular}{| p{4cm} | p{2cm} | p{1.2cm} | p{1.4cm} | p{1.3cm} |p{5.5cm}|}
\hline
\hline
 System      & Company        & Channel type        & Approach   & Accuracy  & Applications\\ \hline
Underwater acoustic LBL positioning \cite{evo}. & Evo Logics & Acoustic & LBL & 1.5 cm & Offshore positioning, navigation, cartography, geodesy, and sensors tracking.\\
HiPAP - Acoustic underwater positioning and navigation systems \cite{KONGSBERG}. & KONGSBERG & Acoustic & SBL & 2 cm & Seabed positioning of vessels, sub-sea meteorology, and telemetry.\\
Mini-Ranger 2 Underwater Positioning (USBL) system \cite{RANGER}. & Sonardyne & Acoustic & Ultra SBL & 2 cm & Oil and gas exploration, marine robotics, and marine security.\\
USBL positioning systems \cite{ixblue} & iXblue & Acoustic & Ultra SBL & 6 cm & Hydrography, maritime vessels, ocean science, and defense.\\
VideoRay ROV Positioning Systems \cite{kcf} & KCF Technologies & Acoustic & Ultra SBL & 150 cm & Navigation, tracking, search and rescue, and target detection.\\
TrackLink Acoustic Tracking Systems \cite{linkquest} & LinkQuest Inc & Acoustic & Ultra SBL & 0.5 cm & Navigation, tracking, underwater surveys, and oil and gas exploration.\\

Teledyne Benthos underwater acoustic systems \cite{teledyne} & Teledyne Marine & Acoustic & LBL/Ultra SBL & 5 cm & Navigation, tracking, underwater surveys, and oil and gas exploration.\\

Bluecomm UOWC \cite{blucom}& Sonardyne & Optical & Not defined & - & High speed underwater wireless communication.\\

Anglerfish UOWC \cite{stm} & STM & Optical & Not defined & - & High speed underwater wireless communication.\\
\hline
\hline        
\end{tabular}
\end{table*}

\subsection{Localization Techniques for UOWNs}
UOWNs localization is one of the major research areas nowadays because of the development of high-speed UOWC systems. Localization in terrestrial wireless networks has been studied widely and detailed surveys are presented on this topic \cite{Santosh2006, MAO2007, Kulaib2011, Nabil2013, Kuriakose2014, Khan2017}. Nevertheless, GPS and all of these RF-based localization schemes cannot work in the underwater environment. Thus, many researchers developed localization schemes for the underwater environment based on acoustic waves. Localization of underwater acoustic networks have also been studied widely in the past and number of surveys are presented on this subject \cite{Chandrasekhar2006, kantarci2011survey, TAN20111663, Tuna2017, Luo2018}.  Since the localization techniques used for terrestrial wireless networks and underwater acoustic networks cannot be directly applied to UOWNs, novel localization schemes have recently been developed for UOWNs. We divide these localization schemes into two categories as distributed and centralized schemes. In distributed localization schemes, every underwater optical sensor node localizes itself by communicating with multiple anchor nodes. In centralized localization schemes, the underwater optical sensor nodes do not localize themselves but the location information is sent to them by the surface buoy or sink node periodically.
\subsubsection{Distributed Localization Schemes for UOWNs}
In this section, we summarize two distributed UOWNs localization schemes where one is based on ToA ranging and the other is based on RSS based ranging.
\begin{itemize}
\item ToA Based Scheme: In \cite{Akhoundi2017underwater}, the authors have proposed for the first time a ToA based underwater  optical wireless positioning system. The authors considered an OBS placed in an underwater hexagonal cell and a number of users with transceivers capable of UOWC. Each OBS consists of 60 green LEDs forming an underwater OCDMA network where the modulation scheme of OOK is considered. For the ToA scheme, first, the distance is estimated between the users and the OBS by using the relationship of the transmission time of an optical signal, speed, and the distance. It is assumed that all the OBSs and the users are synchronized, and all the OBSs transmit the beacon signals at $\tau = \tau_0$. The users receive multiple beacon signals from multiple OBSs at different times namely $\tau_1,~\tau_2~\tau_,...,\tau_m$, where $m$ are the number of OBSs. Different underwater channel impairments such as turbulence, current, and multipath lead to the distance estimation error for ToA ranging. Once the ToA based estimated distances are available from at least three OBSs, the user was able to locate itself in two dimensions by using linear least square solution.
\item  RSS based Scheme: As the optical signal from the OBSs to the user passes through the underwater environment it suffers from attenuation, absorption, and scattering. The underwater user requires the RSS signals from at least three OBSs in this case as well. RSS scheme has low cost because every transceiver is able to estimate the received signal power. However, the RSS based distance estimation requires precise modeling of the channel \cite{Korotkova2012, Jamali2016}.  The RSS based distance estimation strongly depends on different parameters such as characteristics of the underwater optical channel, divergence angle of the transmitter, field of view of the receiver, transmitted power, and trajectory angle. The widening and attenuation of the underwater optical signals are dependent on the wavelength. 
Monte Carlo simulations were used in \cite{Akhoundi2017underwater} to find out the RSS based distances. Once the RSS based distances were estimated to at least three OBSs, the user was able to locate itself in two dimensions by using linear least square solution.
\end{itemize}

\begin{figure*}
    \centering
    \begin{subfigure}[b]{0.48\textwidth}
    \includegraphics[width=1\columnwidth]{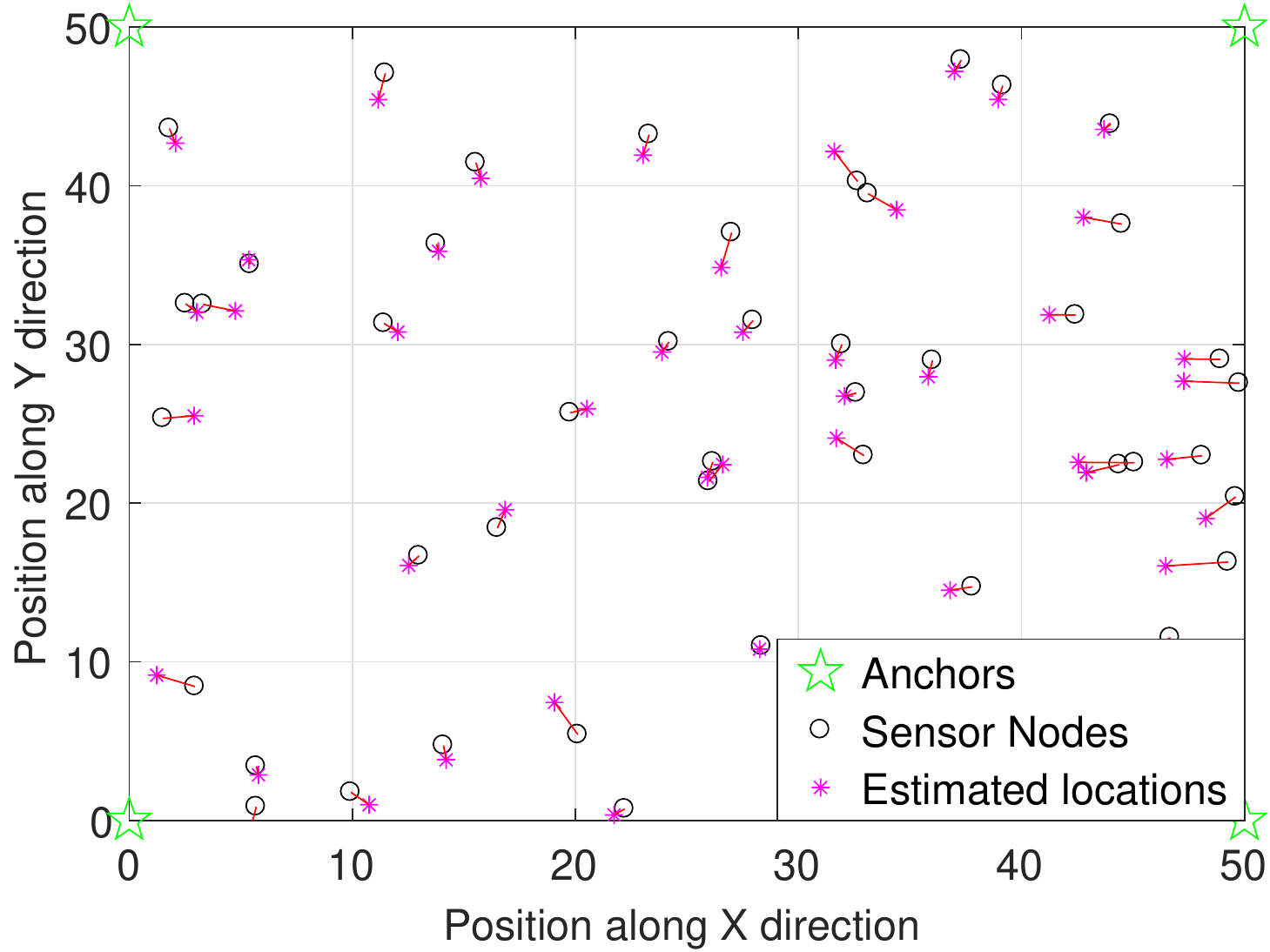}  
\caption{Localization performance of ToA based distributed UOWNs \cite{Akhoundi2017underwater}\label{fig:comparedtoa}}  
    \end{subfigure}
    \begin{subfigure}[b]{0.48\textwidth}
\includegraphics[width=1\columnwidth]{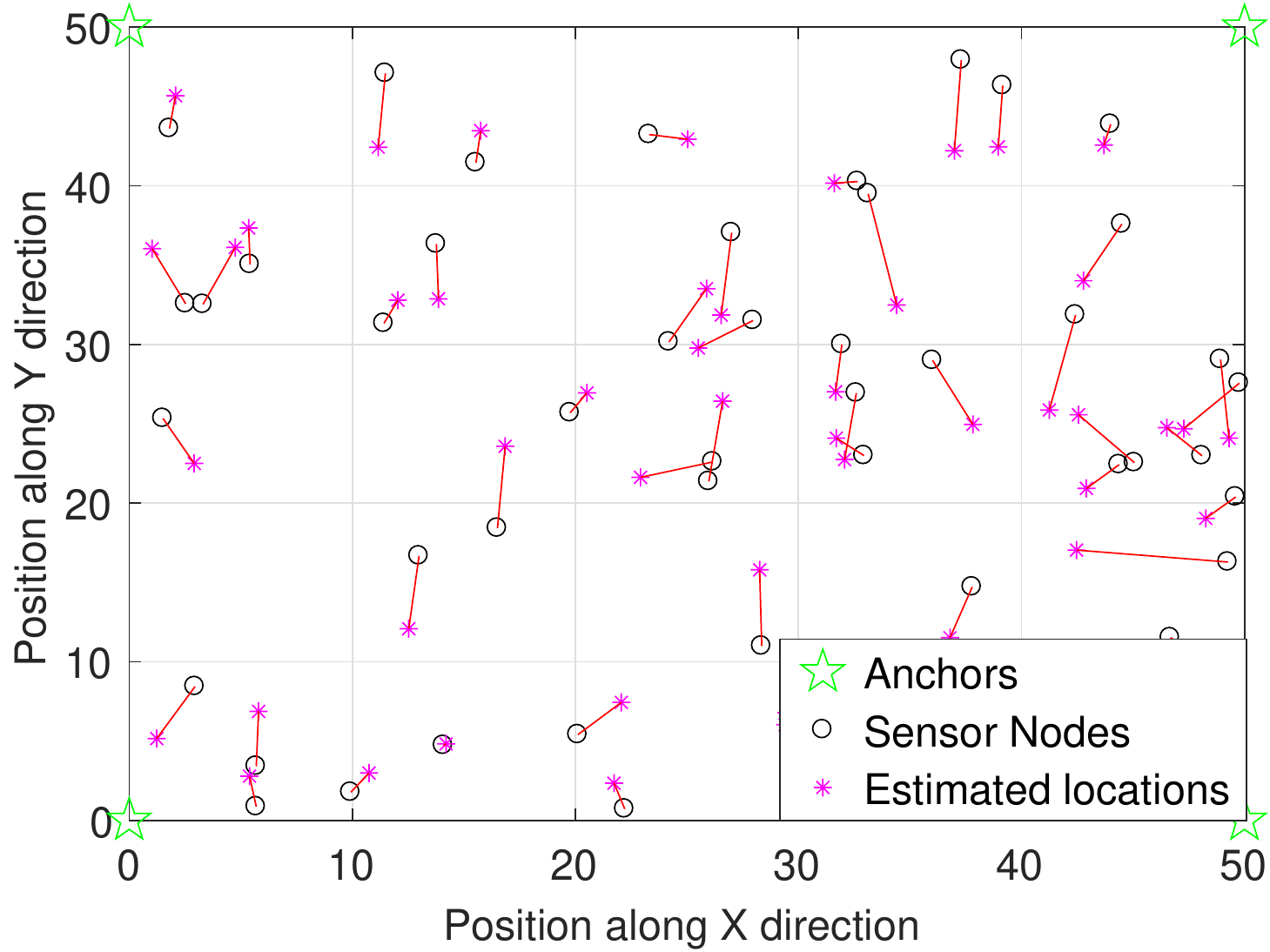}  
\caption{Localization performance of RSS based distributed UOWNs \cite{Akhoundi2017underwater}\label{fig:comparedrss}}  
    \end{subfigure}
    
    \begin{subfigure}[b]{0.48\textwidth}
\includegraphics[width=1\columnwidth]{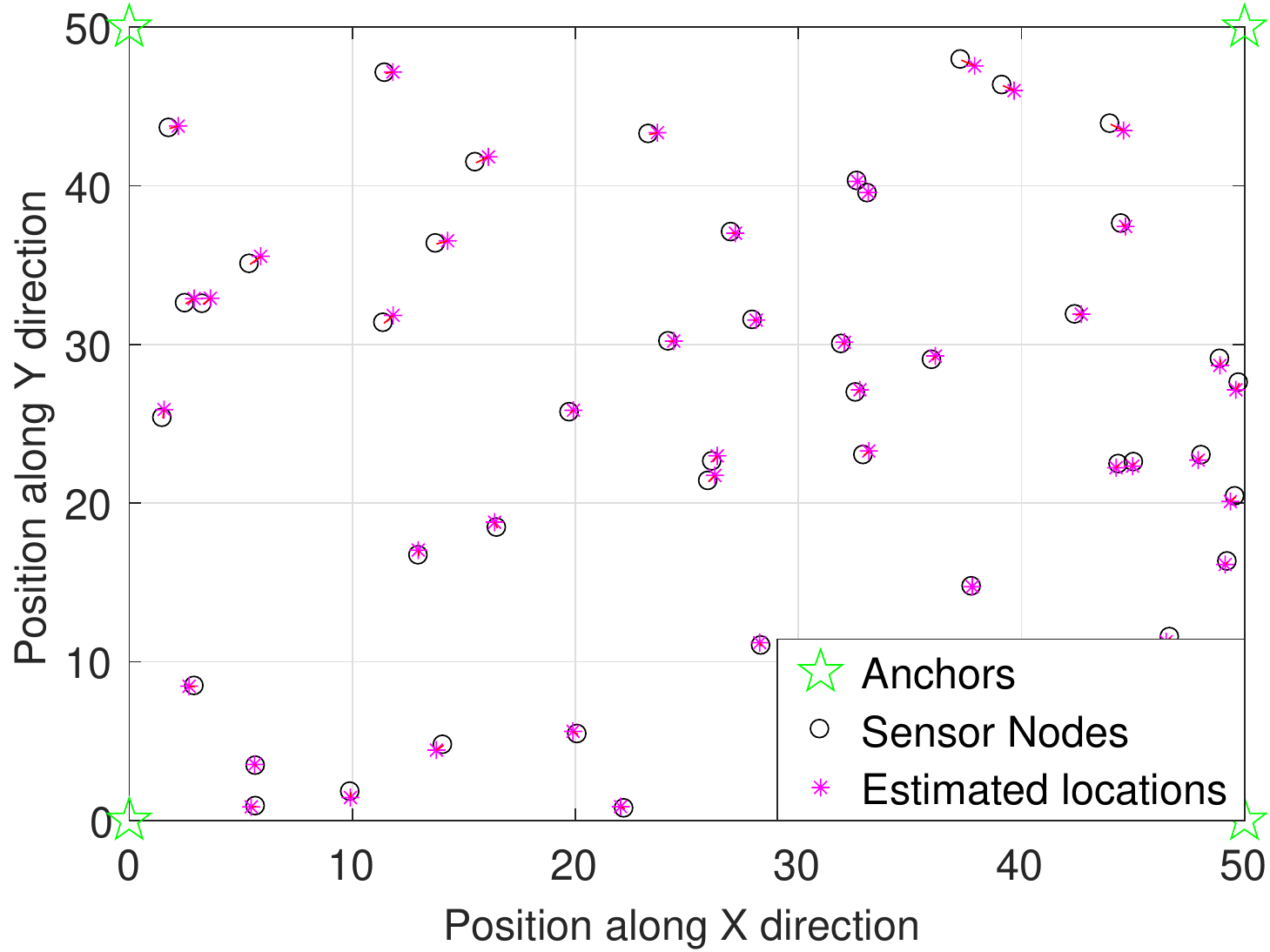}  
\caption{Localization performance of ToA based centralized UOWNs \cite{Nasir2018limited, Nasir2018twc}\label{fig:comparectoa}} 
    \end{subfigure}
    \begin{subfigure}[b]{0.48\textwidth}
\includegraphics[width=1\columnwidth]{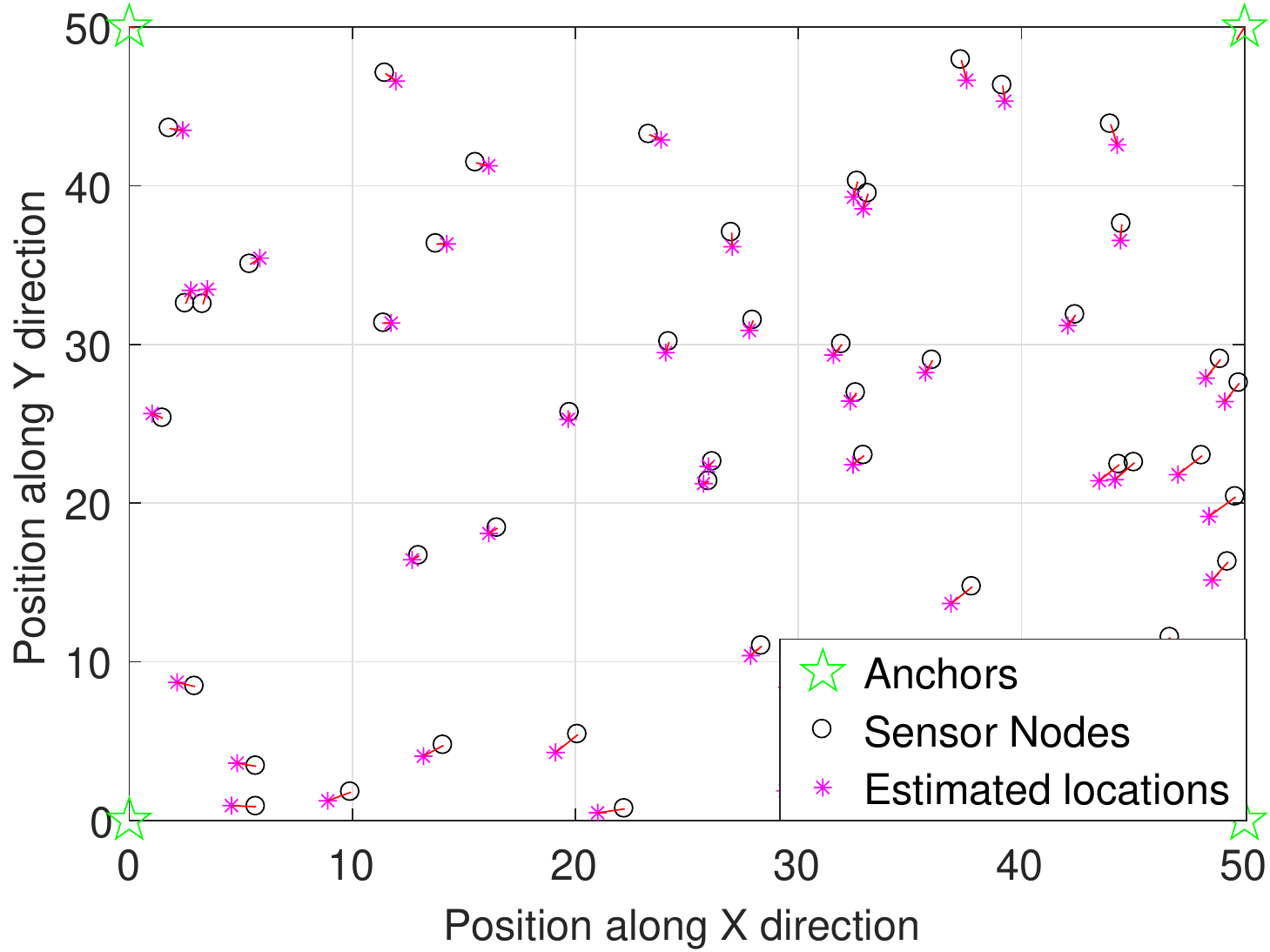}  
\caption{Localization performance of RSS based centralized UOWNs \cite{Nasir2018limited, Nasir2018twc}\label{fig:comparecrss}}  
    \end{subfigure}    
\caption{Localization performance of ToA and RSS based distributed and centralized UOWNs.}
\label{fig:toa_rss}
\end{figure*}


\begin{table*}[htb!]
\centering
\caption{Comparison  of different UOWNs localization schemes.}
\label{TableUOWNsloc}
\begin{tabular}{| p{8cm} | p{2cm} | p{2cm} | p{4cm} |}
\hline
\hline
Scheme              & Ranging Method        & Computation   & Architecture  \\ \hline
Underwater optical positioning systems \cite{Akhoundi2017underwater}                      & ToA     &       Distributed & Optical \\
Underwater optical positioning systems \cite{Akhoundi2017underwater} & RSS     &      Distributed & Optical \\
UOWNs localization with limited connectivity \cite{Nasir2018limited}                        & RSS  & Centralized  & Optical  \\
Energy harvesting empowered UOWNs localization \cite{Nasir2018twc}                        & RSS  & Centralized   & Optical \\
 Energy harvesting hybrid acoustic/optical UOWNs localization \cite{Saeed2017}                        & RSS  & Centralized  & Hybrid acoustic/optical  \\
\hline
\hline        
\end{tabular}
\end{table*}
\subsubsection{Centralized Localization Schemes for UOWNs}
In centralized UOWNs localization schemes, the underwater user does not localize itself but the location is sent to the user by the surface buoy or sink node periodically. To the best of our knowledge, only RSS based centralized localization schemes are developed for UOWNs. In \cite{Nasir2018limited}, we have proposed a localization scheme for UOWNs with limited connectivity. As the transmission range of users in UOWNs is limited, a multihop UOWNs is considered and the single neighborhood distances are computed by using RSS. Using these single neighborhood distances a novel distance completion strategy was used by the surface station to get the global view of the whole network. In \cite{Nasir2018twc}, we have presented an energy harvesting empowered underwater optical localization scheme where the underwater sensor nodes are able to harvest the energy from ambient underwater sources. As the nodes can harvest energy from the underwater environment, it helps to increase the localization accuracy and lifetime of the network. Based on the harvested energy availability, the sensor nodes communicate
with its neighbor nodes and computes the RSS ranges. A closed form localization
technique was developed to find the location of every optical sensor node in UOWNs. The
proposed localization technique accurately minimizes the error function by partitioning the kernel
matrix into smaller block matrices. Furthermore, a novel matrix completion strategy was introduced to
complete the missing elements in block matrices, which results in better approximation. In \cite{Saeed2017}, we have proposed an energy harvesting localization scheme for hybrid acoustic and optical underwater wireless communication system. A weighting strategy was used in \cite{Saeed2017} to give more preference to accurate measurements. 

\subsubsection{Comparison of Localization Schemes for UOWNs}
In order to compare the different localization schemes for UOWNs, we have simulated two different scenarios where the actual locations of the sensor nodes and anchor nodes are kept same in both scenarios for fair comparison.
To evaluate the performance of distributed ToA and RSS based UWONs localization schemes proposed in \cite{Akhoundi2017underwater}, we considered 50 optical sensor nodes deployed randomly in $50~m \times 50~m$ square area and 4 anchor nodes deployed at each corner of the considered area. The optical sensor nodes are able to communicate with at least three anchor nodes directly and localize itself using linear least square solution. Fig.~\ref{fig:comparedtoa} and Fig.~\ref{fig:comparedrss} shows the localization performance of the two schemes with root mean square error of 0.8 m and 1.6 m respectively.
To evaluate the performance of centralized ToA and RSS based UWONs localization schemes, we have considered the same scenario of 50 optical sensor nodes deployed randomly in $50~m \times 50~m$ square area and 4 anchor nodes deployed at each corner of the area. But here the limited transmission range of optical sensor nodes is taken into account which leads to multi-hop UOWNs. In this case, the internode single hop distances are measured by the optical sensor nodes and sent to the surface station via surface buoys. The surface station then finds out the location of each optical sensor node by using dimensionality reduction techniques and linear transformations \cite{Nasir2018limited, Nasir2018twc}.  Fig.~\ref{fig:comparectoa} and Fig.~\ref{fig:comparecrss} shows the localization performance of the two schemes with root mean square error of 0.3 m and 0.9  m respectively. Table \ref{TableUOWNsloc} summarizes the UOWNs localization schemes.

\section{Future Perspectives of UOWNs}
\label{sec:challenges}
In the following, we will advise some potential future UOWNs research directions.
\subsection{Future Directions in UOWNs Research}
\subsubsection{UOWC Channel Modeling}
To model the UOWC channel, there is still a need to further investigate and analyze new theoretical models which can either be developed analytically or computationally. The analytic models for UOWC channel are quite simple because of simplifying the complex nature of photon propagation but these models are either analytically intractable or hard to evaluate computationally. On the other hand, computational models are complex and their time complexity may not be suitable to employ in a large scale network. Therefore, modeling and performance analysis of UOWNs necessitates accurate and simple UOWC channel models as they are building blocks of UOWNs.

\subsubsection{Novel Network Protocols}
The current research on UOWNs is highly concentrated on physical layer problems, which tines out toward the higher layers. To the best of author's knowledge, the networking aspects are studied only in few papers so far \cite{Mora2013, Vavoulas2014, Akhoundi2016cellular, Jamali2016, Jamali2017, Jamali2017Photonics, celik2018modeling, Akhoundi2017underwater, Saeed2017, Nasir2018limited, Nasir2018icc, TABESHNEZHAD2017, Wang22017, SONG2017}. Noting that the limited communication range and directivity of UOWCs yield limited network connectivity, implementing UOWNs in real life necessitates adequate protocols and network architectures. 

First and foremost, UOWNs requires effective routing  algorithms in order to increase the network connectivity and performance by extending the communication range and expanding the coverage. Even though some of the potential routing protocols are highlighted in Section \ref{sec:routing}, there is no sufficient efforts toward UOWN routing techniques except \cite{celik2018modeling} which only considers a centralized routing scenario to show impact of multihop communication on network performance. Therefore, it is quite of interest to develop distributed and dynamic routing algorithms which adapt itself according to environmental and network changes. Furthermore, a novel transport layer protocols are required because UOWC channels are quite different from terrestrial and underwater acoustic wireless networks. 

\subsubsection{Cross Layer Design Issues}
Even though we have surveyed the UOWNs following a strictly layered perspective, which is traditionally employed for wired networks, considering a cross layer design could improve the overall system performance in a great extent. Indeed, QoS requirements of application layer can only be satisfied by mapping them into the lower-layer metrics such as end-to-end data rate, delay, energy efficiency, packet loss, etc. Accordingly, it is interesting to investigate a cross-layer optimization framework which adapts physical layer parameters (e.g., divergence angle, transmission power, communication range, etc.) to channel conditions and dynamically change the routing paths to satisfy QoS requirements, avoid congestion, increase the reliability, and maintain the network connectivity. 

\subsubsection{Localization in UOWNs}
Localization is of utmost importance for UOWNs where it can be used for node tracking, intruder detection, and data tagging.  A greater number of underwater applications demands for distributed localization schemes as they can provide online monitoring. Although few research work is carried out to develop distributed \cite{Akhoundi2017underwater} and centralized \cite{Nasir2018limited, Saeed2017, Nasir2018twc} localization schemes for  UOWNs.  Due to the severe UOWC channel conditions distributed localization schemes for UOWNs are challenging and needs further investigation. Limited range of UOWC links and higher energy consumption of distributed schemes led to the development of centralized UOWNs localization schemes where the localization is performed at the surface station. Centralized localization schemes are good to get the overall global view of the UOWN. Moreover, the impact of localization schemes on location-based routing and clustering for UOWNs still need to be investigated. Also, the cross-layer schemes such as the impact of link quality, connectivity, transmission range, and energy on localization performance are open issues.

\subsubsection{Practical Implementations of UOWNs}
Research on implementation of UOWNs is limited and further need to be studied. There is a dire need to develop underwater optical transceivers which can overcome the problem of link misalignment, low transmission range, low bandwidth, energy consumption, and compactness. There is a great potential to develop more advance low cost and low power underwater transmission light sources, receiving nodes, and energy harvesting systems. Testing of the UOWNs also needs to be carried out in the real underwater environment. Hybrid systems comprising of both acoustic and optical underwater wireless communication system have been introduced in \cite{Han2014, Johnson20142, Moriconi2015}. Authors in \cite{Diamant2017} have explored a statistical analogy between underwater acoustic and optical wireless links for predicting the signal to noise ratio of underwater optical links. The research on developing hybrid systems is still in its infancy and needs proper analysis. Also, the adaptive switching between an acoustic and optical mode for different operations need extra attention.
\subsubsection{Energy Harvesting for UOWNs}
Underwater optical sensor nodes have limited energy resources, which has a substantial impact on the network lifetime. Taking the engineering hardship and monetary cost of battery replacement into account, an energy self-sufficient UOWN is essential to maximize the network lifetime. In this regard, energy harvesting is a promising solution to collect energy from the ambient sources in the aquatic environment and storing it in an energy buffer. As surveyed in  \cite{Sudevalayam2011}, ongoing research efforts on terrestrial communications have shown that energy harvesting plays a significant role in enhancing the performance.  However, most of the energy harvesting techniques are designed for outdoor environments and not applicable in the aquatic environment. Recently, acoustics resonators are used in \cite{Li2016} to acquire acoustic energy from the underwater environment and harvest to the sensor nodes. A muti-source energy harvesting system was proposed in \cite{Saeed2017, Nasir2018twc} which harvest energy from multiple underwater sources such as acoustic resonators and microbial fuel cells (MFCs) \cite{Wang2015}, and harvested to the sensor nodes. Moreover, albeit the notable research body on designing different protocols for underwater communication networks, no significant research is carried out on the energy harvesting methods for UOWNs.
\subsubsection{Towards Internet of Underwater Things (IoUTs)}
There has recently been a growing interest in developing internet of underwater things (IoUTs) which can lead to enabling various underwater applications \cite{Chien2017}. In the recent past, several attempts have been made to develop routing \cite{Zhou2016, Javaid2017}, scheduling \cite{Xu2017}, and data analytic \cite{Berlian2016} techniques for IoUTs. The IoUTs research is still in its infancy and need to be more explored. Multi-hop UOWNs can be a potential technology to implement the IoUTs because of its low power consumption and higher data rate.

\section{Conclusions}
\label{sec:conc}
In this paper, we have presented a comprehensive survey of  underwater optical wireless networks (UOWNs) research. This survey covers different aspects of cutting-edge UOWNs from a layer by layer perspective. Firstly, each layer of UOWNs such as physical, data link, networking, transport, and application layers are briefly presented and then localization techniques for UOWNs are surveyed. We started with defining different possible architectures for UOWNs and then the issues related to each layer are thoroughly discussed.  Besides providing the technical background on UOWNs, we have also provided details on the challenges to design a practical UOWN.  Additionally, localization is an important task where the location of the underwater optical sensor node can be used for node tracking, intruder detection, and data tagging. Conventional terrestrial and underwater acoustic localization schemes do not meet the requirements of UOWNs where the unfavorable behavior of UOWC asks for novel localization schemes. Even though we have surveyed the state of the art localization schemes for UOWNs, the subject still remains open and requires to develop accurate and practical localization schemes. To reach this goal,  communication, networking, and localization in UOWNs require more research efforts.  In short, this survey can help the novice readers to get an insight of each layer and localization of UOWNs which can lead to the development of practical UOWNs.

\bibliographystyle{../bib/IEEEtran}
\bibliography{../bib/IEEEabrv,../bib/nasir_ref}

\begin{IEEEbiography}[{\includegraphics[width=1in,height=1.25in]{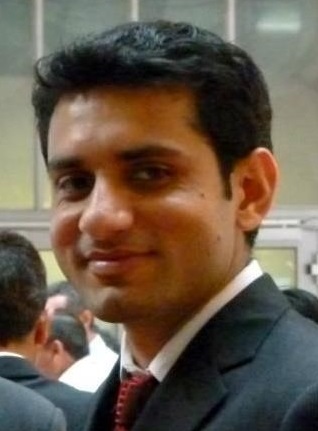}}]{Nasir Saeed}(S'14-M'16) received his Bachelors of Telecommunication degree from N.W.F.P University of Engineering and Technology, Peshawar, Pakistan, in 2009 and received Masters degree in satellite navigation from Polito di Torino, Italy, in 2012. He received his Ph.D. degree in electronics
and communication engineering from Hanyang University, Seoul, South Korea in 2015. He was an assistant professor at the
Department of Electrical Engineering, Gandhara Institute of Science and IT, Peshawar, Pakistan from August 2015 to September 2016. Dr. Saeed worked as an assistant professor at IQRA National University, Peshawar, Pakistan from October 2017 to July 2017. He is currently a postdoctoral research fellow at Communication Theory Lab, King Abdullah University of Science and Technology (KAUST).   His current areas of interest include cognitive radio networks, underwater optical wireless networks, and localization.
\end{IEEEbiography}

\begin{IEEEbiography}[{\includegraphics[width=1in,height=1.25in]{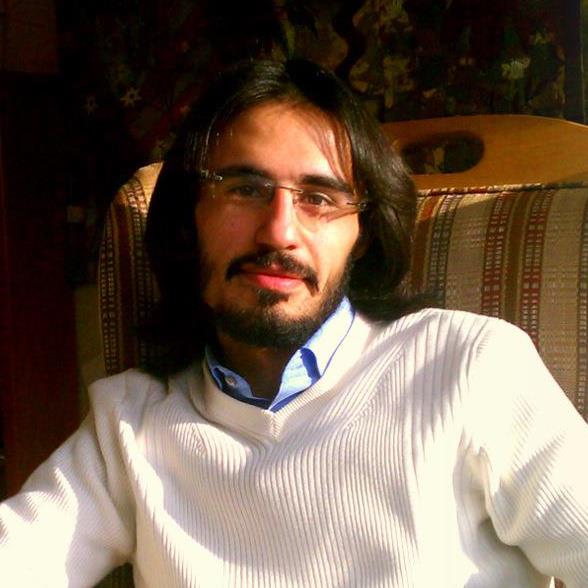}}]{Abdulkadir Celik}
(S'14-M'16) received the B.S. degree in electrical-electronics engineering from Selcuk University, Turkey [2009], the M.S. degree in electrical engineering [2013], the M.S. degree in computer engineering [2015], and the Ph.D. degree in co-majors of electrical engineering and computer engineering [2016], all from Iowa State University, Ames, IA. Dr. Celik is currently a postdoctoral research fellow at Communication Theory Laboratory of King Abdullah University of Science and Technology (KAUST). His current research interests include but not limited to cognitive radio networks, green communications, non-orthogonal multiple access, D2D communications, heterogeneous networks, and optical wireless communications and networking for data centers and underwater sensor networks. 
\end{IEEEbiography}

\begin{IEEEbiography}[{\includegraphics[width=1in,height=1.25in]{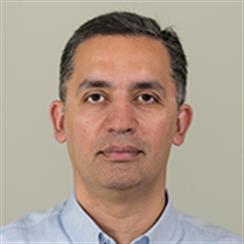}}]{Tareq Y. Al-Naffouri }
(M’10) Tareq Al-Naffouri received the B.S. degrees in mathematics and electrical engineering (with first honors) from King Fahd University of Petroleum and Minerals, Dhahran, Saudi Arabia, the M.S. degree in electrical engineering from the Georgia Institute of Technology, Atlanta, in 1998, and the Ph.D. degree in electrical engineering from Stanford University, Stanford, CA, in 2004. He was a visiting scholar at California Institute of Technology, Pasadena, CA, from January to August 2005 and during summer 2006. He was a Fulbright Scholar at the University of Southern California from February to September 2008. He has held internship positions at NEC Research Labs, Tokyo, Japan, in 1998, Adaptive Systems Lab, University of California at Los Angeles in 1999, National Semiconductor, Santa Clara, CA, in 2001 and 2002, and Beceem Communications Santa Clara, CA, in 2004. He is currently an Associate professor at the Electrical Engineering Department, King Abdullah University of Science and Technology (KAUST). His research interests lie in the areas of sparse, adaptive, and statistical signal processing and their applications and in network information theory. He has over 150 publications in journal and conference proceedings, 9 standard contributions, 10 issued patents, and 6 pending. Dr. Al-Naffouri is the recipient of the IEEE Education Society Chapter Achievement Award in 2008 and Al-Marai Award for innovative research in communication in 2009. Dr. Al-Naffouri has also been serving as an Associate Editor of Transactions on Signal Processing since August 2013.
\end{IEEEbiography}

\begin{IEEEbiography}[{\includegraphics[width=1in,height=1.25in]{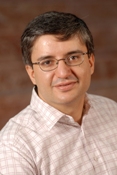}}]{Mohamed-Slim Alouini}
(S'94-M'98-SM'03-F'09) was born in Tunis, Tunisia. He received the Ph.D. degree in Electrical Engineering from the California Institute of Technology (Caltech), Pasadena, CA, USA, in 1998. He served as a faculty member in the University of Minnesota, Minneapolis, MN, USA, then in the Texas A\&M University at Qatar, Education City, Doha, Qatar before joining King Abdullah University of Science and Technology (KAUST), Thuwal, Makkah Province, Saudi Arabia as a Professor of Electrical Engineering in 2009. His current research interests include the modeling, design, and performance analysis of wireless communication systems.
\end{IEEEbiography}

\end{document}